\documentclass[fleqn]{mn}

\usepackage{amsfonts}
\usepackage{amsmath}
\usepackage{graphicx}
\usepackage{subfigure}
\usepackage{multicol}

\font\eightrm=cmr8 scaled 1000

\def\gsim{\;\lower4pt\hbox{${\buildrel\displaystyle >\over\sim}$}\;}
\def\lsim{\;\lower4pt\hbox{${\buildrel\displaystyle <\over\sim}$}\;}
\def\grls{\;\lower4pt\hbox{${\buildrel\displaystyle >\over <}$}\;}

\begin{document}

\title[Stationary Configurations of Two-Fluid Scale-Free Discs]
{Gravitationally coupled scale-free discs}
\author[Y. Shen and Y.-Q. Lou]
{Yue Shen$^{1}$ and Yu-Qing Lou$^{1,2,3}$ \\
$^{1}$Physics Department, The Tsinghua Center for
Astrophysics, Tsinghua University, Beijing 100084, China\\
$^{2}$National Astronomical Observatories, Chinese Academy
of Sciences, A20, Datun Road, Beijing 100012, China\\
$^{3}$Department of Astronomy and Astrophysics, The University
of Chicago, 5640 S. Ellis Ave., Chicago, IL 60637 USA }
\date{Accepted 2004 ... Received 2004 ...;
in original form 2003 ... } \maketitle

\begin{abstract}
In a composite fluid system of two gravitationally coupled barotropic
scale-free discs bearing a rotation curve $v\propto r^{-\beta}$ and
a power-law surface mass density $\Sigma_0\propto r^{-\alpha}$ with
$\alpha=1+2\beta$, we construct coplanar stationary aligned and
spiral perturbation configurations in the two discs. Due to the
mutual gravitational interaction, there are two independent classes
of perturbation modes with surface mass density disturbances in the
two coupled discs being either in-phase or out-of-phase. We derive
analytical criteria for such perturbation
modes to exist and show numerical examples. We compute the aligned and
spiral perturbation modes systematically to explore the entire parameter
regime. For the axisymmetric $m=0$ case with radial oscillations, there
are two unstable regimes of ring-fragmentation and collapse corresponding
to short and long radial wavelengths, respectively. Only within a certain
range of the rotation parameter $D_s^2$ (square of the effective Mach
number for the stellar disc), can a composite disc system be stable
against all axisymmetric perturbations. Compared with a single-disc
system, the coupled two-disc system becomes less stable against such
axisymmetric instabilities. Our investigation generalizes the previous
work of Syer \& Tremaine on the single-disc case and of Lou \& Shen on
two coupled singular isothermal discs (SIDs). Non-axisymmetric instabilities
are briefly discussed. These stationary models for various large-scale
patterns and morphologies may be useful in contexts of disc galaxies.
\end{abstract}


\section{Introduction}

Rotating disc systems on various spatial scales are of broad
astrophysical interest since most spiral galaxies, various
binary accretion systems, and proto-stellar and proto-planetary
systems appear grossly in disc shape, which is believed to be a
key intermediate stage that many astrophysical processes may
attain (e.g., the phase of gas accretion onto a central black
hole that drives an active galactic nucleus). It is therefore
important to study the dynamical processes in disc systems for
theoretical understanding and for astrophysical applications.
Lin, Shu and co-workers pioneered the classic density wave
theory in a differentially rotating disc system (Lin \& Shu
1964, 1966; Lin 1987) and achieved a great deal of success in
understanding the basic physics and dynamics of spiral galaxies
(Binney \& Tremaine 1987; Bertin \& Lin 1996). The disc
perturbation theory (linear or even nonlinear) has proven to
show broad potentials in dealing with problems of shapes and
shaping, large-scale structures, instabilities in spiral
galaxies and in other disc systems involving self-gravity,
differential rotation and magnetic fields (e.g., Binney \&
Tremaine 1987; Bertin \& Lin 1996; Balbus \& Hawley 1998), as
well as problems of angular momentum transfer in accretion
discs or planetary discs (e.g., Lynden-Bell \& Kalnjas 1972;
Goldreich \& Tremaine 1978). In many cases, perturbations
developed in earlier stages prior to the moment when a system
experiences more dramatic or violent processes (e.g., collapses),
are crucial for dynamical evolution and are therefore worthwhile
to pursue for their physical consequences.

Among various problems in disc dynamics, a scale-free disc is often
picked up by theorists for its relative simplicity and is explored
as a powerful vehicle for a possible global analytical analysis. The
term `scale-free' here means that all physical quantities in the disc
system vary as powers of cylindrical radius $r$ (e.g., the linear
velocity of disc rotation $v\propto r^{-\beta}$ and the equilibrium
surface mass density $\Sigma_0\propto r^{-\alpha}$)
with $\alpha$ and $\beta$ being two related exponents. The two
examples in mind are the rigidly rotating Maclaurin discs
and Kalnajs discs (Kalnajs 1972; Binney \& Tremaine 1987), where
the angular rotation speed $\Omega$ remains constant with
$v\propto r$. These discs are also known to have analytical normal
mode spectrum, but are thought to rarely exist in nature. In
contrast, thin discs with more or less flat rotation curves (i.e.,
$v=\hbox{constant}$) are common in most normal spiral galaxies as
an important evidence for unseen masses of dark matter haloes
associated with spiral galaxies. Besides these two limiting
classes of discs with rigid and flat rotations, differentially
rotating discs may have a rotation curve $\propto r^{-\beta}$
with a rotation index $\beta$ satisfying $-1<\beta<1/2$ and
with $\beta=1/2$ corresponding to a well-known Keplerian disc
system.\footnote{Scale-free disc solutions do exist for
$\beta$ in the range of $-1/4<\beta<1/2$ for warm discs
according to our analysis. }

Discs with complete flat rotation curves are usually referred to
as singular isothermal discs (SIDs), which form the simplest class
in the family of scale-free discs. Since the introduction of SIDs
by Mestel (1963), this idealized theoretical model has attracted
a considerable attention in various astrophysical contexts of disc
dynamics (e.g., Zang 1976; Toomre 1977; Lemos, Kalnajs \&
Lynden-Bell 1991; Lynden-Bell \& Lemos 1993; Syer \& Tremaine
1996; Goodman \& Evans 1999; Shu et al. 2000; Lou 2002; Lou \&
Fan 2002; Lou \& Shen 2003; Shen \& Lou 2003; Lou \& Zou 2004;
Lou \& Wu 2004). In the SID model, both the angular rotation
speed $\Omega$ and the equilibrium surface mass density
$\Sigma_0$ scale as $r^{-1}$, which is a scale-free condition.

There has long been a paradox or controversy regarding stability
analyses of scale-free discs because of the singularity as
$r\rightarrow 0$. Starting from Zang (1976), who investigated a
stellar SID numerically and argued that a scale-free disc can
support no normal modes unless central cut-outs were introduced
to remove the central singularity and to prescribe inner
boundary conditions.
Evans \& Read (1998a, b) adopted Zang's approach to construct
power-law discs with central cut-outs and examined numerically
discrete growing normal modes in an `isothermal' stellar disc
(i.e. with a constant velocity dispersion).
In contrast, Lynden-Bell \& Lemos (1993) claimed the presence
of a continuum of unstable normal modes for an unmodified SID.
By specifying the phase of a postulated reflection of spiral
waves from the origin $r=0$, Goodman \& Evans (1999) could
define discrete normal modes for an unmodified gaseous SID.
More recently, Shu et al. (2000) investigated spiral density
wave transmission and reflection at the corotation circle and
speculated that the swing amplification process (Goldreich
\& Lynden-Bell 1965; Toomre 1981; Binney \& Tremaine 1987;
Fan \& Lou 1997) across corotation allows a continuum of normal
modes while proper `boundary conditions' may select from this
continuum a discrete spectrum of unstable normal modes.

Besides normal modes analyses, stationary perturbation
configurations or zero-frequency neutral modes are emphasized
as marginal instability modes in scale-free discs (e.g.,
Lemos, Kalnajs \& Lynden-Bell 1991; Syer \& Tremaine 1996;
Shu et al. 2000; Lou \& Shen 2003; Shen \& Lou 2003; Lou \&
Zou 2004). It is believed that axisymmetric instabilities set
in through transitions of such neutral modes (Lynden-Bell \&
Ostriker 1967; Lemos, Kalnajs \& Lynden-Bell 1991; Shu et al.
2000). By using properties of zero-frequency modes, Shu et al.
(2000) further claimed that logarithmic spiral modes of
stationary configurations also signal onsets of non-axisymmetric
instabilities, a result compatible with the criterion of Goodman
\& Evans (1999) for instabilities in their normal mode treatment.
Recently, Lou \& Shen (2003) extended results of Shu et al.
(2000) in a gravitationally coupled composite disc system of
one gaseous SID and one stellar SID in a two-fluid formalism.
Stationary coplanar configurations were readily constructed
in such a composite SID system.

The objective of this paper is to construct scale-free stationary
configurations in a two-fluid stellar and gaseous disc system with
a more general rotation curve $v\propto r^{-\beta}$ and equilibrium
surface mass density profile $\Sigma_0\propto r^{-\alpha}$ using
a barotropic equation of state. The departure from the SID model
will cause stationary configurations to vary significantly from
those derived previously (Lou \& Shen 2003) in some circumstances.

This paper is organized in the following way. In Section 2,
we describe the theoretical formalism, obtain the background
rotational equilibrium and present linear coplanar perturbation
equations. In Section 3, we discuss aligned and spiral
disturbances in details and derive analytical criteria for
stationary configurations. We summarize our results and give
discussions in Section 4. Specific technical details are
contained in the Appendices for the convenience of references.

\section{Two-fluid formalism}

We adopt the two-fluid formalism sufficient for large-scale
stationary aligned and unaligned coplanar disturbances (Kalnajs
1973) in a background rotational equilibrium with axisymmetry (Jog
\& Solomon 1984a, b; Elmegreen 1995; Jog 1996; Lou \& Shen 2003;
Shen \& Lou 2003). 
In this section, we
provide the governing equations for a two-fluid composite disc
system, composed of a stellar disc and a gaseous disc presumed
to be razor-thin. Given qualifications and assumptions,
equilibrium properties of both barotropic discs characterized
by rotation curves $v\propto r^{-\beta}$ and power-law surface
mass densities $\Sigma_0\propto r^{-\alpha}$ with different
proportional constants are described. We then derive linear
coplanar perturbation equations in such a composite disc system.

It is important to note that the fluid formalism is well suited
for large-scale dynamical behaviours in a gaseous disc but is an
approximation to describe the large-scale dynamics in a stellar
disc. The latter would be more appropriately modelled by the
coupled collisionless Boltzmann equation (i.e., Vlasov equation)
and Poisson equation (e.g. Julian \& Toomre 1966; Lin \& Shu 1966;
Zang 1976; Nicholson 1983; Binney \& Tremaine 1987; Evans \& Read
1998a, b), where an equilibrium distribution function (DF) is
perturbed and the coupled Vlasov-Poisson equations are linearized
to give the density wave dispersion relation. The introduction of
an isotropic effective pressure term to mimic random stellar
motions in a collisionless system is justified by the qualitative
agreement between a hydrodynamic formalism and a DF approach for
treating collective particle dynamical behaviours (e.g., Berman \&
Mark 1977). One illustrating analogy between fluid and DF
approaches is perhaps the density wave dispersion relation in the
WKBJ regime, namely, $(m\Omega-\omega)^2=\kappa^2-2\pi
G\Sigma_0|k|+k^2v_s^2$ in a gaseous disc and
$(m\Omega-\omega)^2=\kappa^2-2\pi G\Sigma_0|k|F$
in a stellar disc where $F$ is the so-called reduction factor. In
general, $F$ is determined by the specific form of the DF, tends
to reduce the gravity response and functions as an extra pressure
term. The $Q$ parameter in the Toomre criterion for local
axisymmetric stability (Safronov 1960) also bears strikingly
similar forms for both gaseous and stellar discs where for the
latter the radial stellar velocity dispersion mimics the sound
speed. This provides an empirical rationale for treating the
stellar disc by a simpler fluid approximation when dealing with
the global axisymmetric stability for a composite disc system
although the results may be quantitatively modified when we treat
a stellar disc using the more exact (and more formidable) DF
approach. The major deviation of a DF approach from the fluid
formalism may occur in handling the corotation and Lindblad
resonances. Therefore, for behaviours near the resonances,
we need to rely on the DF approach for a stellar disc. In
the present context of constructing large-scale stationary
configurations, such resonances do not arise and the simpler
two-fluid formalism for a composite disc system suffices.

\subsection{Basic Nonlinear Two-Fluid Equations }

We approximate both discs as razor-thin (i.e., infinitesimally
thin) discs and use either superscript or subscript $s$ and $g$ to
denote associations with the stellar and gaseous discs,
respectively. The large-scale dynamic coupling between the two
discs is due to the mutual gravitational interaction. For
large-scale perturbations, we ignore diffusive effects such as
viscosity, resistivity and thermal conduction, etc. Then the set
of coplanar fluid equations for the stellar disc and the gaseous
disc can be written out using the system of cylindrical
coordinates $(r,\ \theta,\ z)$ in the $z=0$ plane, such as
\begin{equation}
\frac{\partial \Sigma^{i}}{\partial t} +\frac{1}{r} \frac{\partial
}{\partial r}(r\Sigma^{i}u^{i}) +\frac{1}{r^{2}}\frac{\partial
}{\partial \theta } (\Sigma^{i}j^{i})=0\ ,
\end{equation}
\begin{equation}
\frac{\partial u^{i}}{\partial t} +u^{i}\frac{\partial u^{i}}
{\partial r} +\frac{j^{i}}{r^{2}}\frac{\partial u^{i}} {\partial
\theta }-\frac{j^{i2}}{r^{3}} =-\frac{1}{\Sigma^{i}}\frac{\partial
\Pi^i}{\partial r} -\frac{\partial \phi }{\partial r}\ ,
\end{equation}
\begin{equation}
\frac{\partial j^{i}}{\partial t}+u^{i}\frac{\partial
j^{i}}{\partial r} +\frac{j^{i}}{r^{2}}\frac{\partial
j^{i}}{\partial \theta } =-\frac{1}{\Sigma^{i}}
\frac{\partial\Pi^i } {\partial \theta } -\frac{\partial \phi
}{\partial \theta }\ ,
\end{equation}
where $i=s,\ g$ denotes the stellar or gaseous disc here and
throughout this paper.
The coupling between the two discs is due to the gravitational
potential $\phi$ through Poisson's integral
\begin{eqnarray}\label{fish}
\phi(r,\theta,t)= \oint\!d\psi\!\!\int_0^{\infty}
\!\!\frac{-G\Sigma (r^{\prime },\psi ,t)r^{\prime }
dr^{\prime }}{\left[ r^{\prime 2}+r^{2}-2rr^{\prime }
\cos (\psi -\theta )\right]^{1/2}}\ ,
\end{eqnarray}
where $\Sigma =\Sigma^{s}+\Sigma^{g}$ is the total surface
mass density. In a disc galaxy, the gravitational potential
associated with a dark matter halo plays an important role.
We shall take that into account later in a composite system
of two coupled partial discs.
\footnote{The construction of a composite system of two partial
discs are described in Section 5. In our notation, the potential
ratio ${\cal F}=\phi_0/(\phi_0+\Phi)$ where $\phi_0$ stands for
the equilibrium background potential arising from the two discs
and $\Phi$ stands for that arising from an axisymmetric dark
matter halo. Syer \& Tremaine (1996) used dimensionless ratio
$f=\Phi/\phi_0$ instead.} In equations $(1)-(4)$, $\Sigma^{i}$ is
the disc surface mass density, $u^{i}$ is the radial component of
the fluid velocity, $j^{i}$ is the $z-$component of the specific
angular momentum about the $z-$axis, and $\Pi^i$ is the
two-dimensional effective pressure in the barotropic
approximation, $\phi$ is the total gravitational potential
expressed in terms of Poisson's integral containing the total
surface mass density $\Sigma =\Sigma^{s}+\Sigma^{g}$ in a
composite disc.
Here, we assume that the stellar and gaseous disks
interact primarily through the mutual gravitational coupling on
large scales (Jog \& Solomon 1984a,b; Bertin \& Romeo 1988; Romeo
1992; Elmegreen 1995; Jog 1996; Lou \& Fan 1998; Lou \& Shen 2003;
Shen \& Lou 2003; Lou \& Zou 2004).

A barotropic equation of state assumes the relation
between the pressure $\Pi$ and the surface mass
density $\Sigma$, namely,
\begin{equation}\label{poly}
\Pi=K\Sigma^{n}\ ,
\end{equation}
where $K>0$ (i.e., warm discs) and $n>0$ are two constant
coefficients and subscripts $s$ (stellar disc) and $g$ (gaseous
disc) are implicit. This directly leads to the definition of sound
speed $a$ (in a stellar disc the velocity dispersion mimics the
sound speed) by
\begin{equation}
a^2=\frac{d\Pi_0}{d\Sigma_0}=n K\Sigma_0^{n-1},
\end{equation}
which gives $a\propto\Sigma_0^{(n-1)/2}$ with $n=1$
for an isothermal sound speed.

\subsection{Rotational Equilibrium of Axisymmetry}

It is straightforward to derive properties of the background
rotational equilibrium of axisymmetry for the two gravitationally
coupled discs, with physical variables denoted by a subscript $0$.
Let us first clarify several basic properties of a composite
system of two scale-free discs. The background equilibrium surface
mass densities of the two fluid discs $\Sigma_0^s$ and
$\Sigma_0^g$ take the power-law form of $\propto r^{-\alpha}$ with
a common $\alpha$ exponent yet different proportional
coefficients, while the disc rotation curves $v_s$ and $v_g$ take
the power-law form of $\propto r^{-\beta}$ with a common $\beta$
exponent yet different proportional coefficients. The special case
of $\beta=0$ gives two flat rotation curves with $v_s\neq v_g$
being allowed in general (Lou \& Shen 2003). For the background
equilibrium, we also have $u_0^s=u_0^g=0$, $j_0^s=rv_s$ and
$j_0^g=rv_g$. By imposing these conditions in equations (1)$-$(4)
[particularly radial momentum eqn. (2)], we obtain
\begin{equation}\label{relation0}
v_s^2+\alpha n K_s(\Sigma_0^s)^{n-1}=v_g^2+\alpha n
K_g(\Sigma_0^g)^{n-1}=r\frac{d\phi_0}{dr}\ .
\end{equation}
To compute the gravitational potential $\phi_0$ arising
from the equilibrium total surface mass density
\begin{equation}
\Sigma_0=\Sigma_0^s+\Sigma_0^g
=\sigma_0^sr^{-\alpha}+\sigma_0^gr^{-\alpha}
\end{equation}
where $\sigma_0^s$ and $\sigma_0^g$ are two constant coefficients,
we simply take equation $(A5)$ in Appendix A of Syer \& Tremaine
(1996) and readily obtain
\begin{equation}
\phi_0=-2\pi Gr(\Sigma_0^s+\Sigma_0^g){\cal P}_0\ ,
\end{equation}
where we introduce an auxiliary parameter function
\begin{equation}
{\cal P}_0\equiv \frac{\Gamma(-\alpha/2+1)
\Gamma(\alpha/2-{1}/{2})}
{2\Gamma(-\alpha/2+{3}/{2})\Gamma(\alpha/2)}
\end{equation}
(Kalnajs 1971).

The requirement of radial force balance (\ref{relation0}) for all
radii (i.e., the scale-free condition) implies
\begin{equation}
2\beta=\alpha(n-1)=\alpha-1\ ,
\end{equation}
which gives the relationship among
$\alpha$, $\beta$ and $n$, namely
\begin{equation}
\alpha=1+2\beta\qquad \hbox{ and }
\qquad n=\frac{1+4\beta}{1+2\beta}
\end{equation}
(Syer \& Tremaine 1996). It follows from $n>0$ in
barotropic equation of state (\ref{poly}) for warm
discs that $\beta >-1/4$. For cold discs (i.e.,
$K\rightarrow 0$), this inequality is unnecessary.

As discussed in Syer \& Tremaine (1996), mass distributions with
$\beta>1/2$ ($\alpha>2$) would be unphysical because they contain
infinite point masses. Furthermore, Poisson integral (\ref{fish})
for $\phi_0$ converges for $1<\alpha<2$ and thus $0<\beta<1/2$ for
a system of axisymmetry; the range of $\alpha$ (and thus of
$\beta$) is broader for nonaxisymmetric systems. However, the
total force arising from axisymmetric equilibrium surface mass
densities remains finite in an extended range of
$\beta\in(-1/2,1/2)$ ($0<\alpha<2$).
In summary, we therefore have $-1/4<\beta<1/2$ [the
left bound is implied by $n>0$ for warm discs and the right bound
is required such that the central point mass will not diverge
(Syer \& Tremaine 1996)]. For cold discs (i.e., $K=0$), the
$\beta$ range can be extended to $-1/2<\beta<1/2$. When $\beta=0$
for flat rotation curves, we have surface mass densities
proportional to $r^{-1}$ corresponding to a composite system of
two SIDs (Lou \& Shen 2003; Shen \& Lou 2003; Lou \& Zou 2004; Lou
\& Wu 2004).

According to equilibrium condition (\ref{relation0}), we have
\begin{equation}\label{ADeqn0}
{\cal V}_s^2+A_s^2={\cal V}_g^2+A_g^2=2\pi G(2\beta{\cal
P}_0)r^{1+2\beta}(\Sigma_0^s+\Sigma_0^g)\ ,
\end{equation}
where $v={\cal V}r^{-\beta}$ and $a^2=A^2r^{-2\beta}/(1+2\beta)$
with ${\cal V}$ and $A$ being two constant coefficients. By
introducing ${\cal V}\equiv AD$ to define a dimensionless
parameter $D$, we obtain
\begin{equation}\label{2sigma}
\begin{split}
&\Sigma_0^s=\frac{A_s^2(D_s^2+1)r^{-(1+2\beta)}} {2\pi
G(2\beta{\cal P}_0)(1+\delta)}\ ,
\\
&\Sigma_0^g=\frac{A_g^2(D_g^2+1)\delta r^{-(1+2\beta)}} {2\pi
G(2\beta{\cal P}_0)(1+\delta)}\ ,
\end{split}
\end{equation}
where $\delta\equiv\Sigma_0^g/\Sigma_0^s$ is the ratio of
the surface mass density of the gaseous disc to that of the
stellar disc. We note that the value of $2\beta{\cal P}_0$
falls within $(0,\infty)$ for $\beta\in(-1/4,1/2)$ and is
equal to 1 when $\beta=0$ for the case of SIDs.

An equivalent version of requirement (\ref{ADeqn0}) is
\begin{equation}\label{ADeqn}
A_s^2(D_s^2+1)=A_g^2(D_g^2+1)\ ,
\end{equation}
where $D_s$ and $D_g$ are two dimensionless rotation parameters
and $\eta\equiv A_s^2/A_g^2=a_s^2/a_g^2$ is the square of the
ratio of the velocity dispersion in the stellar disc to the sound
speed in the gaseous disc. Note that $A$ is actually related to
the sound speed $a\propto r^{-\beta}$ [but scaled by a factor
$(1+2\beta)^{1/2}$] and the parameter $D$ is essentially the
effective Mach number for disc rotation. We are going to express
other equilibrium physical variables in terms of $A$ and $D$.
Besides, we have also introduced two dimensionless parameters to
compare properties of the two discs. The first one is the surface
mass density ratio $\delta\equiv\Sigma_0^g/\Sigma_0^s$. The second
parameter is the square of the ratio of the effective sound speeds
in two discs $\eta\equiv A_s^2/A_g^2$. For disc galaxies, ratio
$\delta$ can be either greater or less than $1$ depending on whether
the system is stellar matter dominant or gas material dominant
(in the early universe). Without loss of generality, we may take
$\eta>1$ as the situation is symmetric for $\eta<1$ and typically
the stellar velocity dispersion (mimicked by a sound speed) in the
stellar disc is greater than the sound speed in the gaseous disc.
The special case of $\eta=1$ should give some familiar results of
a single disc except for an additional mode due to gravitational
coupling, as we have already learned from the simpler case of two
coupled SIDs (Lou \& Fan 1998; Lou \& Shen 2003).

The specific $z-$component angular momenta ($j_0^s$
and $j_0^{g}$) and the sound speeds ($a_s$ and $a_g$)
of the two discs in an equilibrium state simply read
\begin{equation}
j_0^i=A_iD_ir^{1-\beta}\ , 
\end{equation}
\begin{equation}
\begin{split}
a_i^2=n
K_i(\Sigma_0^i)^{n-1}=A_i^2/[(1+2\beta)r^{2\beta}]\ .\\
\end{split}
\end{equation}
Similarly, the disc angular rotation speed
$\Omega\equiv j_0/r^2$ and the epicyclic frequency
$\kappa\equiv [(2\Omega/r)d(r^2\Omega)/dr]^{1/2}$
are expressed in terms of two dimensionless
parameters $A$ and $D$ as
\begin{equation}
\begin{split}
\Omega_i=A_iD_ir^{-1-\beta}\ ,\qquad
\kappa_i=[2(1-\beta)]^{1/2}\Omega_i\ ,\\
\end{split}
\end{equation}
and therefore we have $dj_0/dr=(1-\beta)v=r\kappa^2/(2\Omega)$
that simplifies the linear perturbation equations displayed
in the next subsection.

For the convenience of comparison and cross referencing, we note
that our chosen notations for parameters have counterparts in
those adopted by previous authors (Lemos, Kalnajs \& Lynden-Bell
1991; Syer \& Tremaine 1996). In Lemos et al. (1991), their
notations $\sigma$ and $v$ stand for
$$
\sigma^2\rightarrow\frac{A^2}{(1+4\beta)r^{2\beta}}\ ,
\qquad
v^2\rightarrow\frac{A^2D^2}{r^{2\beta}}\ ,
$$
$$
\frac{\sigma}{(\sigma^2+v^2)^{1/2}}
\rightarrow\frac{1}{[1+(1+4\beta)D^2]^{1/2}}\
$$
here. While in Syer \& Tremaine (1996),
their notation $w$ stands for
$$
w\rightarrow\frac{1}{(1+2\beta)D^2}\
$$
for a full disc with their $f=0$. These various adopted
notations are relevant to the case of a single disc.
All authors arrived at the same prescription for an
axisymmetric background in rotational equilibrium.

\subsection{Equations for Linear Coplanar Perturbations }

For small coplanar perturbations in a composite disc system
denoted by subscript $1$ associated with relevant physical
variables, the perturbation equations can be readily derived by
linearizing the basic equations (1)$-$(4), namely
\begin{equation}\label{perturbs0}
\begin{split}
&\frac{\partial \Sigma_1^{i}}{\partial t}
+\frac{1}{r}\frac{\partial }{\partial r}(r\Sigma_0^{i}u_1^{i})
+\Omega_i\frac{\partial \Sigma_1^i}
{\partial\theta}+\frac{\Sigma_0^{i}}{r^2}\frac{\partial
j_1^{i}}{\partial \theta}=0\ ,\\
&\frac{\partial u_1^{i}}{\partial t}+\Omega_i\frac{\partial
u_1^{i}} {\partial\theta}-2\Omega_i\frac{j_1^{i}}{r}
=-\frac{\partial}{\partial r}
\bigg(a_i^2\frac{\Sigma_1^{i}}{\Sigma_0^{i}}+\phi_1\bigg)\ ,\\
&\frac{\partial j_1^{i}}{\partial t}
+\frac{r\kappa_i^2}{2\Omega_i}u_1^{i} +\Omega_i\frac{\partial
j_1^{i}} {\partial \theta}=-\frac{\partial}{\partial
\theta}\bigg(a_i^2\frac{\Sigma_1^{i}}{\Sigma_0^{i}}+\phi_1\bigg)\
,
\end{split}
\end{equation}
for coplanar perturbations in the stellar disc and the gaseous
disc,
with the total gravitational potential perturbation given by
\begin{equation}\label{perturbV0}
\phi_1(r,\theta,t)=
\oint\!d\psi\!\!\int_0^{\infty}\frac{-G(\Sigma_1^{s}
+\Sigma_1^{g})r^{\prime}dr^{\prime }}{\left[ r^{\prime 2}
+r^{2}-2rr^{\prime }\cos (\psi -\theta)\right]^{1/2}}\ .
\end{equation}
Assuming a Fourier component form of exp$[i(\omega t-m\theta)]$
periodic in time $t$ and in azimuthal angle $\theta$ for all
perturbation variables with $m\ge 0$, we write for coplanar
perturbations in the stellar and gaseous discs in the forms of
\begin{equation}\label{fouriers}
\begin{split}
&\Sigma_1^i=\mu^i(r)\hbox{exp}[\hbox{i}(\omega t-m\theta)]\ ,\\
&u_1^i=U^i(r)\hbox{exp}[\hbox{i}(\omega t-m\theta)]\ ,\\
&j_1^i=J^i(r)\hbox{exp}[\hbox{i}(\omega t-m\theta)]\ ,\\
\end{split}
\end{equation}
with the total gravitational potential
perturbation in the form of
\begin{equation}\label{fourierV}
\phi_1=V(r)\hbox{exp}[\hbox{i}(\omega t-m\theta)]
\end{equation}
within the disc plane at $z=0$ (we discriminate the imaginary unit
$\hbox{i}$ and sub- or superscript $i$ for two discs). By
substituting expressions (\ref{fouriers})$-$(\ref{fourierV}) into
equations (\ref{perturbs0})$-$(\ref{perturbV0}), we readily derive
for the stellar and gaseous discs
\begin{equation}\label{perturbs1}
\begin{split}
&\hbox{i}(\omega-m\Omega_i)\mu^i+\frac{1}{r}\frac{d}{dr}(r\Sigma_0^iU^i)
-\hbox{i}m\Sigma_0^i\frac{J^i}{r^2}=0\ ,\\
&\hbox{i}(\omega-m\Omega_i)U^i-2\Omega_i\frac{J^i}{r}
=-\frac{d}{dr}\bigg(a_i^2\frac{\mu^i}{\Sigma_0^i}+V\bigg)\ ,\\
&\hbox{i}(\omega-m\Omega_i)J^i+\frac{r\kappa_i^2}{2\Omega_i}U^i
=\hbox{i}m\bigg(a_i^2\frac{\mu^i}{\Sigma_0^i}+V\bigg)\ ,
\end{split}
\end{equation}
and for the total gravitational potential perturbation
\begin{equation}\label{perturbV1}
V(r)=\oint\!d\psi\!\!\int_0^{\infty}\frac{-G(\mu^{s}+\mu^{g})\cos
(m\psi)r^{\prime }dr^{\prime }}{(r^{\prime 2}+r^{2}-2rr^{\prime }\cos
\psi)^{1/2}}\ .
\end{equation}
We now use the last two equations in (\ref{perturbs1}) to express
$U$ and $J$ in terms of $\Psi\equiv a^2\mu/\Sigma_0+V$ for the
stellar and gaseous discs
\begin{equation}\label{subUJs}
\begin{split}
&U^i=\frac{\hbox{i}}{(\omega-m\Omega_i)^2-\kappa_i^2}
\bigg[-2\Omega_i\frac{m}{r}
+(\omega-m\Omega_i)\frac{d}{dr}\bigg]\Psi^i\ ,\\
&\frac{J^i}{r}=\frac{1}{(\omega-m\Omega_i)^2-\kappa_i^2}
\bigg[(\omega-m\Omega_i)\frac{m}{r}-\frac{\kappa_i^2}
{2\Omega_i}\frac{d}{dr}\bigg]\Psi^i\ .
\end{split}
\end{equation}
%
Substitution of expressions (\ref{subUJs}) into the first equation
of (\ref{perturbs1}) leads to
\begin{eqnarray}\label{perturbs2}
& &\!\!\!\!\!\!\!\!\! 0=(\omega -m\Omega_i )\mu^{i}
+\frac{1}{r}\frac{d}{dr} \nonumber \\ & & \!\!\!\!\!\!\!\!\!\!\!\!
\times\bigg\{\frac{r\Sigma_{0}^{i}} {(\omega -m\Omega_i
)^{2}-\kappa_i^{2}}\bigg[-2\Omega_i\frac{m}{r}+(\omega -m\Omega_i
) \frac{d}{dr}\bigg]\Psi^i\bigg\}
\nonumber \\ & & \!\!\!\!\!\!\!\!\!\!\!\!
-\frac{m\Sigma_{0}^{i}}{r[(\omega -m\Omega_i )^{2}-\kappa_i^{2}]}
\bigg[(\omega -m\Omega_i )\frac{m}{r} -\frac{\kappa_i
^{2}}{2\Omega_i }\frac{d}{dr}\bigg]\Psi^i
\end{eqnarray}
for the stellar and gaseous discs.

Based on equation (\ref{perturbs2}), we construct stationary
perturbation solutions with $\omega=0$. With the axisymmetric
background equilibrium conditions derived in Section 2.2, we
rewrite equation (\ref{perturbs2}) by setting $\omega=0$, namely
\begin{equation}\label{stationarys}
\begin{split}
&m\bigg\{-\mu ^{s}
+\\
&\frac{1}{D_{s}^{2}(1+2\beta)(m^{2}-2+2\beta)}
\bigg(\frac{m^{2}+2\beta}{r}-2\frac{d}{dr}-r
\frac{d^{2}}{dr^{2}}\bigg)\\
&\times\bigg[r\mu^{s}+\frac{(1+2\beta)(1+D_s^2)}
{2\pi G(2\beta{\cal P}_0)}
\frac{V}{1+\delta}\bigg]\bigg\}=0\ ,
\end{split}
\end{equation}
for the stellar disc and
\begin{equation}\label{stationaryg}
\begin{split}
&m\bigg\{-\mu ^{g}
+\\
&\frac{1}{D_{g}^{2}(1+2\beta)(m^{2}-2+2\beta)}
\bigg(\frac{m^{2}+2\beta}{r}-2\frac{d}{dr}-r
\frac{d^{2}}{dr^{2}}\bigg)\\
&\times\bigg[r\mu^{g}+\frac{(1+2\beta)(1+D_{g}^{2})}
{2\pi G(2\beta{\cal P}_0)}
\frac{V\delta}{1+\delta}\bigg]\bigg\}=0\ ,
\end{split}
\end{equation}
for the gaseous disc, respectively. The above two equations
(\ref{stationarys}) and (\ref{stationaryg}) are to be solved
together with Poisson's integral (\ref{perturbV1}). Note that
equations (\ref{stationarys}) and (\ref{stationaryg}) are valid
only for $m\neq 0$. In order to investigate the axisymmetric $m=0$
case, we should take a different limiting procedure by first
setting $m=0$ in equation (\ref{perturbs2}) before letting
$\omega\rightarrow 0$ (Lou \& Zou 2004).

\section{Aligned and Spiral Cases }

We investigate in this section stationary density wave patterns in
an inertial frame of reference using equations (\ref{stationarys})
and (\ref{stationaryg}) coupled with Poisson integral
(\ref{perturbV1}). We distinguish two types of coplanar
disturbances, that is, aligned and unaligned (logarithmic spiral)
coplanar perturbations. Aligned perturbation patterns correspond
to distorted streamlines with the maximum and minimum radii at
different radial locations lined up in the azimuth, while
unaligned or spiral perturbations correspond to distorted
streamlines with the maximum and minimum radii shifted
systematically in azimuth at different radial locations (Kalnajs
1973). Aligned perturbations relate to purely azimuthal
propagations of density waves (Lou 2002; Lou \& Fan 2002) while
the spiral perturbations relate to both azimuthal and radial
propagations (Lou 2002; Lou \& Fan 2002; Lou \& Shen 2003; Lou \&
Zou 2004). Furthermore for aligned cases, we consider
perturbations that carry the same density power-law dependence as
that of the background equilibrium disc does. In contrast, we
consider logarithmic spiral perturbations for nonaxisymmetric
spiral cases (Kalnajs 1971; Syer \& Tremaine 1996; Shu et al.
2000; Lou 2002; Lou \& Fan 2002; Lou \& Shen 2003; Lou \& Zou
2004; Lou \& Wu 2004).

\subsection{Aligned Perturbation Configurations }

The aligned $m=0$ case is somewhat trivial in the sense of a
re-scaling of the axisymmetric background equilibrium state (Shu
et al. 2000; Lou 2002; Lou \& Shen 2003; Lou \& Zou 2004). For
$m\ge 1$, we consider aligned perturbations that carry the same
radial power-law dependence as that of the axisymmetric background
equilibrium disc does. The perturbed surface mass densities and
the total gravitational potential read\footnote{For aligned
coplanar perturbations with a radial variation different from
that of the background equilibrium state, the perturbation
potential-density pair consistent with the Poisson integral
(\ref{perturbV0}) will be $\mu^i=\sigma^ir^{-\lambda}$,
$V=-2\pi Gr(\mu^s+\mu^g){\cal P}_m(\lambda)$ where the
superscript $i=s,g$ for stellar and gaseous discs,
respectively, numerical factor ${\cal P}_m(\lambda)\equiv
\Gamma(m/2-\lambda/2+1)\Gamma(m/2+\lambda/2-1/2)/
[2\Gamma(m/2-\lambda/2+3/2)\Gamma(m/2+\lambda/2)]$ and the
$\lambda$ range of $-m+1<\lambda<m+2$ is required. Following
the same procedure of analysis, we can construct a more broad
class of stationary coplanar aligned perturbation solutions.}
\begin{equation}\label{alignedPertur}
\begin{split}
&\mu^s=\sigma^sr^{-(1+2\beta)}\ ,
\qquad\qquad
\mu^g=\sigma^gr^{-(1+2\beta)}\ ,\\
&V=-2\pi Gr(\mu^s+\mu^g){\cal P}_m(\beta)\ ,
\end{split}
\end{equation}
where $\sigma^s$, $\sigma^g$ are small constant coefficients
and the parameter function ${\cal P}_m(\beta)$ is defined by
\begin{equation}\label{Pm}
{\cal P}_m(\beta)\equiv
\frac{\Gamma(m/2 -\beta+1/2)\Gamma(m/2+\beta)}
{2\Gamma(m/2-\beta+1)\Gamma(m/2+\beta+1/2)}\ ,
\end{equation}
with $-m/2<\beta<(m+1)/2$ (Qian 1992; Syer \& Tremaine 1996). The
prescribed ranges of $\beta\in (-1/4, 1/2)$ for warm discs and
of $\beta\in (-1/2, 1/2)$ for cold discs happen to satisfy this
requirement for $m\ge 1$. Note that for the isothermal case of
$\beta=0$, we simply have ${\cal P}_m=1/m$
which \footnote{We have assumed $m\ge 0$,
otherwise $|m|$ should be used instead.}
is just the case of Shu et al. (2000), Lou
(2002), Lou \& Fan (2002), Lou \& Shen (2003), and Lou \& Zou
(2004). One can readily derive the recursion relation in $m$
of ${\cal P}_m(\beta)$ for a fixed $\beta$ value, namely
\begin{equation}\label{recurP}
{\cal P}_{m+1}(\beta){\cal P}_m(\beta)
=[(m+2\beta)(m+1-2\beta)]^{-1}\ .
\end{equation}
In both ranges of $\beta\in(-1/4,1/2)$ and $\beta\in(-1/2,1/2)$,
it is also useful to derive the asymptotic expression of
${\cal P}_m(\beta)$
\begin{equation}\label{asymP}
{\cal P}_m(\beta)\approx(m^2+2\beta-4\beta^2)^{-1/2}
\end{equation}
for $m\ge 2$ with an accuracy better than $2\%$.
Larger values of $m$ would lead to higher accuracies.

By imposing condition (\ref{alignedPertur}) with $m\ge 1$,
equations (\ref{stationarys}) and (\ref{stationaryg})
can be cast into the following forms, namely
\begin{equation}\label{hoho}
\begin{split}
\mu^s=\bigg(\frac{m^2+2\beta}{r}-2\frac{d}{dr}
-r\frac{d^2}{dr^2}\bigg)(H_1r\mu^s+G_1r\mu^g)\ ,\\
\mu^g=\bigg(\frac{m^2+2\beta}{r}-2\frac{d}{dr}
-r\frac{d^2}{dr^2}\bigg)(H_2r\mu^g+G_2r\mu^s)\ ,
\end{split}
\end{equation}
where the four coefficients $H_1$, $H_2$, $G_1$
and $G_2$ explicitly involved are defined by
\begin{equation}\label{alignedHG}
\begin{split}
&H_1\equiv\frac{1}{D_s^2(1+2\beta)(m^{2}-2+2\beta)} \\
&\qquad\qquad\qquad\times
\bigg[1-\frac{(1+2\beta)(1+D_s^2)}{2\beta{\cal P}_0}
\frac{{\cal P}_m}{(1+\delta)}\bigg]\ ,\\
&H_2\equiv\frac{1}{D_g^2(1+2\beta)(m^{2}-2+2\beta)} \\
&\qquad\qquad\qquad\times
\bigg[1-\frac{(1+2\beta)(1+D_g^2)}{2\beta{\cal P}_0}
\frac{{\cal P}_m\delta}{(1+\delta)}\bigg]\ ,\\
&G_1\equiv-\frac{1+D_s^2}{D_s^2(2\beta{\cal P}_0)
(m^{2}-2+2\beta)}\frac{{\cal P}_m}{(1+\delta)}\ ,\\
&G_2\equiv-\frac{1+D_g^2}{D_g^2(2\beta{\cal P}_0)
(m^{2}-2+2\beta)}\frac{{\cal P}_m\delta}{(1+\delta)}\ ,\\
\end{split}
\end{equation}
where ${\cal P}_0$ appears due to background variables (\ref{2sigma}).
Note that ${\cal P}_0$ diverges at $\beta=0$ and $\beta=1/2$,
and approaches zero as $\beta\rightarrow -1/2$. Meanwhile,
$2\beta{\cal P}_0=1$ is regular at $\beta=0$. By expressions
(\ref{alignedPertur}), equations (\ref{hoho}) can be rearranged into
\begin{equation}\label{hoho1}
\begin{split}
&[1-H_1(m^2+4\beta-4\beta^2)]\mu^s
=G_1(m^2+4\beta-4\beta^2)\mu^g\
,\\
&[1-H_2(m^2+4\beta-4\beta^2)]\mu^g
=G_2(m^2+4\beta-4\beta^2)\mu^s\
.\\
\end{split}
\end{equation}
We introduce below several handy notations for parameters
that depend only on $m$ and $\beta$ to greatly simplify
analytical expressions, namely
\begin{equation}\label{ABCH}
\begin{split}
&{\cal A}_m(\beta)\equiv m^2+4\beta-4\beta^2\ ,\\
&{\cal B}_m(\beta)\equiv (1+2\beta)(m^2-2+2\beta)\ ,\\
&{\cal C}(\beta)\equiv   (1+2\beta)/(2\beta{\cal P}_0)\ , \\
&{\cal H}_m(\beta)\equiv {\cal C}(\beta){\cal P}_m(\beta)
{\cal A}_m(\beta)+{\cal B}_m(\beta) \ .
\end{split}
\end{equation}
From these expressions, one immediately knows that
parameter ${\cal C}(\beta)$ decreases monotonically with
increasing $\beta$ and $0<{\cal C}<\pi^2/\{4[\Gamma(3/4)]^4\}
\cong 1.094$ for $\beta\in(-1/4,1/2)$ and ${\cal C}(0)=1$ by
taking the limit of $\beta\rightarrow 0$.

A straightforward combination of the two conditions in
equation (\ref{hoho1}) leads to the following stationary
dispersion relation for nonaxisymmetric aligned coplanar
perturbations
\begin{equation}\label{SDRformal}
\begin{split}
(1-H_1{\cal A}_m)(1-H_2{\cal A}_m)=G_1G_2{\cal A}_m^2\ .
\end{split}
\end{equation}
By substituting the expressions of $H_1$, $H_2$, $G_1$ and $G_2$
in equation (\ref{alignedHG}) into equation (\ref{SDRformal}) and
using the background relation $D_g^2=\eta(D_s^2+1)-1$ where
$\eta\equiv A_s^2/A_g^2$, we obtain a quadratic equation in terms
of $y\equiv D_s^2$ for the stationary dispersion relation
(\ref{SDRformal})
\begin{equation}\label{aligned}
C_2y^2+C_1y+C_0=0\ ,
\end{equation}
where coefficients $C_2$, $C_1$ and $C_0$ are functions of $m$,
$\beta$, $\delta$ and $\eta$, and are explicitly defined by
\begin{equation}\label{alignedcoeff}
\begin{split}
C_2\equiv &{\cal B}_m{\cal H}_m\eta\ ,\\
C_1\equiv &\bigg[({\cal B}_m-{\cal A}_m){\cal H}_m
+\frac{({\cal A}_m
+{\cal B}_m)({\cal H}_m-{\cal B}_m)}{(1+\delta)}\bigg]\eta\\
&\qquad\qquad  -\frac{({\cal A}_m+{\cal B}_m)
({\cal H}_m+{\cal B}_m\delta)}{(1+\delta)}\ ,\\
C_0\equiv &\bigg[-{\cal A}_m{\cal H}_m+\frac{({\cal A}_m
+{\cal B}_m)({\cal H}_m-{\cal B}_m)}{(1+\delta)}\bigg]\eta\\
&\quad +({\cal A}_m+{\cal B}_m)^2-\frac{({\cal A}_m+{\cal B}_m)
({\cal H}_m+{\cal B}_m\delta)}{(1+\delta)}\ .
\end{split}
\end{equation}
Given specified parameters of $\beta$, $\delta$ and $\eta$,
equation (\ref{aligned}) can be readily solved analytically
for different values of $m$ as
\begin{equation}
\begin{split}
y_{1,2}=\frac{-C_1\pm (C_1^2-4C_2C_0)^{1/2}}{2C_2}\ ,\nonumber
\end{split}
\end{equation}
where the determinant $\Delta\equiv C_1^2-4C_2C_0$ remains always
positive for $m\ge 1$ as shown in Appendix A so that stationary
dispersion relation (\ref{aligned}) has two distinct real
solutions for $D_s^2$. To illustrate the procedure, we choose
several sets of parameters to numerically solve equation
(\ref{aligned}).

The $\beta=0$ (or equivalently, $n=1$) case has been thoroughly
studied as a composite system of two coupled SIDs (Lou \& Shen
2003; Shen \& Lou 2003; Lou \& Zou 2004) where the surface mass
density profiles scale as $r^{-1}$ with flat rotation curves.
In the present model, $\beta$ is allowed to take on values in
the range $(-1/4,1/2)$ for warm discs. For two representative
examples, we choose $\beta=-1/8$ and $\beta=1/4$ corresponding
to barotropic index $n=2/3$ and $n=4/3$, respectively. There
are two more free parameters to be chosen: the first is the
ratio of the surface mass density of the gaseous disc to that
of the stellar disc $\delta$. We choose three trial values of
$\delta=1/4$, $1$ and $4$ for three different cases. We simply
assume $\eta>1$ as the stellar disc has a relatively higher
`temperature' (i.e., higher velocity dispersion). We note when
$\eta=1$, the three coefficients $C_2$, $C_1$ and $C_0$ in
equation (\ref{aligned}) turn out to be independent of $\delta$.
For $\eta=1$ and $\beta=0$, the situation is reduced to two SIDs
with the same sound speed (Lou \& Shen 2003; Shen \& Lou 2003).

The representative solutions of equation (\ref{aligned}) for
different sets of parameters $\beta$, $\delta$, $\eta$ and $m$
will be discussed more specifically later on. We here offer
several remarks. Mathematically, two eigen-solutions for
coplanar perturbations can be found due to the gravitational
coupling between the two discs, but they may not always
satisfy the physical requirement $D_s^2\ge 0$ simultaneously.

It is important to realize that the two branches of $D_s^2$
solution of equation (\ref{aligned}), $y_1$ and $y_2$, are
both monotonic functions of $\eta$ (either monotonically
increasing or monotonically decreasing) once other parameters
are specified. This rule also applies to the spiral cases
discussed later(see the proof in Appendix B). Therefore, once
we know the solutions at $\eta=1$ and $\eta\rightarrow\infty$
the entire solution ranges are qualitatively determined. More
fortunately, the solutions at the two boundaries of $\eta$
have explicit analytical forms as shown below.

For $\eta=1$, we obtain two real solutions of
quadratic equation (\ref{aligned}) in the forms of
\begin{equation}\label{alisoleta1}
\begin{split}
Y_1^A=\frac{{\cal A}_m}{{\cal B}_m}\
\qquad\hbox{ and }\qquad
Y_1^B=\frac{(1-{\cal C}{\cal P}_m){\cal A}_m}{{\cal H}_m}\ .
\end{split}
\end{equation}
This case is special since the second expression $Y_1^B$ is the
result of one single disc and the first expression $Y_1^A$ is
extra due to the gravitational coupling. For the physical
solution branch with $D_s^2>0$, Table 1 contains information
of phase relationship between $\mu^g$ and $\mu^s$ for the
$m=1$ case with different values of $\beta\in (-1/4,1/2)$.
For $Y_1^B$ being physical, $\mu^g$ and $\mu^s$ are in-phase as
a counterpart of a single disc, while for $Y_1^A$ being physical,
$\mu^g$ and $\mu^s$ are out-of-phase and this has no counterpart
in the case of a single disc.

Meanwhile in the limit of $\eta\rightarrow\infty$, we
solve quadratic equation (\ref{aligned}) to derive
the following two solutions
\begin{equation}\label{alisoletainf}
\begin{split}
Y_{\infty}^A&=\frac{{\cal A}_m[{\cal H}_m\delta+(1-{\cal C}{\cal P}_m)
{\cal B}_m]} {{\cal B}_m{\cal H}_m(1+\delta)}\\
&=-1+\frac{({\cal A}_m+{\cal B}_m)({\cal H}_m\delta+{\cal B}_m)}
{{\cal B}_m{\cal H}_m(1+\delta)}
\\
&\!\!\!\!\hbox{\!\!\!\! and}\qquad\qquad
\\
Y_{\infty}^B&=-1\ .
\end{split}
\end{equation}
Apparently, the latter $Y_{\infty}^B$
is unphysical for being negative.

For the phase relationship between the two surface mass
density perturbations $\mu^s$ and $\mu^g$, we make use
of equation (\ref{hoho1}) to derive
\begin{equation}\label{alignedPhaseR}
\begin{split}
\frac{\mu^g}{\mu^s}&=\frac{1-H_1{\cal A}_m}{G_1{\cal A}_m}
=-1-\frac{(D_s^2{\cal B}_m-{\cal A}_m)(1+\delta)}
{{\cal C}{\cal P}_m{\cal A}_m(1+D_s^2)}\\
&=-1-\frac{{\cal B}_m(1+\delta)}{{\cal C}{\cal P}_m{\cal A}_m}
+\frac{({\cal A}_m+{\cal B}_m)(1+\delta)}{{\cal C}
{\cal P}_m{\cal A}_m(1+D_s^2)} \ ,
\end{split}
\end{equation}
and then to calculate the phase relationship of surface
mass density perturbations for stationary perturbation
modes by inserting the value of $D_s^2$ solution obtained.

\begin{figure}
\centering
\includegraphics[scale=0.42]{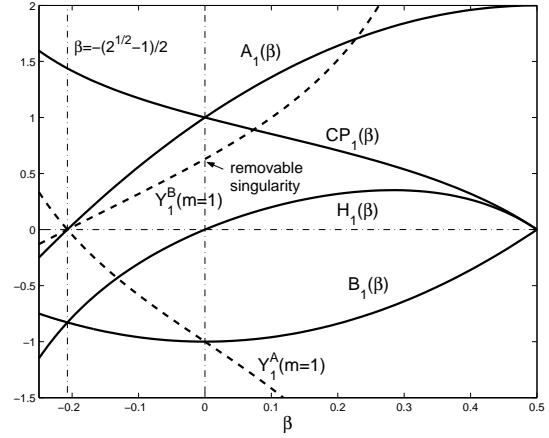}
\caption{Variations of coefficients ${\cal A}_1(\beta)$, ${\cal
B}_1(\beta)$, ${\cal C}{\cal P}_1(\beta)$, ${\cal H}_1(\beta)$ as
defined in equation (\ref{ABCH}) and $D_s^2$ solutions of
$Y_1^A|_{m=1}$ and $Y_1^B|_{m=1}$ with $\eta=1$ as obtained in
equation (\ref{alisoleta1}) in the $\beta$ range of
$\beta\in(-1/4,1/2)$.}
\end{figure}

\begin{table}\label{table01}
\centering
\begin{tabular}{cccc}
\hline    $\beta$  &$(-1/4,-(2^{1/2}-1)/2)$ &
$(-(2^{1/2}-1)/2,0)$ & $(0,1/2)$\\
\hline      $C_2$                  &  +   &  +   &  --\\
            $1-{\cal C}{\cal P}_1$ &  --  &  --  &  + \\
            ${\cal A}_1(\beta)$    &  --  &  +   &  + \\
            ${\cal B}_1(\beta)$    &  --  &  --  &  -- \\
            ${\cal H}_1(\beta)$    &  --  &  --  &  + \\
            $Y_1^A|_{m=1}$         &  +   &  --  &  -- \\
            $Y_1^B|_{m=1}$         &  --  &  +   &  + \\
            $\mu^g/\mu^s$          &  --  &  +   &  + \\
            \hline
\end{tabular}
\caption{Signs for various parameters when $\beta$ falls within
the three contiguous subintervals of $\beta\in(-1/4,1/2)$ for the
$m=1$ case. The phase relationship between $\mu^g$ and $\mu^s$ is
only given for the physical solution branch with $D_s^2>0$ and
remains valid for all combinations of $\delta$ and $\eta$.
Specifically, $Y_1^A|_{m=1}$ is positive for $\beta\in (-1/4,\
-(2^{1/2}-1)/2)$, while $Y_1^B|_{m=1}$ is positive for $\beta\in
(-(2^{1/2}-1)/2,\ 1/2)$.}
\end{table}

\subsubsection{The $m=1$ Aligned Case}

The $m=1$ case is somewhat special corresponding to
$1<{\cal C}{\cal P}_1<\pi^4/\{12[\Gamma(3/4)]^8\}\cong 1.596$
for $\beta\in (-1/4,\ 0)$ and to $0<{\cal C}{\cal P}_1<1$ for
$\beta\in (0,\ 1/2)$, respectively. We know that for a composite
system of coupled two SIDs with $\beta=0$ (i.e. flat rotation
curves), the aligned $m=1$ case imposes no restriction on the
dimensionless rotation parameter $D_s^2$ (Lou \& Shen 2003).
This situation changes qualitatively for $\beta\neq 0$. The
additional freedom of $\beta$ parameter rules out that equation
(\ref{aligned}) be automatically satisfied for arbitrary $D_s^2$.
We will see that to a certain extent, the aligned $m=1$ case is
very similar to the spiral $m=1$ case in form. The dependence of
solutions $y_1$ and $y_2$ on the square ratio of sound speeds
$\eta$ and on the ratio of surface mass densities $\delta$ is
distinctly different from those for the $m\ge 2$ cases. First
we rewrite equations (\ref{alisoleta1}) and (\ref{alisoletainf})
for $m=1$ as
\begin{equation}
\begin{split}
&Y_1^A|_{m=1}=\frac{{\cal A}_1}{{\cal B}_1}\ ,\quad\qquad\qquad
Y_1^B|_{m=1}=\frac{(1-{\cal C}{\cal P}_1){\cal A}_1}{{\cal H}_1}\ ,\\
&Y_{\infty}^A|_{m=1}=-1+\frac{4\beta({\cal H}_1\delta+{\cal B}_1)}
{{\cal B}_1{\cal H}_1(1+\delta)}\ ,\qquad Y_{\infty}^B|_{m=1}=-1\ ,
\end{split}\nonumber
\end{equation}
where except for a removable singularity\footnote{When $\beta=0$,
we have ${\cal H}_1=0$ while $Y_1^B|_{m=1}$ given by equation
(\ref{alisoleta1}) remains finite because the numerator also
vanishes. } at $\beta=0$ for $Y_1^B|_{m=1}$,
the coefficients ${\cal A}_1$, ${\cal B}_1$, ${\cal C}{\cal P}_1$
and ${\cal H}_1$ as well as the $D_s^2$ solutions $Y_1^A$ and
$Y_1^B$ as functions of $\beta$ are displayed in Fig. 1 within
the open interval $\beta\in(-1/4,1/2)$. The sign variations of
each parameters within the open interval $\beta\in(-1/4,1/2)$
are further summarized in Table 1 for reference.

For the $m=1$ case, the lower branch (either $y_1$ or
$y_2$)\footnote{When $m=1$, $y_1$ is the upper branch and $y_2$
is the lower branch for $\beta<0$ ($C_2>0$), while the situation
is reversed for $\beta>0$ ($C_2<0$). However when $m\ge 2$, $y_1$
always remains to be the upper branch and $y_2$ always remains to
be the lower branch as $C_2>0$ within the range of all
$\beta\in(-1/4,1/2)$.} of the solutions to equation (\ref{aligned})
is always negative as can be seen later on and is therefore
unphysical. It is possible for the upper branch of $D_s^2$ solution
to be positive for a specific range of $\eta$, depending on the
parameters $\beta$ and $\delta$ (see the critical $\eta_c$ discussed
below). The variation within different parameter regimes is subtle
for the special $m=1$ case as well as as for the phase relationship.
And we only need to consider the positive portion in the upper branch
of $D_s^2$ solution.

As indicated in Fig. 1 and Table 1, we divide the open interval
of $-1/4<\beta<1/2$ into three subintervals to analyze properties
of aligned coplanar perturbations with $m=1$.
\begin{enumerate}

\item[{\it $\star$ Case I}]
for $-1/4<\beta<-(2^{1/2}-1)/2$ when
$y_1>0$ is the upper branch

As solutions $y_1$ and $y_2$ of equation (\ref{aligned}) are
both monotonic functions of $\eta$ (see the proof in Appendix B),
we can use the explicit $y_1$ and $y_2$ solutions at $\eta=1$
[see solutions (\ref{alisoleta1})] and $\eta\rightarrow\infty$
[see solutions (\ref{alisoletainf})] to well bracket the ranges
of $D_s^2$ value. For $\eta=1$, we have
\begin{equation}
\begin{split}
&y_1=Y_1^A|_{m=1}=\frac{{\cal A}_1}{{\cal B}_1}>0\ ,\\
&y_2=Y_1^B|_{m=1}=\frac{(1-{\cal C}{\cal P}_1)
{\cal A}_1}{{\cal H}_1}<0\ ,
\end{split}
\end{equation}
while for $\eta\rightarrow\infty$, we have
\begin{equation}\label{etainfm01}
\begin{split}
&y_1=Y_{\infty}^A|_{m=1}=-1+\frac{4\beta({\cal H}_1\delta
+{\cal B}_1)}{{\cal B}_1{\cal H}_1(1+\delta)}>-1\ ,\\
&y_2=Y_{\infty}^B|_{m=1}=-1\ .
\end{split}
\end{equation}
In this case, the lower $y_2$ branch remains always negative and
thus unphysical, while the upper $y_1$ branch increases
monotonically with increasing $\delta$ and decreases monotonically
with increasing $\eta$.
For solutions
(\ref{etainfm01}), there exists a critical $\eta_c$ beyond which
$y_1$ becomes negative for fixed values of $\beta$ and $\delta$.
This $\eta_c$ can be explicitly determined by an analytical
expression
\begin{equation}\label{alignetac1}
\begin{split}
\eta_c =1+\frac{(1-{\cal C}{\cal P}_1){\cal A}_1(1+\delta)}
{{\cal H}_1\delta+(1-{\cal C}{\cal P}_1){\cal B}_1}\ .
\end{split}
\end{equation}
This expression (\ref{alignetac1}) remains valid
only when $\eta_c>1$ which further requires
\begin{equation}\label{aligndeltac1}
0<\delta<\frac{({\cal C}{\cal P}_1-1){\cal B}_1}
{{\cal H}_1}<{\cal C}{\cal P}_1-1\ ,
\end{equation}
that in turn defines a critical value
$\delta_c\equiv ({\cal C}{\cal P}_1-1){\cal B}_1/{\cal H}_1$.
For $\delta>\delta_c$, the upper $y_1$ branch remains
always positive and physical.

By equation (\ref{alignedPhaseR}), the phase relationship for
surface mass density perturbations with $m=1$ corresponding
to the physical portion of the $y_1$ branch is given by
\begin{equation}\label{aliPhRm1}
\frac{\mu^g}{\mu^s}=-1-\frac{{\cal B}_1(1+\delta)}
{{\cal C}{\cal P}_1{\cal A}_1 }
+\frac{4\beta(1+\delta)}{{\cal C}{\cal P}_1{\cal A}_1(1+D_s^2)}\ ,
\end{equation}
which decreases monotonically with increasing $D_s^2$ in the
$\beta$ subinterval of case (I) and thus increases monotonically
with increasing $\eta$ for the $y_1$ branch. We therefore have
\begin{equation}
-1<\frac{\mu^g}{\mu^s}<\bigg\{
\begin{array}{l l}
(1+\delta-{\cal C}{\cal P}_1)/({\cal C}{\cal P}_1)<0,
&\hbox{if $0<\delta<\delta_c$}\\
-{\cal C}{\cal P}_1{\cal A}_1\delta
/({\cal H}_1\delta+{\cal B}_1)<0,
&\hbox{if $\delta>\delta_c$}
\end{array}
\end{equation}
where the left-hand bound corresponds to $\eta=1$ and the
right-hand bound corresponds either to $\eta=\eta_c$
when condition (\ref{aligndeltac1}) is met or to
$\eta\rightarrow\infty$ when $\delta>\delta_c$. At any rate, the
phase relationship of surface mass density perturbations in the
two coupled discs for the upper $y_1$ solution branch remains
always out-of-phase.

\begin{figure}
\centering \subfigure[Case I: $-1/4<\beta<-(2^{1/2}-1)/2\ .$]{
\includegraphics[scale=0.42]{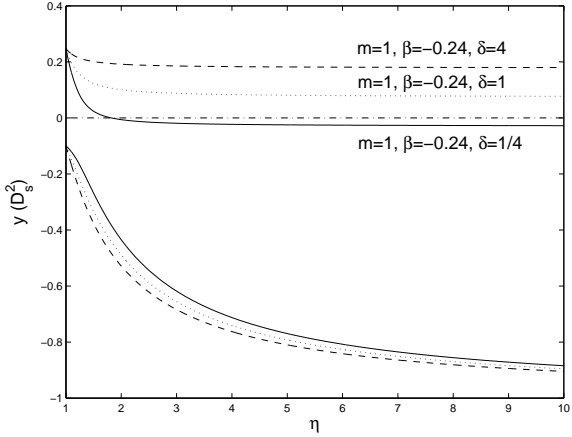}}
\subfigure[Case II:  $-(2^{1/2}-1)/2<\beta<0\ .$]{
\includegraphics[scale=0.42]{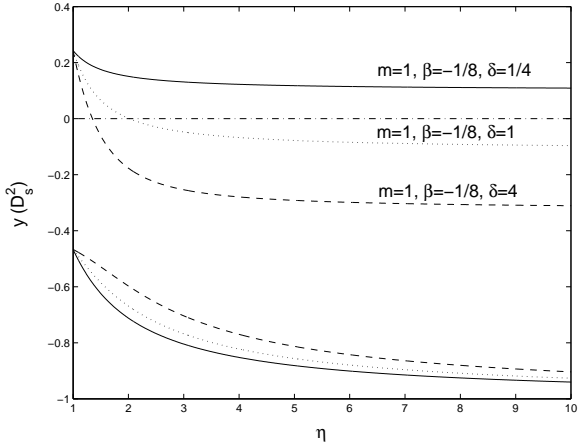}}
\subfigure[Case III:  $0<\beta<1/2\ .$]{
\includegraphics[scale=0.42]{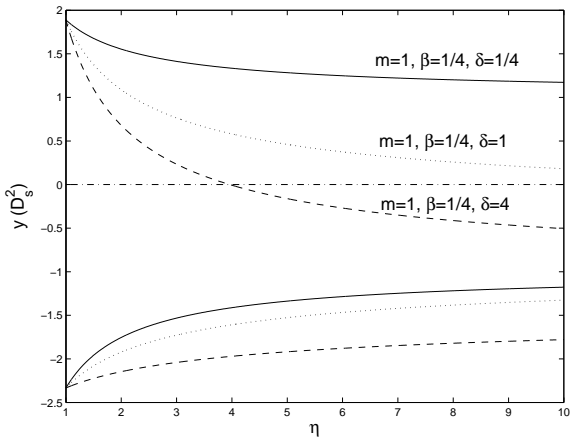}}
\caption{Two $D_s^2$ solution branches of equation (\ref{aligned})
as functions of $\eta$ with combinations of $m=1$, $\beta=-0.24,\
-1/8,\ 1/4$ and $\delta=1/4,\ 1,\ 4$. For the same set of
parameters in each panel (a), (b), (c), each linetype corresponds
to two solutions of $D_s^2$ for a range of $1<\eta<10$.}
\end{figure}

\item[{\it $\star$ Case II}]
for $-(2^{1/2}-1)/2<\beta<0$ when
$y_1>0$ is the upper branch

For $\eta=1$, we have
\begin{equation}
\begin{split}
&y_1=Y_1^B|_{m=1}=\frac{(1-{\cal C}{\cal P}_1){\cal A}_1}
{{\cal H}_1}>0\ ,\\
&y_2=Y_1^A|_{m=1}=\frac{{\cal A}_1}{{\cal B}_1}<0\ ;
\end{split}
\end{equation}
while for $\eta\rightarrow\infty$, we have
\begin{equation}
\begin{split}
&y_1=Y_{\infty}^A|_{m=1}=-1+\frac{4\beta({\cal H}_1\delta
+{\cal B}_1)}{{\cal B}_1{\cal H}_1(1+\delta)}>-1\ ,\\
&y_2=Y_{\infty}^B|_{m=1}=-1\ .
\end{split}
\end{equation}
In this case again, the lower $y_2$ branch remains always
negative and thus unphysical. The upper $y_1$ branch
decreases monotonically with increasing either $\delta$ or
$\eta$. The critical value of $\eta$ beyond which $y_1$
becomes negative for fixed values of $\beta$ and $\delta$
is again determined by the same expression (\ref{alignetac1}).
This criterion remains valid only when $\eta_c>1$ that
further requires
\begin{equation}\label{aligndeltac2}
\delta>\delta_c\equiv\frac{({\cal C}{\cal P}_1-1)
{\cal B}_1}{{\cal H}_1}>{\cal C}{\cal P}_1-1\ .
\end{equation}
This then implies that for $0<\delta<\delta_c$, the upper
$y_1$ branch of $D_s^2$ solution remains always physical
for being positive.

Similarly for the phase relationship between $\mu^g$ and
$\mu^s$, we note that by expression (\ref{aliPhRm1}),
$\mu^g/\mu^s$ increases monotonically with increasing $D_s^2$
in the $\beta$ subinterval of case (II) and thus decreases
monotonically with increasing $\eta$. We therefore have
\begin{equation}
\delta>\frac{\mu^g}{\mu^s}>\bigg\{
\begin{array}{l l}
(1+\delta-{\cal C}{\cal P}_1)/({\cal C}{\cal P}_1)>0\ ,
&\qquad\hbox{if $\delta>\delta_c$}\\
-{\cal C}{\cal P}_1{\cal A}_1\delta
/({\cal H}_1\delta+{\cal B}_1)>0\ ,
&\hbox{if $0<\delta<\delta_c$}
\end{array}
\end{equation}
where the left-hand bound corresponds to $\eta=1$ and the
right-hand bound corresponds either to $\eta=\eta_c$
when condition (\ref{aligndeltac2}) is satisfied or to
$\eta\rightarrow\infty$ when $0<\delta<\delta_c$. At any rate, the
phase relationship between $\mu^g$ and $\mu^s$ for the upper $y_1$
branch solution remains always in-phase with $\mu^g/\mu^s$ smaller
than $\delta$.

\item[{\it $\star$ Case III}]
for $0<\beta<1/2$ when $y_2$ becomes the upper branch

For $\eta=1$, we have
\begin{equation}
\begin{split}
&y_1=Y_1^A|_{m=1}=\frac{{\cal A}_1}{{\cal B}_1}<0\ ,\\
&y_2=Y_1^B|_{m=1}=\frac{(1-{\cal C}
{\cal P}_1){\cal A}_1}{{\cal H}_1}>0\ ;
\end{split}
\end{equation}
while for $\eta\rightarrow\infty$, we have
\begin{equation}
\begin{split}
&y_1=Y_{\infty}^A|_{m=1}=-1+\frac{4\beta({\cal H}_1\delta
+{\cal B}_1)}{{\cal B}_1{\cal H}_1(1+\delta)}<-1,\\
&y_2=Y_{\infty}^B|_{m=1}=-1\
\qquad\qquad\qquad\hbox{if }
\delta>-\frac{{\cal B}_1}{{\cal H}_1}\ ,
\\
&\hbox{or else, }\\
&y_1=Y_{\infty}^B|_{m=1}=-1\ ,\\
&y_2=Y_{\infty}^A|_{m=1}=-1
+\frac{4\beta({\cal H}_1\delta+{\cal B}_1)}
{{\cal B}_1{\cal H}_1(1+\delta)} >-1\\
&\hbox{if }0<\delta<-\frac{{\cal B}_1}{{\cal H}_1}\ .
\end{split}
\end{equation}
In this case, the lower $y_1$ branch remains always negative and
thus unphysical. The upper $y_2$ branch decreases monotonically
with increasing either $\delta$ or $\eta$. The critical value of
$\eta$ beyond which $y_2$ becomes negative for fixed values of
$\beta$ and $\delta$ is also determined by expression
(\ref{alignetac1}). This criterion is valid only when $\eta_c>1$
which further requires
\begin{equation}\label{aligndeltac3}
\delta>\delta_c\equiv\frac{({\cal C}{\cal P}_1-1)
{\cal B}_1}{{\cal H}_1}\ ;
\end{equation}
this inequality in turn implies that for
$0<\delta<\delta_c$ the upper $y_2$ branch
remains always physical for being positive.

Similarly by equation (\ref{aliPhRm1}) for the phase
relationship between $\mu^g$ and $\mu^s$, the ratio
$\mu^g/\mu^s$ decreases monotonically with increasing
$D_s^2$ in the $\beta$ subinterval of case (III) and
thus increases monotonically with increasing $\eta$.
We therefore have
\begin{equation}
\delta<\frac{\mu^g}{\mu^s}<\bigg\{
\begin{array}{l l}
(1+\delta-{\cal C}{\cal P}_1)/({\cal C}{\cal P}_1)>0\ ,
&\hbox{if $\delta>\delta_c$}\\
-({\cal C}{\cal P}_1{\cal A}_1\delta)
/({\cal H}_1\delta+{\cal B}_1)>0\ ,
&\hbox{if $0<\delta<\delta_c$}
\end{array}
\end{equation}
where the left-hand bound corresponds to $\eta=1$ and
the right-hand bound corresponds either to $\eta=\eta_c$
when condition (\ref{aligndeltac3}) is satisfied or to
$\eta\rightarrow\infty$ when $0<\delta<\delta_c$. At any
rate, the phase relationship between $\mu^g$ and $\mu^s$
for the upper $y_2$ branch solution remains always
in-phase with $\mu^g/\mu^s$ greater than $\delta$.

\end{enumerate}

Relevant details of illustrating examples for three cases of
$\beta=-0.24$, $\beta=-1/8$ and $\beta=1/4$ respectively are
shown in three panels (a), (b) and (c) of Fig. 2 and are also
summarized in Table 2.

\begin{table}
\centering
\begin{tabular}{cccccccc}
\hline\hline
&&\multicolumn{2}{c}{$\beta=-0.24$}&\multicolumn{2}{c}{$\beta=-1/8$} &
\multicolumn{2}{c}{$\beta=1/4$}\\
$m$ &$\delta$ &$\delta_c$&$\eta_c$&$\delta_c$&$\eta_c$&$\delta_c$&$\eta_c$\\
\hline 1 & 1/4  &0.401 & 1.821  & 0.522  &/     &0.809 & /      \\
         & 1    &0.401 & /      & 0.522  &2.018 &0.809 & 20.812 \\
         & 4    &0.401 & /      & 0.522  &1.350 &0.809 & 3.959  \\
       2 & 1/4  &/     & 2.514  & /      &2.375 & /    & 2.033  \\
         & 1    &      & 1.813  &        &1.803 &      & 1.809  \\
         & 4    &      & 1.556  &        &1.567 &      & 1.664  \\
       3 & 1/4  &      & 2.641  &        &2.278 &      & 1.782  \\
         & 1    &      & 2.147  &        &1.947 &      & 1.681  \\
         & 4    &      & 1.881  &        &1.752 &      & 1.604  \\
\hline
\end{tabular}
\caption{The critical values of $\eta_c$ and $\delta_c$ for the
three cases of $\beta=-0.24$, $-1/8$ and $1/4$ with different
values of $m=1,2,3$ and $\delta$. A slash ``/'' means that the
critical $\eta_c$ or $\delta_c$ do not exist. Note that only
for the $m=1$ case does there exist such a critical $\delta_c$
that depends only on $\beta$.}
\end{table}

\begin{figure*}
\centering \subfigure[$m=2,\ \beta=-1/8,\ \delta=4,\ 1,\ 1/4$]{
\includegraphics[scale=0.42]{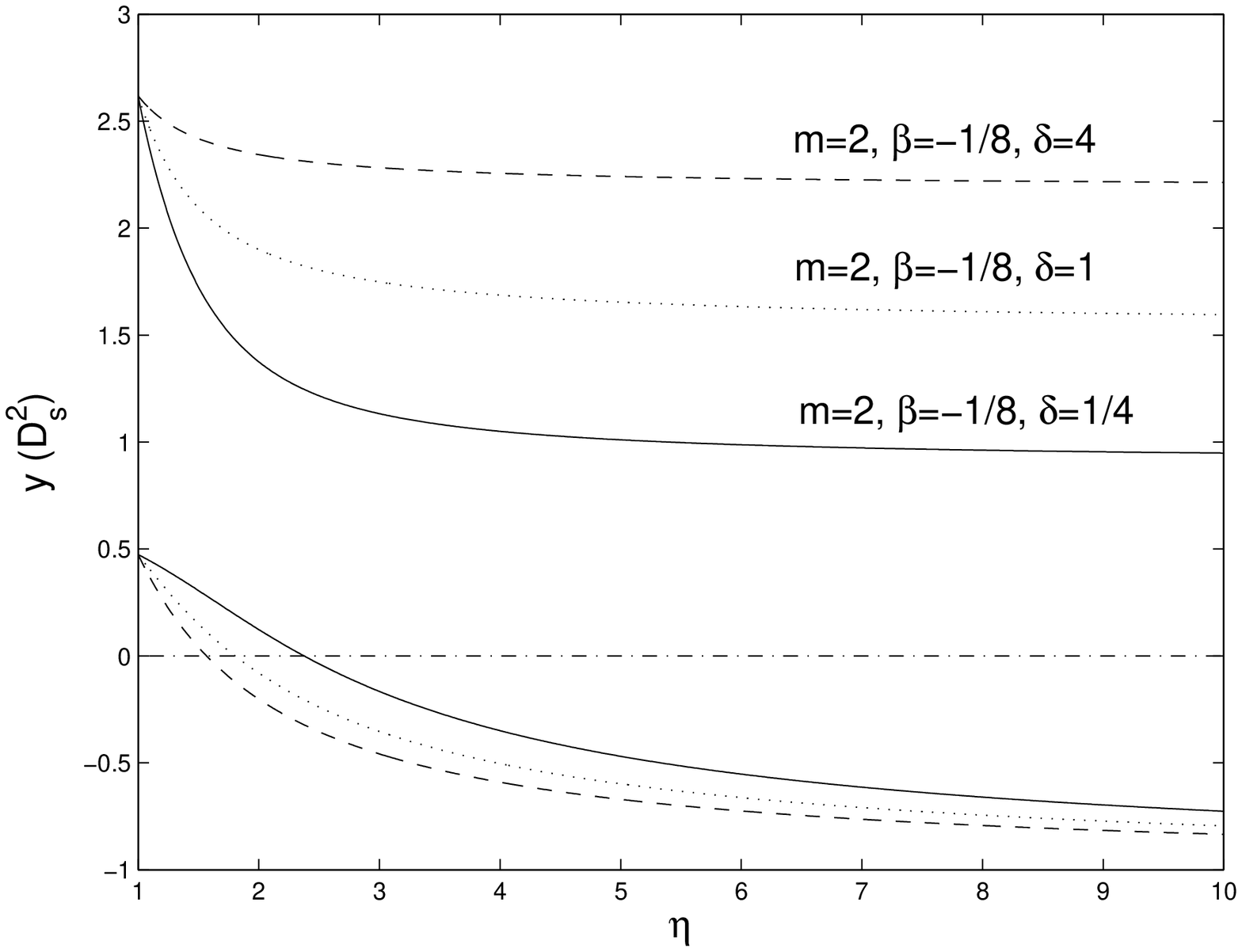}}
\subfigure[$m=2,\ \beta=1/4,\ \delta=4,\ 1,\ 1/4$]{
\includegraphics[scale=0.42]{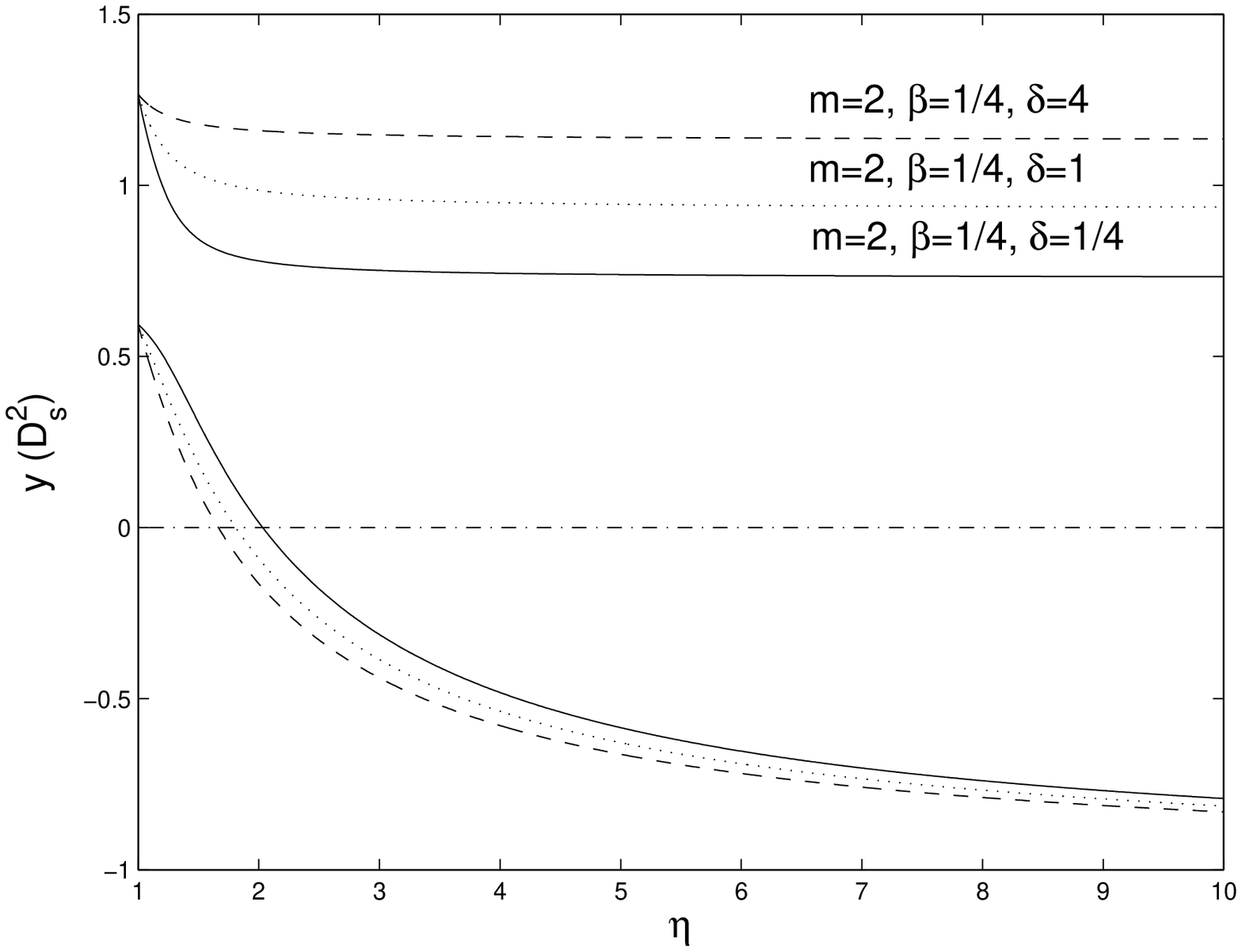}}
\subfigure[$m=3,\ \beta=-1/8,\ \delta=4,\ 1,\ 1/4$]{
\includegraphics[scale=0.42]{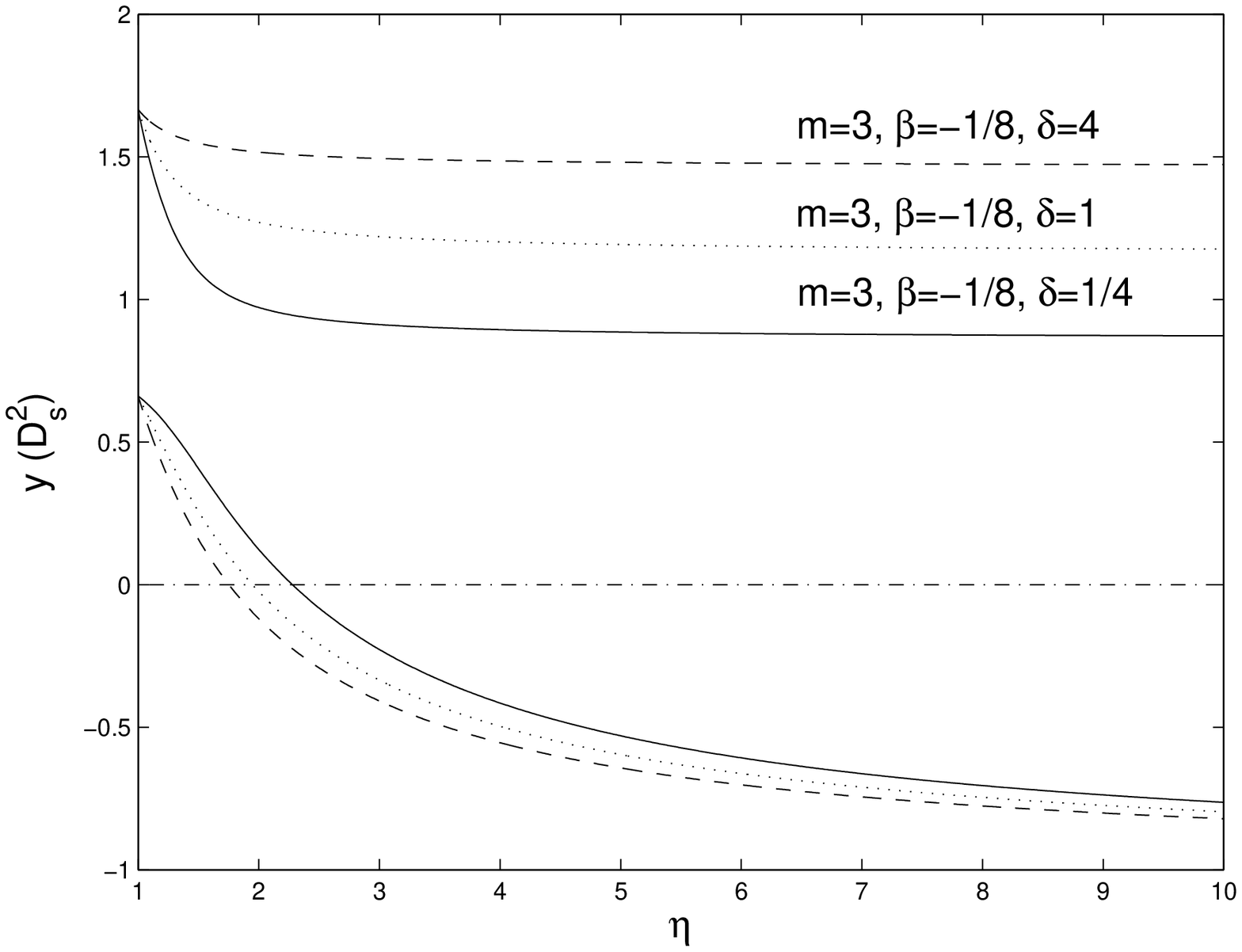}}
\subfigure[$m=3,\ \beta=1/4,\ \delta=4,\ 1,\ 1/4$]{
\includegraphics[scale=0.42]{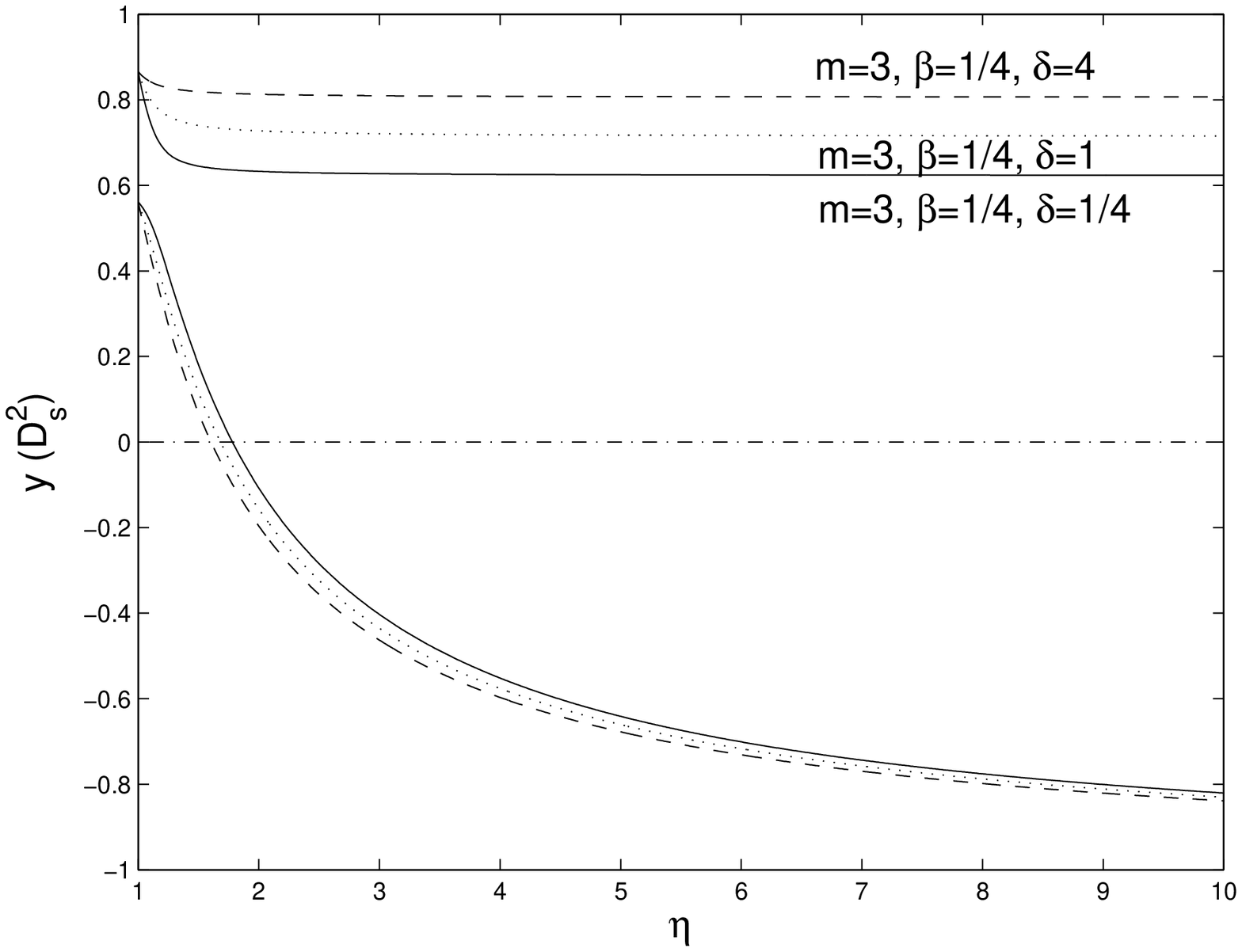}}
\caption{Aligned $D_s^2$ solution curves $y_1$ and $y_2$ as
functions of $\eta$ for different azimuthal periodicities $m=2,3$
and surface mass density ratio $\delta=1/4,1,4$ with two fixed
values of $\beta=-1/8$ and $\beta=1/4$. For the same set of
parameters in each panel (a), (b), (c), (d), each linetype
corresponds to two solutions of $D_s^2$ for a range of
$1<\eta<10$.}
\end{figure*}

\subsubsection{The $m\ge 2$ Aligned Cases}

When $m\ge 2$, we have $C_2>0$, $0<{\cal C}{\cal P}_m<1$,
${\cal A}_m >0$, ${\cal B}_m>0$, ${\cal H}_m>0$ in the open
$\beta$ interval $\beta\in(-1/4,1/2)$. Therefore in this case
of $m\ge 2$, $y_1$ and $y_2$ remain always upper and lower
branches, respectively. By solutions (\ref{alisoleta1}) for
$\eta=1$, we have
\begin{equation}
\begin{split}
&y_1=Y_1^A=\frac{{\cal A}_m}{{\cal B}_m}>0\ ,\\
&y_2=Y_1^B=\frac{(1-{\cal C}{\cal P}_m){\cal A}_m}
{{\cal H}_m}>0\ ,\\
\end{split}
\end{equation}
while by solutions (\ref{alisoletainf})
for $\eta\rightarrow\infty$, we have
\begin{equation}
\begin{split}
&y_1=Y_{\infty}^A=\frac{{\cal A}_m
[{\cal H}_m\delta+(1-{\cal C}{\cal P}_m){\cal B}_m]}
{{\cal B}_m{\cal H}_m(1+\delta)}>0\ ,\\
&y_2=Y_{\infty}^B=-1\ .
\end{split}
\end{equation}
As the limiting situations for $\eta=1$ and $\eta\rightarrow\infty$
well bracket possible ranges of $y_1$ and $y_2$ branches (see
Appendix B), it is obvious that the upper $y_1$ branch remains always
positive and the value of $y_1$ increases with increasing $\delta$
and decreases with increasing
either $m$ or $\eta$. Meanwhile, the lower $y_2$ branch has a
specific critical value $\eta_c$ of $\eta$ beyond which the
$y_2$ solution becomes unphysical for being negative. This
critical value $\eta_c$ is given by
\begin{equation}\label{alignetacm2}
\eta_c=1+\frac{(1-{\cal C}{\cal P}_m){\cal A}_m(1+\delta)}
{{\cal H}_m\delta+(1-{\cal C}{\cal P}_m){\cal B}_m}\ ,
\end{equation}
where we have ${\cal H}_m$ always greater than $m$ for all
$m\ge 2$ within $\beta\in(-1/4,1/2)$. Such a critical $\eta_c$
always exists for the $y_2$ branch for all values of $\delta$.
In other words, there does not exist a critical value
$\delta_c$ for $\delta$ as in the aligned $m=1$ case.

This critical value $\eta_c$ of $\eta$ decreases with increasing
$\delta$, much like the case of two coupled SIDs investigated
recently (Lou \& Shen 2003). For physical regimes of $y_1$ and
$y_2$, they both decrease monotonically with increasing $\eta$,
while $y_1$ branch increases monotonically and $y_2$ branch
decreases monotonically with increasing $\delta$.

For the phase relationship between $\mu^g$ and $\mu^s$, it is
straightforward to show that the ratio $\mu^g/\mu^s$ decreases
monotonically with increasing $D_s^2$ and thus increases
monotonically with increasing $\eta$ for $m\ge 2$. For the upper
$y_1$ branch, we obtain
\begin{equation}
\begin{split}
-1&<\frac{\mu^g}{\mu^s}
<-\frac{{\cal C}{\cal P}_m{\cal A}_m\delta}
{{\cal H}_m\delta+{\cal B}_m}\ ,
\end{split}
\end{equation}
where the left-hand bound corresponds to $\eta=1$ and the
right-hand bound corresponds to $\eta\rightarrow\infty$. In the
specified range of $\beta\in(-1/4,1/2)$, the ratio $\mu^g/\mu^s$
remains always negative for the upper $y_1$ branch, indicating
surface mass density perturbations in two discs are out-of-phase.

For the lower $y_2$ branch in parallel, we derive
\begin{equation}
\delta<\frac{\mu^g}{\mu^s}
<\frac{1+\delta}{{\cal C}{\cal P}_m}-1\ ,
\end{equation}
where the left-hand bound corresponds to $\eta=1$ and the
right-hand bound corresponds to the critical $\eta_c$ which
makes $D_s^2=y_2=0$. Apparently, the lower $y_2$ branch
(if physical) means surface mass density perturbations in
the two coupled discs are in-phase.

\subsection{Spiral Coplanar Perturbation Configurations}

Stationary surface mass density perturbations in both discs scale
in the forms of $\propto\mu e^{-\hbox{i}m\theta}$ in azimuthal
angle $\theta$. For aligned perturbations, we have further taken
$\mu\propto r^{-\varepsilon}$ where $\varepsilon$ is a
positive/negative constant exponent. For example in subsection
3.1, we have chosen $\varepsilon=\alpha=1+2\beta$ for coplanar
perturbations carrying the same radial power-law dependence of the
background equilibrium disc system. On the other hand, for
$\varepsilon$ being a complex constant exponent, perturbations
would appear in spiral forms, namely, the so-called logarithmic
spiral $\mu\propto r^{-\Re(\varepsilon)}\exp
[-\hbox{i}\Im(\varepsilon)\ln r]$ where $\Re(\varepsilon)$ and
$\Im(\varepsilon)$ are the real and imaginary parts of
$\varepsilon$. To ensure the gravitational potential perturbation
arising from this perturbed surface mass density as computed
by Poisson integral (\ref{fish}) being finite requires
$-m+1<\Re(\varepsilon)<m+2$ (Qian 1992). Without loss of generality,
we assume a set of logarithmic spiral density perturbations and the
resulting gravitational potential perturbation in a mathematically
consistent manner\footnote{In parallel with the aligned case of
coplanar perturbations, potential-density pair (\ref{spirPerturb})
is not the only available potential-density pair that satisfies the
Poisson integral (\ref{perturbV0}). For coplanar logarithmic spiral
perturbations, a more general class of allowed potential-density
pairs that are consistent with the Poisson integral (\ref{perturbV0})
is $\mu^j=\sigma^jr^{-\lambda}\exp(\hbox{i}\xi\ln r)$,
$V=-2\pi Gr(\mu^s+\mu^g){\cal L}_m(\xi,\lambda)$ where the
superscript $j=s$ and $g$, respectively, the numerical
factor ${\cal L}_m(\xi,\lambda)\equiv \Gamma(m/2-\lambda/2
+\hbox{i}\xi/2+1)\Gamma(m/2+\lambda/2-\hbox{i}\xi/2-1/2)/
[2\Gamma(m/2-\lambda/2+\hbox{i}\xi/2+3/2)
\Gamma(m/2+\lambda/2-\hbox{i}\xi/2)]$ and the $\lambda$
range of $-m+1<\lambda<m+2$ is required. Following the same
procedure of analysis, we can construct a more broad class of
stationary coplanar perturbation solutions for logarithmic
spiral configurations in a composite disc system.}
(Kalnajs 1971; Syer \& Tremaine; Shu et al. 2000; Lou 2002;
Lou \& Fan 2002; Lou \& Shen 2003; Lou \& Zou 2004; Lou \&
Wu 2004). Specifically, we write
\begin{equation}\label{spirPerturb}
\begin{split}
&\mu^s=\sigma^sr^{-3/2}\exp(\hbox{i}\xi\ln r)\ ,\qquad
\mu^g=\sigma^gr^{-3/2}\exp(\hbox{i}\xi\ln r)\ ,\\
&V=-2\pi Gr(\mu^s+\mu^g){\cal N}_m(\xi)\ ,
\end{split}
\end{equation}
where $\sigma^s$ and $\sigma^g$ are
small constant coefficients,
\begin{equation}\label{calNm}
{\cal N}_m(\xi)=\frac{\Gamma(m/2+\hbox{i}\xi/2+1/4)
\Gamma(m/2-\hbox{i}\xi/2+1/4)}
{2\Gamma(m/2+\hbox{i}\xi/2+3/4)\Gamma(m/2-\hbox{i}\xi/2+3/4)}
\end{equation}
is the Kalnajs function (Kalnajs 1971) and $\xi$
is a kind of radial `wavenumber'. We refresh a few
properties of ${\cal N}_m(\xi)$. First, ${\cal N}_m(\xi)$
is an even function of $\xi$ so only $\xi\ge 0$ will be
considered later. Secondly, ${\cal N}_m(\xi)$ decreases
monotonically with increasing $\xi>0$. Thirdly,
$0<{\cal N}_m<1$ for $m\ge 1$ while ${\cal N}_0$ is
positive and can be greater than 1 for a sufficiently
small $\xi$.

The choice of such form of perturbations is different from that of
Syer \& Tremaine (1996), whose spiral perturbations were taken to
be $\mu\propto r^{-1-2\beta}\exp(\hbox{i}m\xi\ln r)$ for $m>0$
(analysis before subsection 3.4 in their paper) in our notations.
For axisymmetric stability analysis in their subsection 3.4, they
adopted the same spiral perturbations in the form of
(\ref{spirPerturb}) (see also Lemos et al. 1991).
We note
that our background equilibria as well as the adopted form of
logarithmic spiral perturbations are themselves scale-free,
separately, whereas combinations of the background equilibrium
and perturbations are not scale-free except for the special
$\beta=1/4$ case (see Lynden-Bell \& Lemos 1993).

Parallelling with ${\cal P}_m$ for the case of aligned
perturbations, there are two useful formulae for
${\cal N}_m(\xi)$ for logarithmic spiral perturbations.
The first one is the recursion relation
\begin{equation}\label{recurN}
{\cal N}_{m+1}(\xi){\cal N}_m(\xi)=[(m+1/2)^2+\xi^2]^{-1}\ ,
\end{equation}
and the second one is the asymptotic expression for
${\cal N}_m(\xi)$
\begin{equation}\label{asymN}
{\cal N}_m(\xi)\approx(m^2+\xi^2+1/4)^{-1/2}
\end{equation}
for $m^2+\xi^2\gg 1$. For $m\ge 2$, this asymptotic
expression (\ref{asymN}) is accurate enough to compute
values of ${\cal N}_m(\xi)$.

Using potential-density set of (\ref{spirPerturb}) for logarithmic
spirals, we rearrange stationary coplanar perturbation equations
(\ref{stationarys}) and (\ref{stationaryg}) into the following
forms with $m>0$.
\begin{equation}\label{hoho3}
\begin{split}
&\mu^s=\bigg(\frac{m^2+2\beta}{r}-2\frac{d}{dr}
-r\frac{d^2}{dr^2}\bigg)(H_1r\mu^s+G_1r\mu^g)\ ,\\
&\mu^g=\bigg(\frac{m^2+2\beta}{r}-2\frac{d}{dr}
-r\frac{d^2}{dr^2}\bigg)(H_2r\mu^g+G_2r\mu^s)\ ,\\
\end{split}
\end{equation}
where the four relevant coefficients $H_1$, $H_2$,
$G_1$ and $G_2$ are defined here explicitly by
\begin{equation}\label{spiralHG}
\begin{split}
&H_1\equiv\frac{1}{D_s^2(1+2\beta)(m^{2}-2+2\beta)}\\
&\qquad\qquad\qquad\times
\bigg[1-\frac{(1+2\beta)(1+D_s^2)}{2\beta{\cal P}_0}
\frac{{\cal N}_m}{(1+\delta)}\bigg]\ ,\\
&H_2\equiv\frac{1}{D_g^2(1+2\beta)(m^{2}-2+2\beta)}\\
&\qquad\qquad\qquad\times
\bigg[1-\frac{(1+2\beta)(1+D_g^2)}{2\beta{\cal P}_0}
\frac{{\cal N}_m\delta}{(1+\delta)}\bigg]\ ,\\
&G_1\equiv-\frac{(1+D_s^2)}{D_s^2(2\beta{\cal P}_0)
(m^{2}-2+2\beta)}\frac{{\cal N}_m}{(1+\delta)}\ ,\\
&G_2\equiv-\frac{(1+D_g^2)}{D_g^2(2\beta{\cal P}_0)
(m^{2}-2+2\beta)}\frac{{\cal N}_m\delta}{(1+\delta)}\ .\\
\end{split}
\end{equation}
Rewriting equations (\ref{hoho3}) with expressions
(\ref{spirPerturb}) for $\mu^s$ and $\mu^g$, we immediately obtain
\begin{equation}\label{hoho4}
\begin{split}
&[1-H_1(m^2+\xi^2+1/4+2\beta)]\mu^s \\
&\qquad\qquad\qquad\qquad =G_1(m^2+\xi^2+1/4+2\beta)\mu^g\ ,\\
&[1-H_2(m^2+\xi^2+1/4+2\beta)]\mu^g \\
&\qquad\qquad\qquad\qquad =G_2(m^2+\xi^2+1/4+2\beta)\mu^s\ .
\end{split}
\end{equation}
As for the aligned case in subsection 3.1, we define
some useful notations for parameter combinations that
will simplify our following derivations, namely
\begin{equation}\label{ABCHspiral}
\begin{split}
&{\cal A}_m(\beta,\xi)\equiv m^2+\xi^2+1/4+2\beta\ ,\\
&{\cal B}_m(\beta)\equiv (1+2\beta)(m^2-2+2\beta)\ ,\\
&{\cal C}(\beta)\equiv (1+2\beta)/(2\beta{\cal P}_0)\ ,\\
&{\cal H}_m(\beta,\xi)\equiv {\cal C}
{\cal N}_m{\cal A}_m+{\cal B}_m\ .
\end{split}
\end{equation}
With convenient notations (\ref{ABCHspiral}), equation (\ref{hoho4})
leads to the following stationary dispersion relation
\begin{equation}\label{SDPspiral}
(1-H_1{\cal A}_m)(1-H_2{\cal A}_m)=G_1G_2{\cal A}_m^2
\end{equation}
for coplanar logarithmic spiral perturbations
in a composite disc system.

Substituting expressions (\ref{spiralHG}) of $H_1$, $H_2$,
$G_1$ and $G_2$ into stationary dispersion relation
(\ref{SDPspiral}) and using the background condition
$D_g^2=\eta(D_s^2+1)-1$, we obtain one quadratic
equation in terms of $y\equiv D_s^2$, namely
\begin{equation}\label{spiral}
C_2y^2+C_1y+C_0=0\ ,
\end{equation}
where coefficients $C_2$, $C_1$ and $C_0$ are functions
of parameters $m$, $\beta$, $\delta$, $\eta$ and $\xi$,
and are defined by
\begin{equation}\label{spiralcoef}
\begin{split}
C_2=&{\cal B}_m{\cal H}_m\eta\ ,\\
C_1=&\bigg[({\cal B}_m-{\cal A}_m){\cal H}_m
+\frac{({\cal A}_m+{\cal B}_m)
({\cal H}_m-{\cal B}_m)}{(1+\delta)}\bigg]\eta\\
&\qquad\qquad\qquad
-\frac{({\cal A}_m+{\cal B}_m)
({\cal H}_m+{\cal B}_m\delta)}{(1+\delta)}\ ,\\
C_0=&\bigg[-{\cal A}_m{\cal H}_m
+\frac{({\cal A}_m+{\cal B}_m)
({\cal H}_m-{\cal B}_m)}{(1+\delta)}\bigg]\eta\\
&\qquad +({\cal A}_m+{\cal B}_m)^2
-\frac{({\cal A}_m+{\cal B}_m)
({\cal H}_m+{\cal B}_m\delta)}{(1+\delta)}\ .
\end{split}
\end{equation}
Given specific values for $m$, $\beta$, $\delta$ and $\eta$, we
readily solve quadratic equation (\ref{spiral}) analytically for
each `radial wavenumber' $\xi$ for stationary logarithmic spirals.
Again for a non-negative determinant as proven in Appendix A,
there are two real $D_s^2$ solutions to equation (\ref{spiral}),
namely
\begin{equation}
y_{1,2}=\frac{-C_1\pm (C_1^2-4C_2C_0)^{1/2}}{2C_2}\ .\nonumber
\end{equation}

In addition to the well-studied isothermal $\beta=0$ case (Lou
\& Shen 2003; Lou \& Zou 2004), we follow a similar procedure
of analyzing the aligned case to construct logarithmic spiral
configurations for the composite disc system in the range of
$\beta\in(-1/4,1/2)$. For astrophysical relevance, we
specifically choose $\beta=-1/8$ and $\beta=1/4$ as
illustrating examples. Values of
$\delta$ are chosen as 1/4, 1 and 4 with $\eta>1$. For $\eta=1$,
coefficients $C_2$, $C_1$ and $C_0$ as defined by expressions
(\ref{spiralcoef}) are again independent of $\delta$.

We now derive $y=D_s^2$ solutions below for the cases
of $\eta=1$ and $\eta\rightarrow\infty$ separately for
the same rationale as explained in the analysis of the
aligned case.
In terms of coefficients (\ref{ABCHspiral}), the two
explicit $D_s^2$ solutions to stationary dispersion
relation (\ref{spiral}) when $\eta=1$ are
\begin{equation}\label{spiralEta1}
Y_1^A=\frac{{\cal A}_m}{{\cal B}_m}\ \
\qquad\hbox{ and }\qquad
Y_1^B=\frac{(1-{\cal C}{\cal N}_m){\cal A}_m}{{\cal H}_m}\ .
\end{equation}
Since the ratio of sound speeds in the two discs are the same
for $\eta=1$, the composite system of two coupled discs may be
treated as one single disc to a certain extent. In fact, the
second expression $Y_1^B$ in equation (\ref{spiralEta1}) is
simply the result for the case of a single disc, while the
first expression $Y_1^A$ in equation (\ref{spiralEta1}) is
additional due to the gravitational coupling between the two
discs. Under some circumstances (e.g., $m=0\hbox{ and }1$) when
$Y_1^A$ remains always negative, we may practically regard the
two-disc system as being identical with the case of a single
disc for $\eta=1$.

The two explicit $D_s^2$ solutions to stationary
dispersion relation (\ref{spiral}) in the limit of
$\eta\rightarrow\infty$ are
\begin{equation}\label{spiralEtaInfty}
Y_{\infty}^A=\frac{{\cal A}_m[{\cal H}_m\delta
+(1-{\cal C}{\cal N}_m){\cal B}_m]}
{{\cal B}_m{\cal H}_m(1+\delta)}\ \
\quad\hbox{and}\quad \
Y_{\infty}^B=-1\ .
\end{equation}

Meanwhile, the phase relationship for surface mass
density perturbations reads from (\ref{hoho4}) as
\begin{equation}\label{spirPhR}
\frac{\mu^g}{\mu^s}=\frac{1-H_1{\cal A}_m}{G_1{\cal A}_m}
=-1-\frac{(D_s^2{\cal B}_m-{\cal A}_m)(1+\delta)}
{{\cal N}_m{\cal A}_m(1+D_s^2)}\ .
\end{equation}
As expected, all the expressions for the logarithmic spiral
case can be obtained from those for the aligned case by
simply replacing ${\cal P}_m$ with ${\cal N}_m$ resulting
from different potential-density pairs. This comes naturally
from the perspective that both aligned and spiral
configurations are coplanar density waves propagating
relative to the discs in either purely azimuthal directions
or both radial and azimuthal directions (Lou 2002; Lou \&
Shen 2003).

\subsubsection{Marginal Stability of Axisymmetric Disturbances}

It is reminded that for the aligned case, axisymmetric $m=0$
perturbations merely represent a rescaling of the background
equilibrium state of axisymmetry. For the $m=0$ case with radial
oscillations, we should not start from equations
(\ref{stationarys}) and (\ref{stationaryg}), but
instead\footnote{It turns out that the outcome of Shu et al.
(2000) with $\beta=0$ was all right in this regard because the two
different limiting procedures happen to give the same dispersion
relation. Lou (2002) made a mistake in this regard because
magnetic field breaks the degeneracy for the two different
limiting procedures (see subsection 3.2 and corrections in
Appendix B of Lou \& Zou 2004).} should use equation
(\ref{perturbs2}) by first setting $m=0$ with $\omega\neq 0$ and
then take the limit of $\omega\rightarrow 0$. It is then
straightforward to obtain
\begin{equation}\label{HGm0}
\begin{split}
&[1-H_1(\xi^2+1/4)]\mu^s=G_1(\xi^2+1/4)\mu^g\ ,\\
&[1-H_2(\xi^2+1/4)]\mu^g=G_2(\xi^2+1/4)\mu^s\ ,
\end{split}
\end{equation}
where coefficients $H_1$, $H_2$, $G_1$ and $G_2$ are evaluated by
simply setting $m=0$ in definitions (\ref{spiralHG}). It is clear
that equation (\ref{HGm0}) for the $m=0$ case with radial
oscillations differs from equation (\ref{hoho4}) for
non-axisymmetric spiral cases.\footnote{For the isothermal
$\beta=0$ case, the two limiting procedures lead to the
same result similar to the SID case of Shu et al. (2000).
One can see this by directly comparing equations
(\ref{HGm0}) and (\ref{hoho4}), which are identical
for $m=0$ if $\beta=0$.}
In definitions (\ref{ABCHspiral}), we only need to replace
${\cal A}_0$ by ${\cal A^{\prime}}_0=\xi^2+1/4$ and redefine
${\cal H}_0={\cal C}{\cal N}_0{\cal A^{\prime}}_0+{\cal B}_0$,
and then use the explicit solutions derived for $m>0$ spiral
cases in Section 3.2 by setting $m=0$. The marginal stability
curves thus obtained are displayed in Fig. 4$-$9. We show the
complete $y\equiv D_s^2$ solution structure versus $\xi$ while
only the portions above $y=0$ are physically sensible.

To analyze the solution properties as parameters $\beta$,
$\delta$, $\eta$ and `radial wavenumber' $\xi$ vary for
the axisymmetric $m=0$ case, we use asymptotic expression
(\ref{asymN}) for ${\cal N}_4(\xi)$ and then use recursion
relation (\ref{recurN}) to derive an approximate analytical
expression to compute the value of ${\cal N}_0(\xi)$, namely
\begin{equation}\label{N0}
\begin{split}
{\cal N}_0(\xi)
&=\frac{1}{2}\frac{\Gamma(1/4+\hbox{i}\xi/2)\Gamma(1/4-\hbox{i}\xi/2)}
{\Gamma(3/4+\hbox{i}\xi/2)\Gamma(3/4-\hbox{i}\xi/2)}\\
&\approx\frac{(9/4+\xi^2)(49/4+\xi^2)}
{(1/4+\xi^2)(25/4+\xi^2)(65/4+\xi^2)^{1/2}}\
\end{split}
\end{equation}
with relative errors less than $0.5\%$.

In the $\beta$ interval of $\beta\in(-1/4,1/2)$, the
quadratic coefficient $C_2$ of stationary dispersion
relation (\ref{spiral}) vanishes when
\begin{equation}\label{diverge}
{\cal H}_0={\cal C}{\cal N}_0{\cal A^{\prime}}_0+{\cal B}_0=0\ ,
\end{equation}
which for a specified $\beta$ determines the value of $\xi=\xi_c$
at which the value of $D_s^2$ will approach infinity. It is found
that in the open interval of $\beta\in(-1/4,1/2)$, there exists
only one such $\xi_c$ where ${\cal H}_0$ vanishes as can be
numerically computed\footnote{${\cal H}_0$ increases monotonically
with increasing $\xi$ and attains its minimum value at $\xi=0$
with ${\cal H}_{0\_min}<0$ for all $\beta\in(-1/4,1/2)$, implying
only one such $\xi_c$ where ${\cal H}_0=0$. See Appendix C for
details. Only for one exceptional case of $\beta\sim -0.130$, this
$\xi_c$ happens not to be a divergent point for $D_s^2$. See
Appendix D for details. }.
Moreover for $0<\xi<\xi_c$, we have ${\cal H}_0<0$ so that $C_2>0$,
while for $\xi>\xi_c$, we have ${\cal H}_0>0$ so that $C_2<0$. This
means that for wavenumber $\xi<\xi_c$, the upper branch is always
$y_1$, while for wavenumber $\xi>\xi_c$, the upper branch is always
$y_2$.

\begin{figure}
\centering
\includegraphics[scale=0.42]{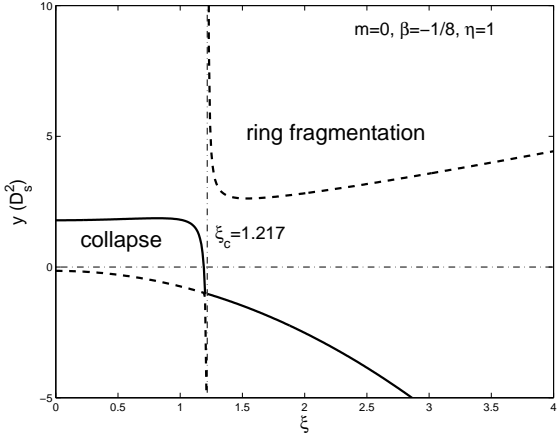}
\caption{Two branches of solutions to equation (\ref{spiral}) for
$m=0$, $\beta=-1/8$ and $\eta=1$. The divergent point is
$\xi_c=1.217$. In this case of $\eta=1$, the value of $\delta$ can
be arbitrary. The $y_1$ branch is plotted in heavy solid line
segments and the $y_2$ branch is plotted in heavy dashed line
segments.}
\end{figure}

First, we closely examine the $\beta=-1/8$ case. This critical
$\xi_c$ is numerically determined from equation (\ref{diverge}) as
$\xi_c=1.217$. We choose $\eta=1$ as a limiting case\footnote{For
$\eta=1$, we find the two-disc case is effectively identical with
the single-disc case as one solution branch (i.e.,
$Y_1^A={\cal A^{\prime}}_0/{\cal B}_0$) remains always negative
and is therefore unphysical. The other solution branch [i.e.,
$Y_1^B=(1-{\cal C}{\cal N}_0){\cal A^{\prime}}_0/{\cal H}_0$]
is simply the same solution of the single-disc case (Syer \&
Tremaine 1996).}
and the two explicit $D_s^2$ solutions
from equation (\ref{spiralEta1}) are
\begin{equation}
\begin{split}
&Y_1^A=\frac{{\cal A^{\prime}}_0}{{\cal B}_0}
=-\frac{16}{27}\xi^2-\frac{4}{27}\ ,\\
&Y_1^B=\frac{(1-{\cal C}{\cal N}_0){\cal A^{\prime}}_0}
{{\cal H}_0}=\frac{(1-{\cal C}{\cal N}_0)(\xi^2+1/4)}
{{\cal C}{\cal N}_0(\xi^2+1/4)-27/16}\ .
\end{split}
\end{equation}
We stress here that $Y_1^A$ and $Y_1^B$ do not correspond to $y_1$
and $y_2$ solutions, respectively, in a simple manner, because
signs of coefficients vary with $\xi$. For instance in Fig. 4,
we display $y_1$ in solid line and $y_2$ in dashed line. Across
$\xi_c$, solution structures changes abruptly. For physically
reasonable marginal stability curves, we only need the portions
above $y=0$. The two unstable regimes shown in Fig. 4 are the
ring fragmentation regime where a composite disc system rotates
too fast to be stable and the collapse regime where a composite
disc system rotates too slowly to be stable against large-scale
Jeans collapse (Lemos et al. 1991; Syer \& Tremaine 1996; Shu
et al. 2000; Lou 2002; Lou \& Fan 2002; Lou \& Shen 2003; Lou
\& Zou 2004). These marginal stability curves can also be derived
from the time-dependent WKBJ analysis by imposing the
scale-free disc conditions (Shen \& Lou 2003), with the more
straightforward $D_s-$criterion equivalent to the effective $Q$
parameters presented by Elmegreen (1995) and Jog (1996). By
varying the sound speed ratio $\eta$ and disc density ratio
$\delta$, we obtain similar marginal stability curves in Fig. 5.
The trends are qualitatively the same as the isothermal $\beta=0$
case (Lou \& Shen 2003; Shen \& Lou 2003). In other words, a
composite disc system is less stable as compared with a single
disc system for overall axisymmetric instabilities but becomes
more difficult for large-scale collapses (Lou \& Shen 2003;
Shen \& Lou 2003).

\begin{figure*}
\centering \subfigure[$m=0$, $\beta=-1/8$, $\delta=1/4$, $\eta=5$]{
\includegraphics[scale=0.42]{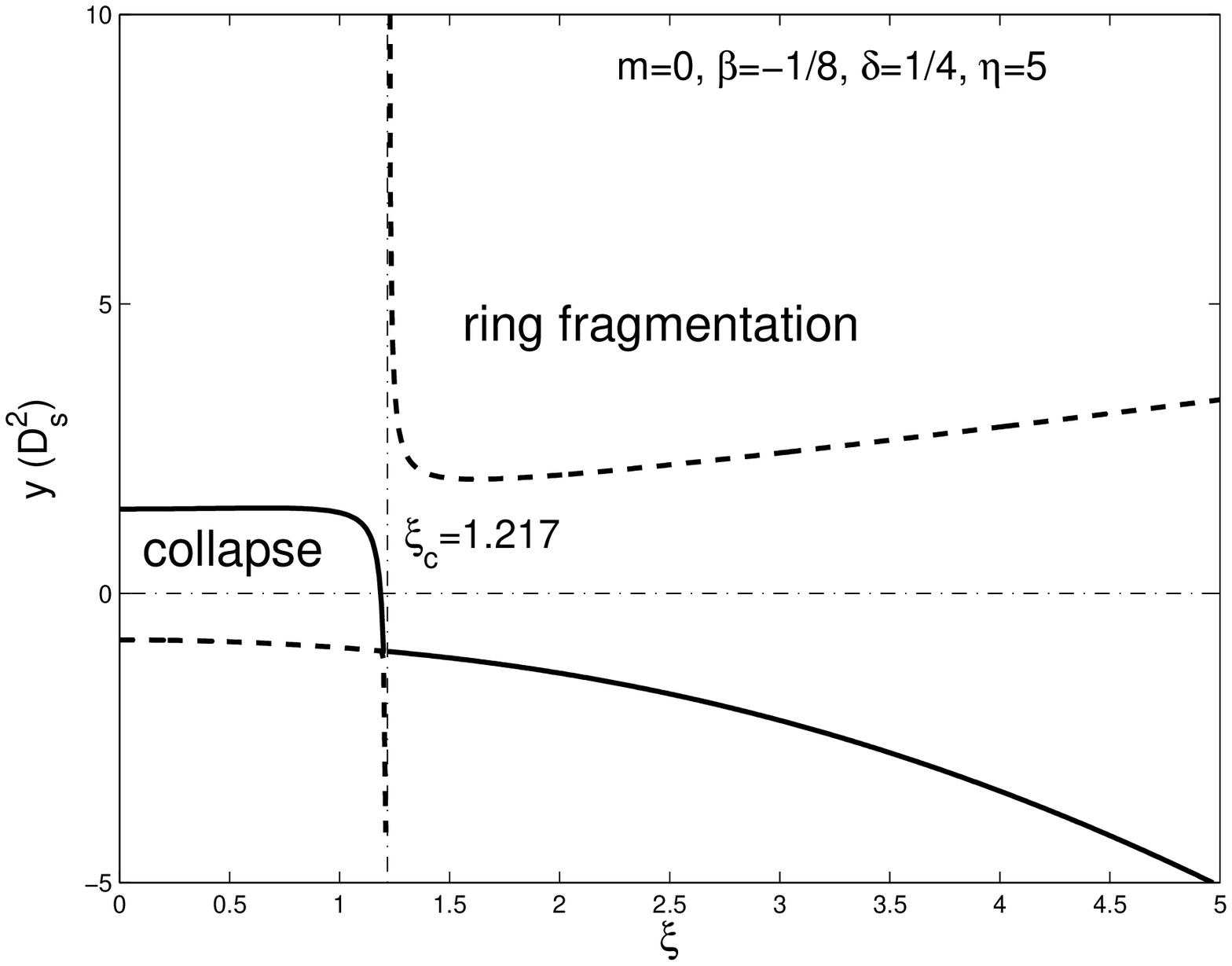}}
\subfigure[$m=0$, $\beta=-1/8$, $\delta=1$, $\eta=5$]{
\includegraphics[scale=0.42]{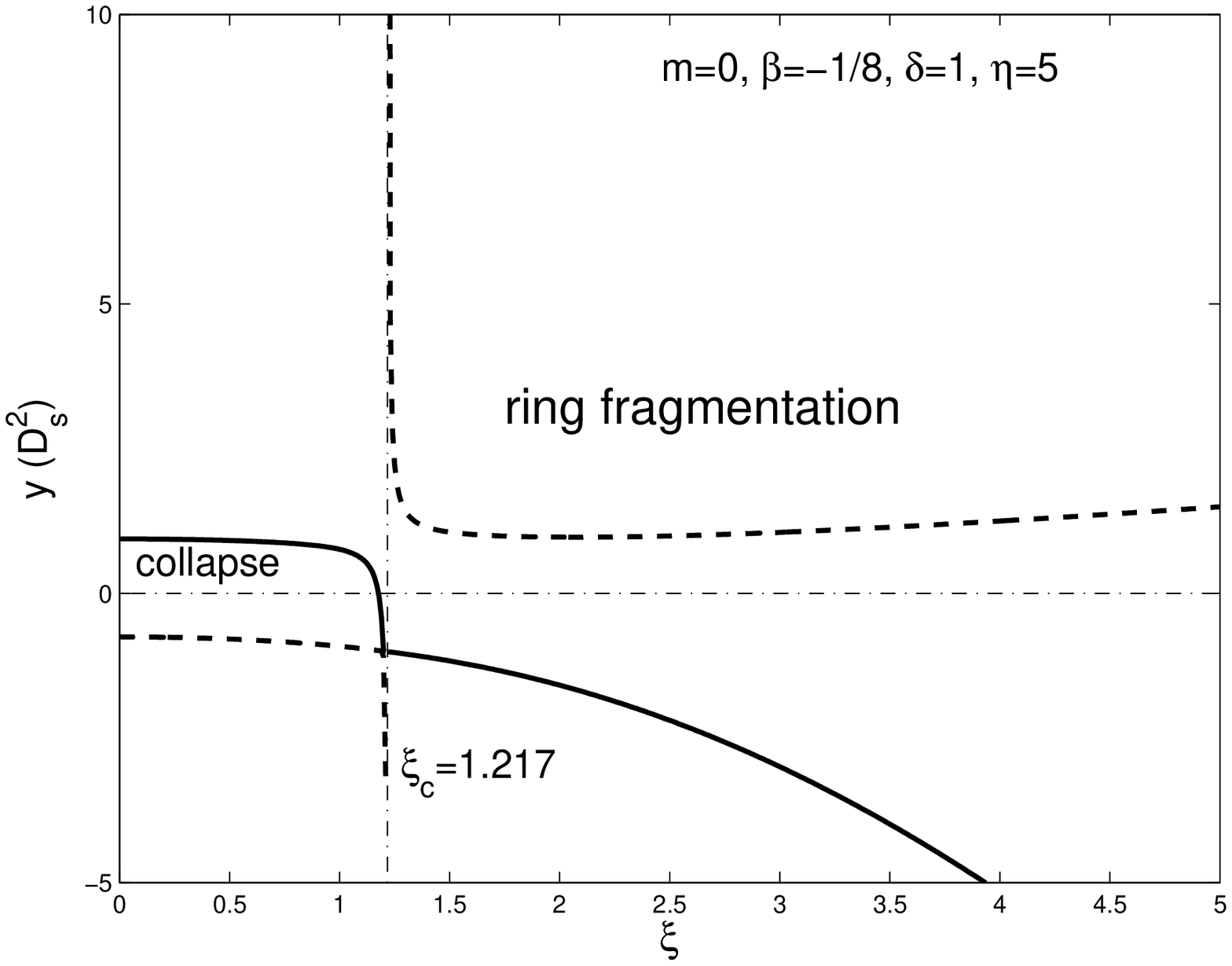}}
\subfigure[$m=0$, $\beta=-1/8$, $\delta=4$, $\eta=5$]{
\includegraphics[scale=0.42]{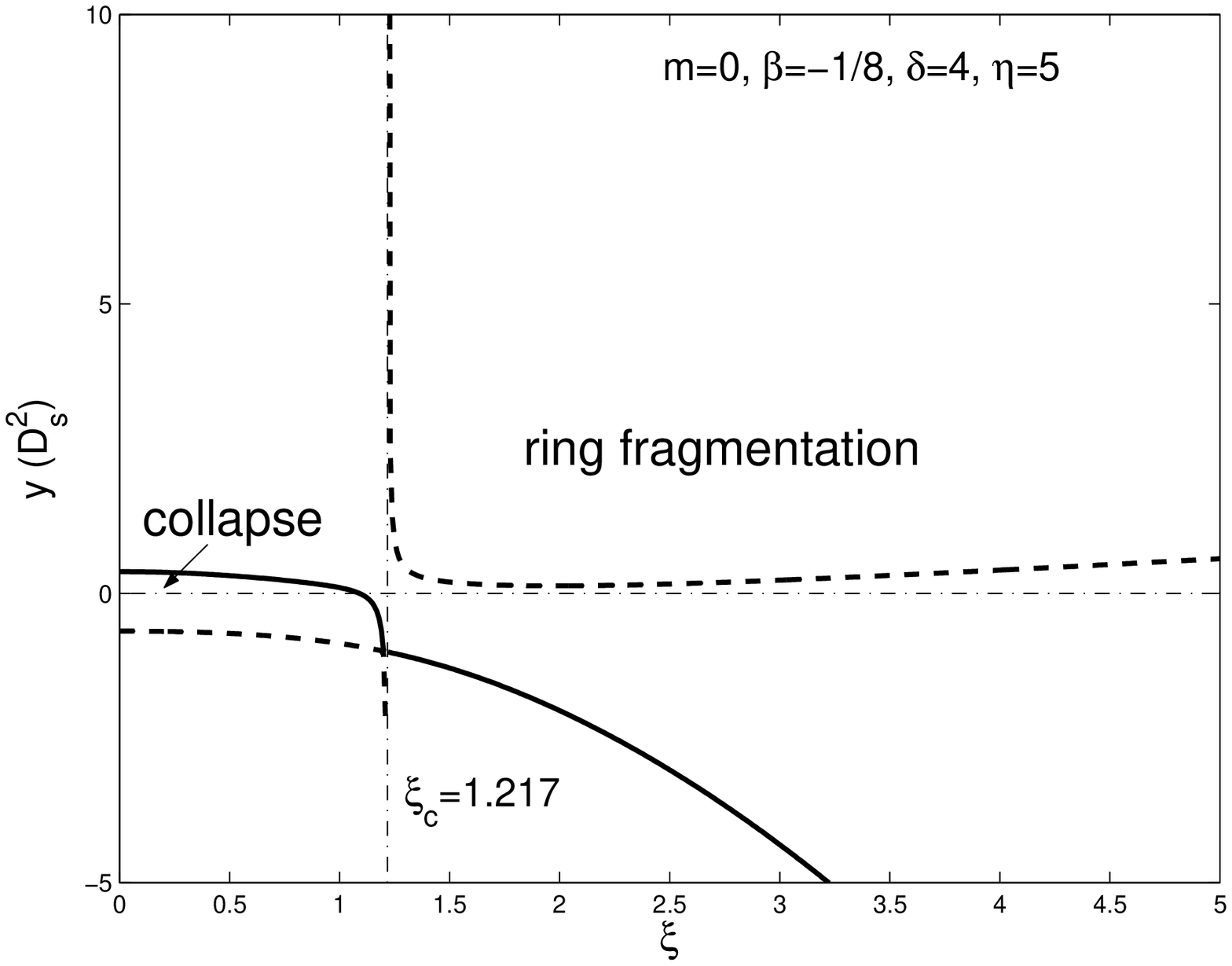}}
\subfigure[$m=0$, $\beta=-1/8$, $\delta=4$, $\eta=10$]{
\includegraphics[scale=0.42]{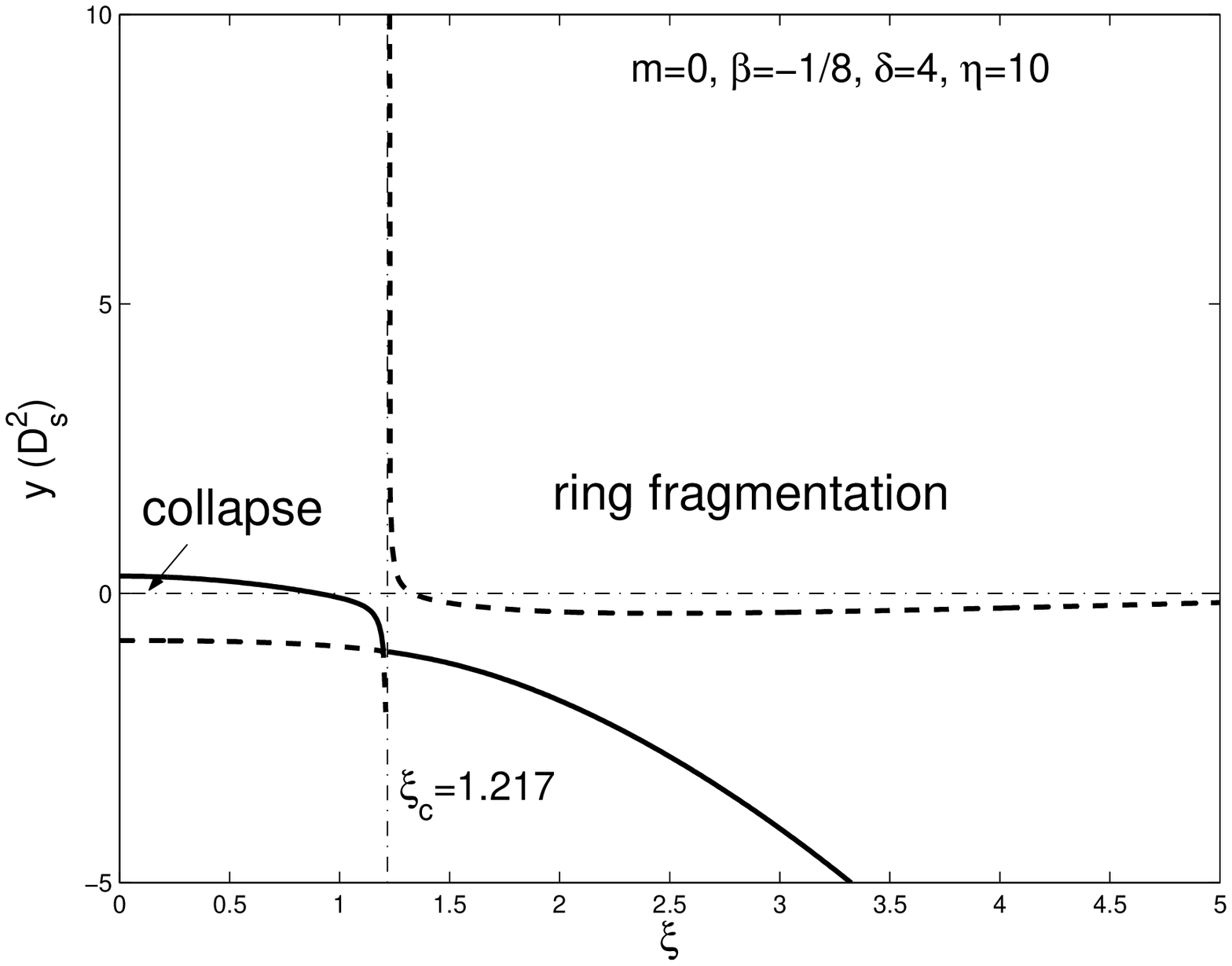}}
\caption{Two $D_s^2$ solution branches of stationary dispersion
relation (\ref{spiral}) with radial oscillations for $m=0$,
$\beta=-1/8$, $\eta=5,\ 10$ and $\delta=1/4,\ 1,\ 4$. In each
panel (a), (b), (c), (d), the segments above $y=0$ constitute the
marginal stability curves as indicated. }
\end{figure*}


Next, we consider the case of $\beta=1/4$. The divergent point now
becomes $\xi_c=3.159$. For qualitative results, we again start from
the special case of $\eta=1$ with the two $D_s^2$ solutions of
stationary dispersion relation (\ref{spiral}) explicitly given by
\begin{equation}
\begin{split}
&Y_1^A=\frac{{\cal A^{\prime}}_0}{{\cal B}_0}
=-4\xi^2/9-1/9\ ,\\
&Y_1^B=\frac{(1-{\cal C}{\cal N}_0){\cal A^{\prime}}_0}
{{\cal H}_0}=\frac{(1-{\cal C}{\cal N}_0)(\xi^2+1/4)}
{{\cal C}{\cal N}_0(\xi^2+1/4)-9/4}\ .
\end{split}
\end{equation}
The corresponding marginal stability curves are displayed
in Fig. 6. We further explored variations of the marginal
stability curves for different sets of parameters in Fig. 7.


\begin{figure}
\centering
\includegraphics[scale=0.42]{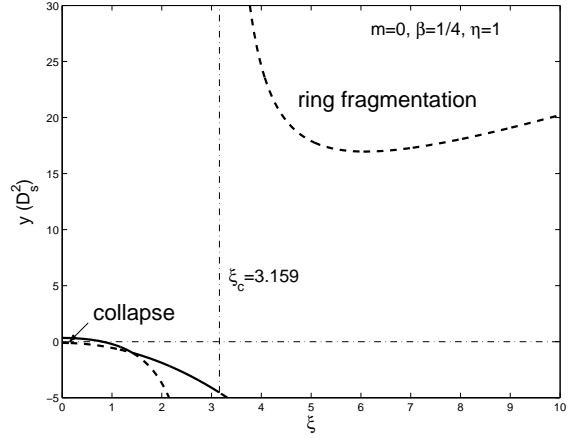}
\caption{Two branches of $D_s^2$ solutions to stationary
dispersion relation (\ref{spiral}) with radial oscillations for
$m=0$, $\beta=1/4$ and $\eta=1$. The divergent point is at
$\xi_{c}=3.159$. In this case of $\eta=1$, $\delta$ can be
arbitrary.}
\end{figure}

\begin{figure*}
\centering \subfigure[$m=0$, $\beta=1/4$, $\delta=1/4$, $\eta=5$]{
\includegraphics[scale=0.42]{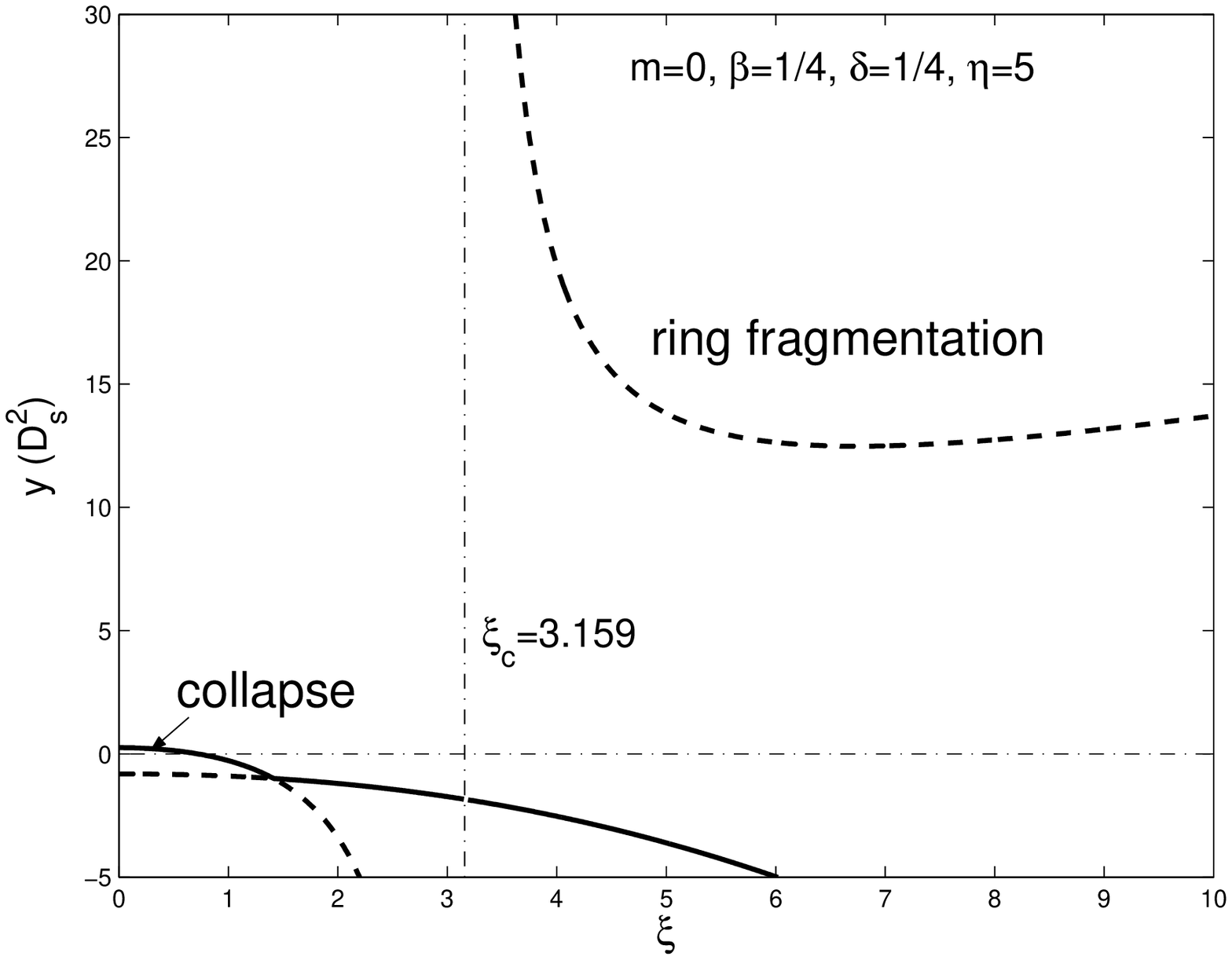}}
\subfigure[$m=0$, $\beta=1/4$, $\delta=1$, $\eta=5$]{
\includegraphics[scale=0.42]{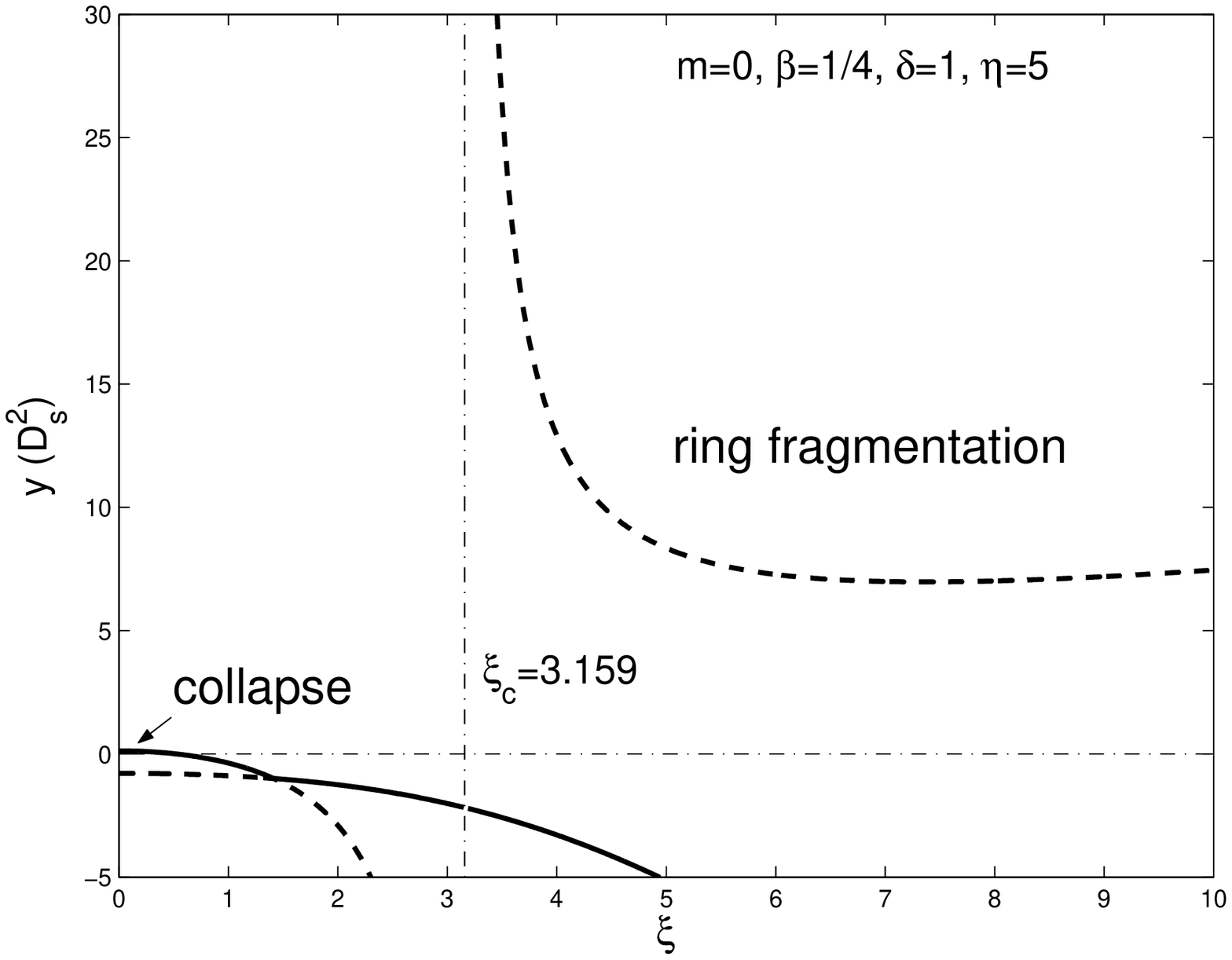}}
\subfigure[$m=0$, $\beta=1/4$, $\delta=4$, $\eta=5$]{
\includegraphics[scale=0.42]{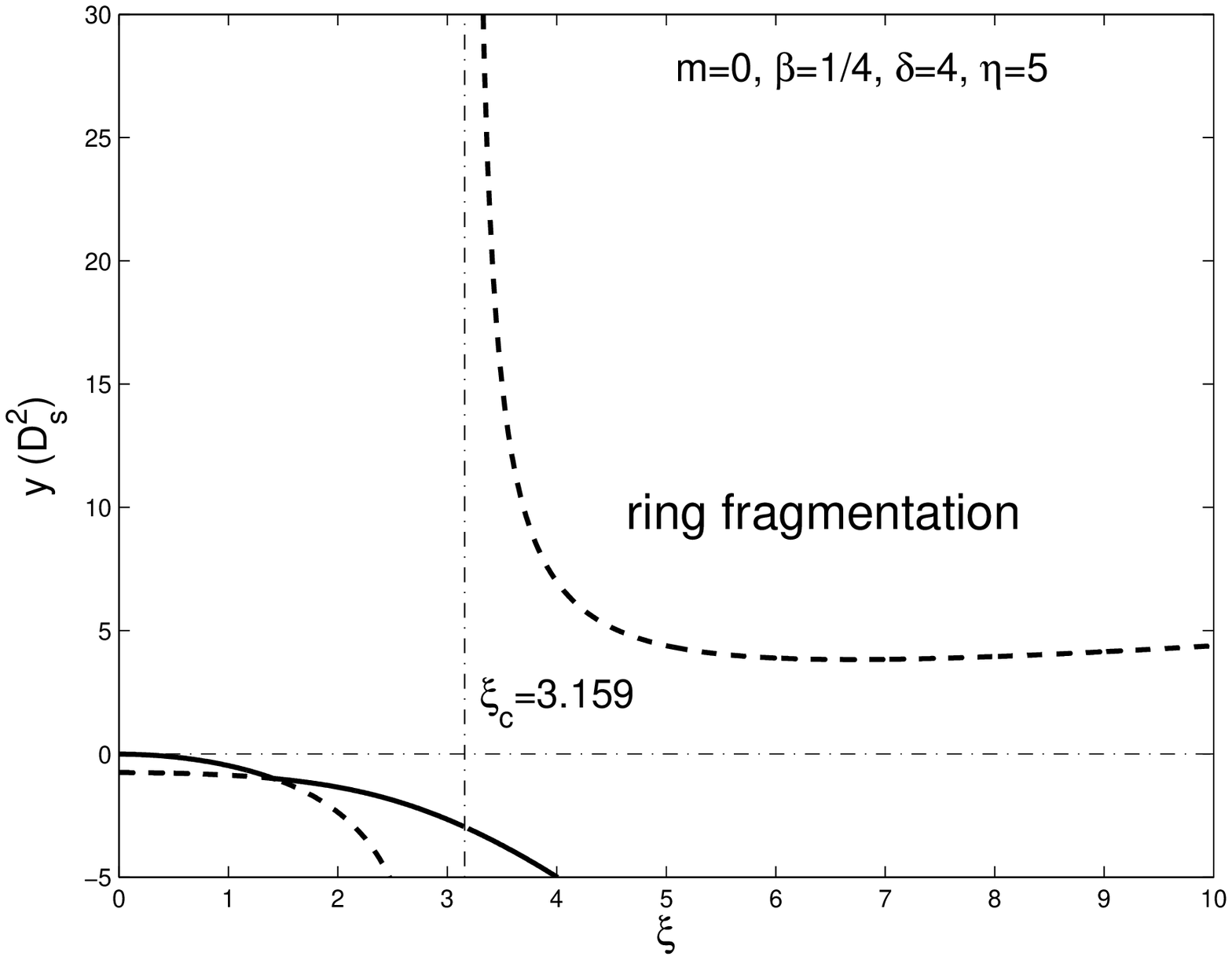}}
\subfigure[$m=0$, $\beta=1/4$, $\delta=4$, $\eta=10$]{
\includegraphics[scale=0.42]{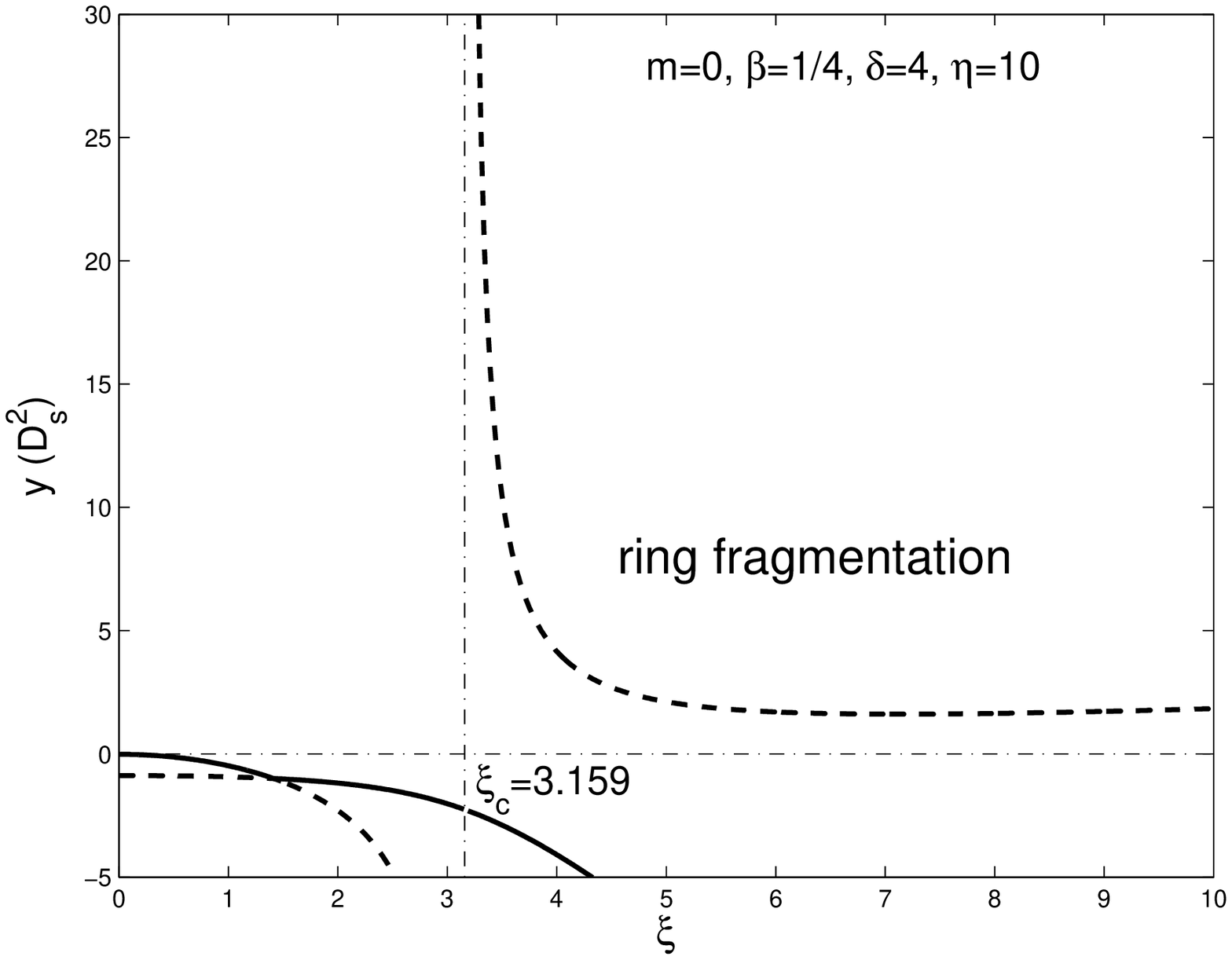}}
\caption{Two $D_s^2$ solution branches of stationary dispersion
relation (\ref{spiral}) with radial oscillations for $m=0$,
$\beta=1/4$, $\eta=5,\ 10$ and $\delta=1/4,\ 1,\ 4$. In each panel
(a), (b), (c), (d), the curve segments above $y=0$ constitute the
marginal stability curves as indicated. }
\end{figure*}

\begin{figure}
\centering
\includegraphics[scale=0.42]{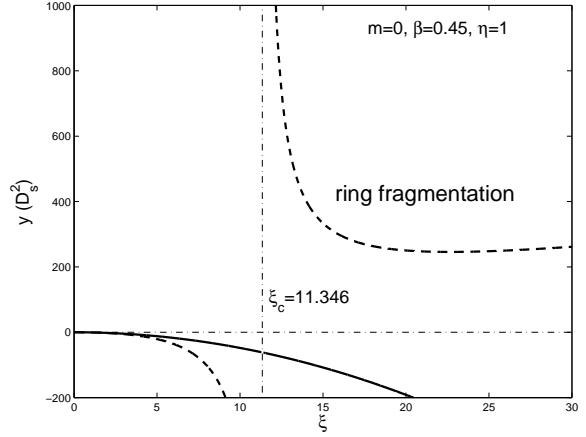}
\caption{Two $D_s^2$ solution branches to stationary dispersion
relation (\ref{spiral}) with radial oscillations for $m=0$,
$\beta=0.45$ and $\eta=1$. The divergent point is at
$\xi_{c}=11.346$. In this case of $\eta=1$, the value of $\delta$
can be arbitrary.}
\end{figure}


There exists a critical $\beta_c$ above which the collapse
regime disappears even for the $\eta=1$ (single disc) case
when the collapse region is largest.
This critical $\beta_c=0.436$ is determined by the condition
of zero collapsed regime for the maximum of the lower-left
branch, namely
\begin{equation}
Y_1^B=\frac{(1-{\cal C}{\cal N}_0){\cal A^{\prime}}_0}
{{\cal H}_0}\bigg|_{\xi=0}=0\ .
\end{equation}
In order to see this clearly, we take $\beta=0.45$ and
obtain marginal stability curves for $\eta=1$ as shown
in Fig. 8 where no collapse regime appears. This result
is consistent with that of Syer \& Tremaine (1996) as can
be seen from their fig. 2 for the marginal axisymmetric
stable curve in terms of their $w=1/[(1+2\beta)D_s^2]$
as noted earlier near the end of subsection 2.2 for
notational correspondences.

With the analysis technique developed by Shen \& Lou (2003),
we can perform time-dependent WKBJ analysis for a composite
system of two coupled scale-free discs described in Section 2.
For the above three cases with $\beta=-1/8$, $1/4$ and $0.45$,
we display contours of frequency $\omega^2$ in terms of
the effective radial wavenumber $\xi\equiv K\equiv|k|r$ and
the rotation parameter $D_s^2$ in the same figure of exact
stationary perturbation configurations in Fig. 9 for the $\eta=1$
case corresponding to the single disc case. The zero-frequency
lines (i.e., marginal stability curves) for both precise and
WKBJ approximation accord well with each other for large
radial wavenumber when the WKBJ approximation is valid; for
small radial wavenumber, the WKBJ approximation breaks down
and the two regimes differ significantly as expected.

In this context, we note the axisymmetric stability analysis by
Lemos et al. (1991) in a single-disc system. Lemos et al. (1991)
derived the same axisymmetric background equilibrium state
for a single disc. For perturbations, they imposed adiabatic
approximation with the adiabatic index $\gamma$ greater than
1 and independent of the barotropic index $n$ used for the
equilibrium state, which is different from Syer \& Tremaine
(1996) and our present analysis.

In the global analysis on axisymmetric $m=0$ instabilities with
radial oscillations, stationary $\omega=0$ wave patterns mark
the onset of instabilities (Lemos et al. 1991; Syer \& Tremaine
1996; Lou 2002; Lou \& Fan 2002; Shu et al. 2000; Lou \& Shen
2003; Shen \& Lou 2003; Lou \& Zou 2004) in a composite disc
system. Apparently, there are two unstable regimes, namely, the
long wavelength collapse regime and the short wavelength ring
fragmentation regime. Therefore, the stability criterion for
$D_s^2$ falls in a range whose width increases with increasing
$\beta$. Both regimes of the collapse instability and the ring
fragmentation instability are reduced for larger values of
$\beta$. As already noted, for $\beta>0.436$, the collapse
regime disappears completely. For sufficiently small values
of $\beta<-0.130$, the stable range of $D_s^2$ does not
exist (see Appendix D for details). As remarked earlier,
a composite system of two coupled discs is less stable than a
single disc system. The introduction of an additional gaseous
disc with larger $\delta$ and $\eta$ will reduce the overall
stable range of $D_s^2$, while tending to suppress the regime
of collapse for large-scale instabilities (Lou \& Shen 2003;
Lou \& Zou 2004).

\begin{figure}
\centering \subfigure[$m=0$, $\beta=-1/8$, $\eta=1$]{
\includegraphics[scale=0.42]{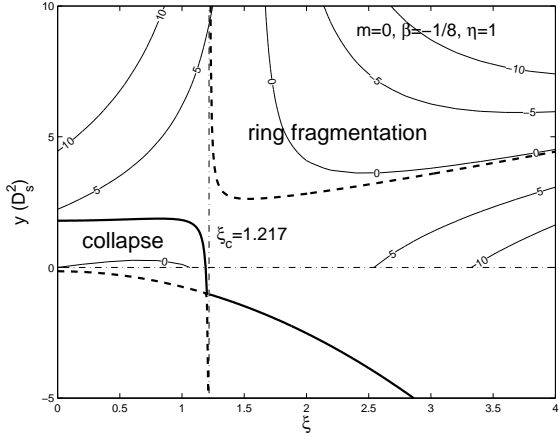}}
\subfigure[$m=0$, $\beta=1/4$, $\eta=1$]{
\includegraphics[scale=0.42]{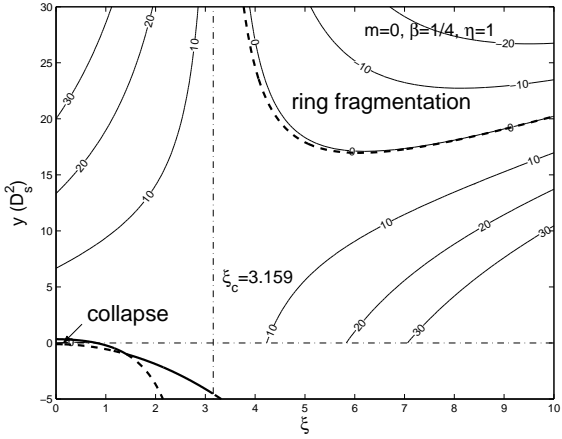}}
\subfigure[$m=0$, $\beta=0.45$, $\eta=1$]{
\includegraphics[scale=0.42]{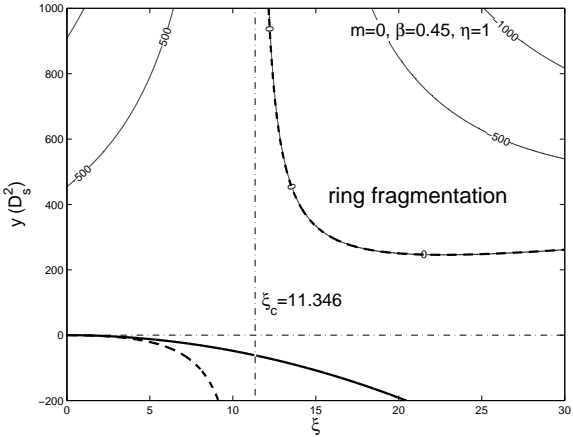}}
\caption{A comparison of stability boundary for exact global
stationary perturbation configurations obtained here for $m=0$
with radial oscillations and of the zero-frequency boundary in the
WKBJ analysis (Shen \& Lou 2003) for three cases: $\beta=-1/8$,
$1/4$ and $0.45$ with $\eta=1$. }
\end{figure}

\subsubsection{The Logarithmic Spiral $m=1$ Case}

The logarithmic spiral $m=1$ case behaves qualitatively
different from the aligned $m=1$ case. The reason is simply
that the additional radial wavenumber parameter $\xi$ will
also alter values of coefficients in equation (\ref{spiral}).

In this case, we again use asymptotic expression (\ref{asymN})
to estimate ${\cal N}_4(\xi)$ and then use recursion relation
(\ref{recurN}) to derive an approximate analytical expression
for ${\cal N}_1(\xi)$, namely
\begin{equation}\label{N1}
\begin{split}
{\cal N}_1&=\frac{1}{2}\frac{\Gamma(3/4+\hbox{i}\xi/2)
\Gamma(3/4-\hbox{i}\xi/2)}
{\Gamma(5/4+\hbox{i}\xi/2)\Gamma(5/4-\hbox{i}\xi/2)}\\
&\approx\frac{(25/4+\xi^2)(65/4+\xi^2)^{1/2}}
{(9/4+\xi^2)(49/4+\xi^2)}\
\end{split}
\end{equation}
with relative error less than $0.5\%$. With ${\cal A}_1
=\xi^2+5/4+2\beta$ and ${\cal B}_1=4\beta^2-1$ according
to definitions (\ref{ABCHspiral}), we immediately obtain
\begin{equation}\label{H1}
{\cal H}_1={\cal C}{\cal N}_1(\xi^2+5/4+2\beta)+4\beta^2-1\ ,
\end{equation}
which increases monotonically with increasing $\xi$ for fixed
$\beta$ values and attains its minimum value\footnote{For
$\beta\in(-1/4,1/2)$, the first order derivative of ${\cal H}_1$
with respect to $\xi$ is 0 at $\xi=0$ and positive for all
$\xi>0$, while the second order derivative of ${\cal H}_1$
with respect to $\xi$ is always positive at $\xi=0$.
Hence $\xi=0$ is a minimum for ${\cal H}_1$. While we have
obtained this result using the exact expression of ${\cal N}_1$,
the approximate expression (\ref{N1}) also works. See Appendix
C for more details.} at $\xi=0$ as
\begin{equation}\label{H1min}
{\cal H}_{1\_\hbox{min}}={\cal H}_1|_{\xi=0}>0\
\qquad\ \hbox{for \ \ \ \ $\beta\in(-1/4,1/2)$}\ .
\end{equation}
This condition (\ref{H1min}) means that no such $\xi$
exists to give ${\cal H}_1=0$. Together with other
relevant inequalities that ${\cal A}_1>0$,
${\cal B}_1<0$, ${\cal A}_1+{\cal B}_1>0$,
$0<{\cal C}{\cal N}_1<1$, ${\cal H}_1>0$ and $C_2<0$
hold for all $\xi>0$ within the open interval of
$\beta\in(-1/4,1/2)$, we know for sure that $y_2$
and $y_1$ are the upper and lower branch $D_s^2$
solutions, respectively. For $\eta=1$, we therefore
determine
\begin{equation}
\begin{split}
&y_1=Y_1^A|_{m=1}=\frac{{\cal A}_1}{{\cal B}_1}<0\ , \\
&y_2=Y_1^B|_{m=1}=\frac{(1-{\cal C}{\cal N}_1){\cal A}_1}
{{\cal H}_1}>0\ ,
\end{split}
\end{equation}
while in the limit of
$\eta\rightarrow\infty$, we obtain
\begin{equation}
\begin{split}
&y_1=Y_{\infty}^A|_{m=1}=-1+\frac{({\cal A}_1
+{\cal B}_1)({\cal H}_1\delta+{\cal B}_1)}
{{\cal B}_1{\cal H}_1(1+\delta)}<-1\ , \\
&y_2=Y_{\infty}^B|_{m=1}=-1 \\
&
\hbox{if inequality }\delta>-{\cal B}_1/{\cal H}_1
\hbox{ holds, or else}\\
&y_1=Y_{\infty}^B|_{m=1}=-1\ ,\\
&y_2=Y_{\infty}^A|_{m=1}=-1+\frac{({\cal A}_1
+{\cal B}_1)({\cal H}_1\delta+{\cal B}_1)}
{{\cal B}_1{\cal H}_1(1+\delta)}>-1\ \\
&\hbox{if inequalties } 0<\delta<-{\cal B}_1/{\cal H}_1
\hbox{ hold otherwise}\ .
\end{split}
\end{equation}
It then follows that the lower $y_1$ branch remains always
negative while the upper $y_2$ branch becomes positive if
\begin{equation}
1<\eta<\eta_c\equiv 1+\frac{(1-{\cal C}{\cal N}_1)
{\cal A}_1(1+\delta)}{[{\cal H}_1\delta
+(1-{\cal C}{\cal N}_1){\cal B}_1]}\ ,
\end{equation}
which further requires the following inequality
\begin{equation}
\delta>\delta_c\equiv\frac{({\cal C}{\cal N}_1-1)
{\cal B}_1}{{\cal H}_1}\
\end{equation}
to be valid.

This case is nearly identical with the case (III) in the aligned
$m=1$ case presented in subsection 3.1.1, except for the
replacement of ${\cal P}_1$ by ${\cal N}_1$ and different
definitions for ${\cal A}_1$, ${\cal H}_1$. The surface mass
density perturbations in the two coupled discs are always in-phase
for such configurations. As examples of illustration, we plot
several cases with $\beta=-1/8$ and $1/4$, $\delta=1/4, 1$ and $4$
and $\eta=1$ and $5$ in Fig. 10 in terms of $y\equiv D_s^2$ versus
$\xi$. Because the lower $y_1$ branch remains negative, we only
show the upper $y_2$ branch in Fig. 10.

\begin{figure*}
\centering \subfigure[$m=1$, $\beta=-1/8$]{
\includegraphics[scale=0.42]{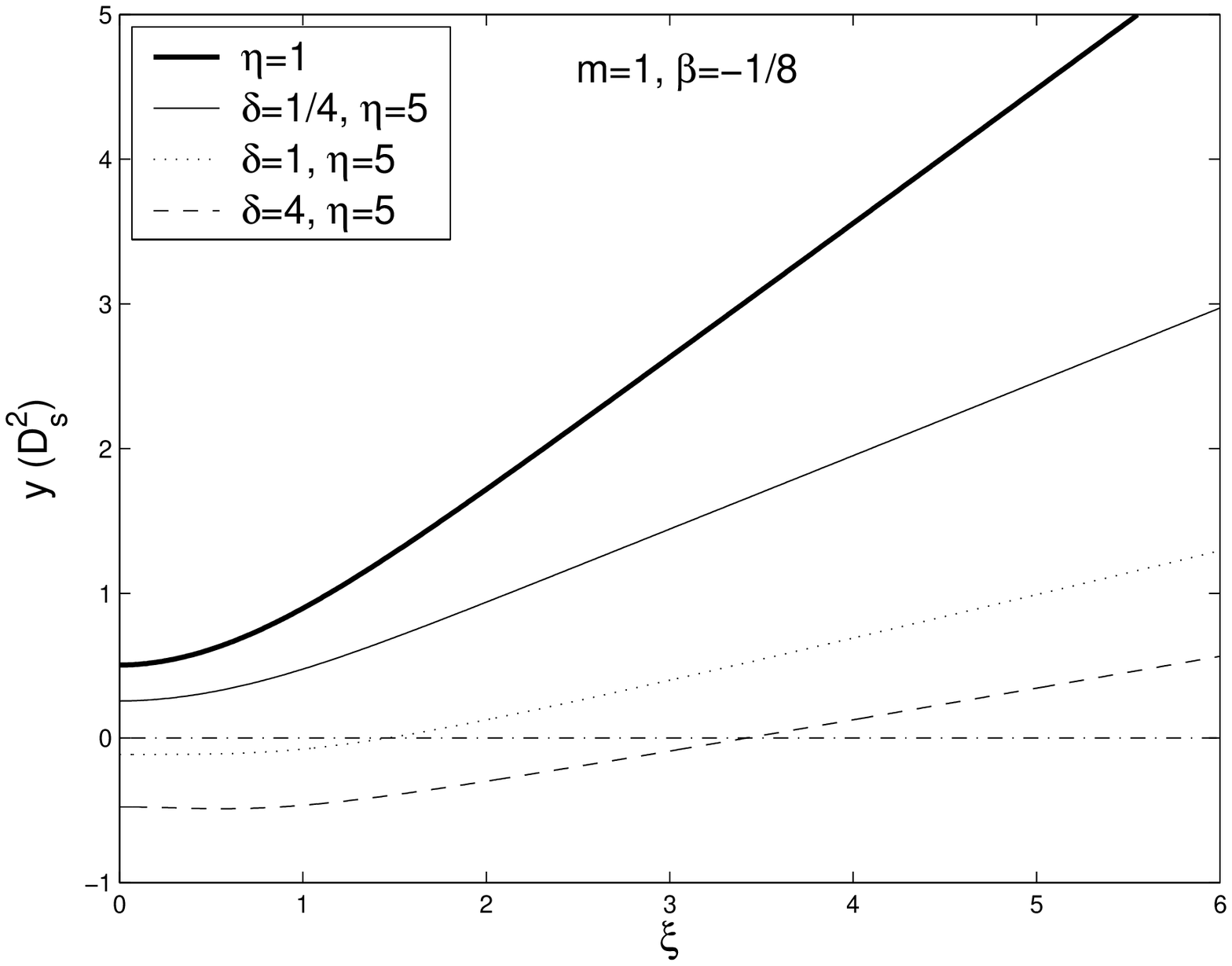}}
\subfigure[$m=1$, $\beta=1/4$]{
\includegraphics[scale=0.42]{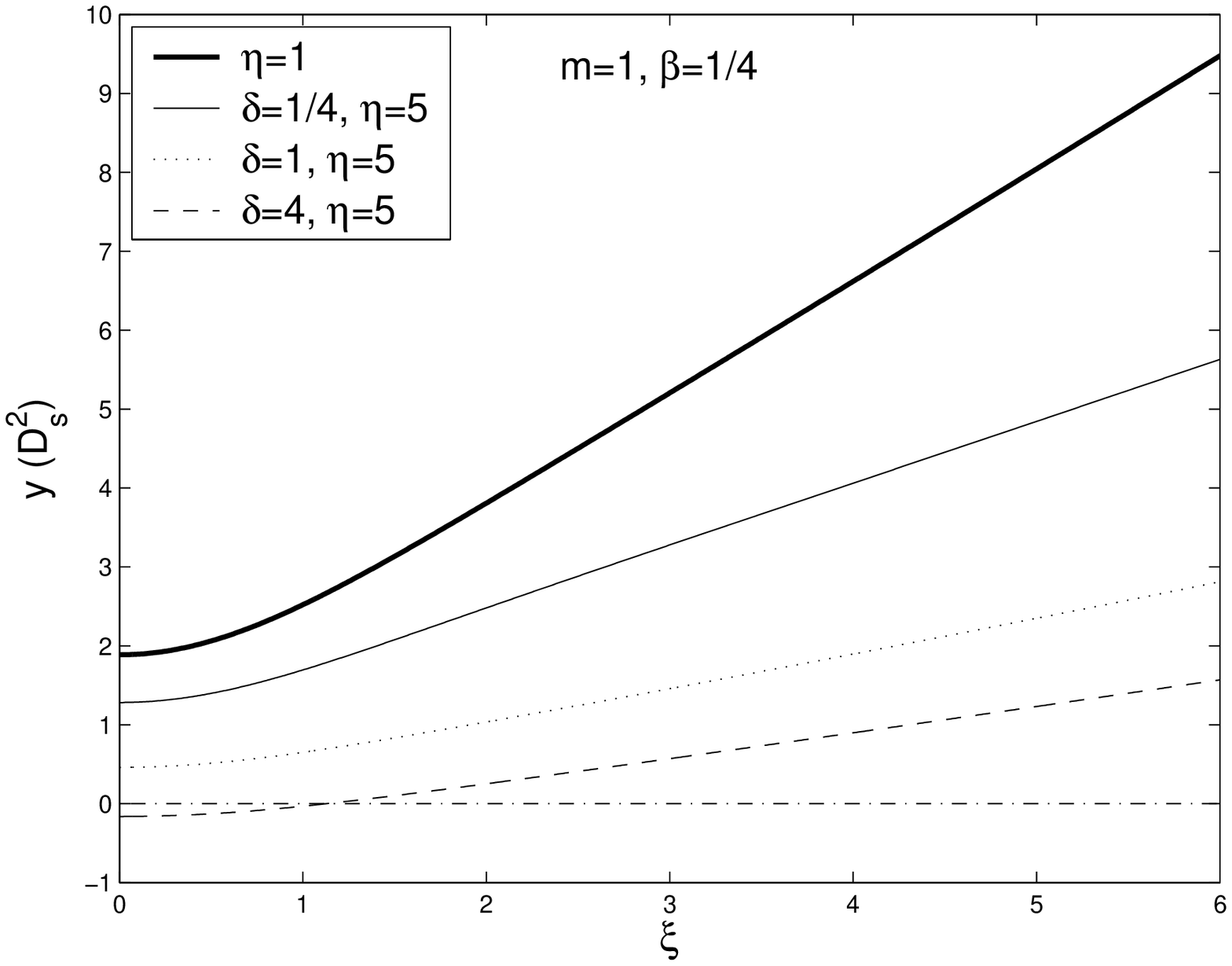}}
\caption{Five solution curves of $D_s^2$ from stationary
dispersion relation (\ref{spiral}) as functions of $\xi$ for
$m=1$, $\beta=-1/8, 1/4$, $\delta=1/4,\ 1,\ 4$ and $\eta=1, 5$.
Only the upper $y_2$ branches are shown here. For $\eta=1$,
solutions $D_s^2$ are independent of $\delta$.}
\end{figure*}
\begin{figure*}
\centering \subfigure[$m=2$, $\beta=-1/8$]{
\includegraphics[scale=0.42]{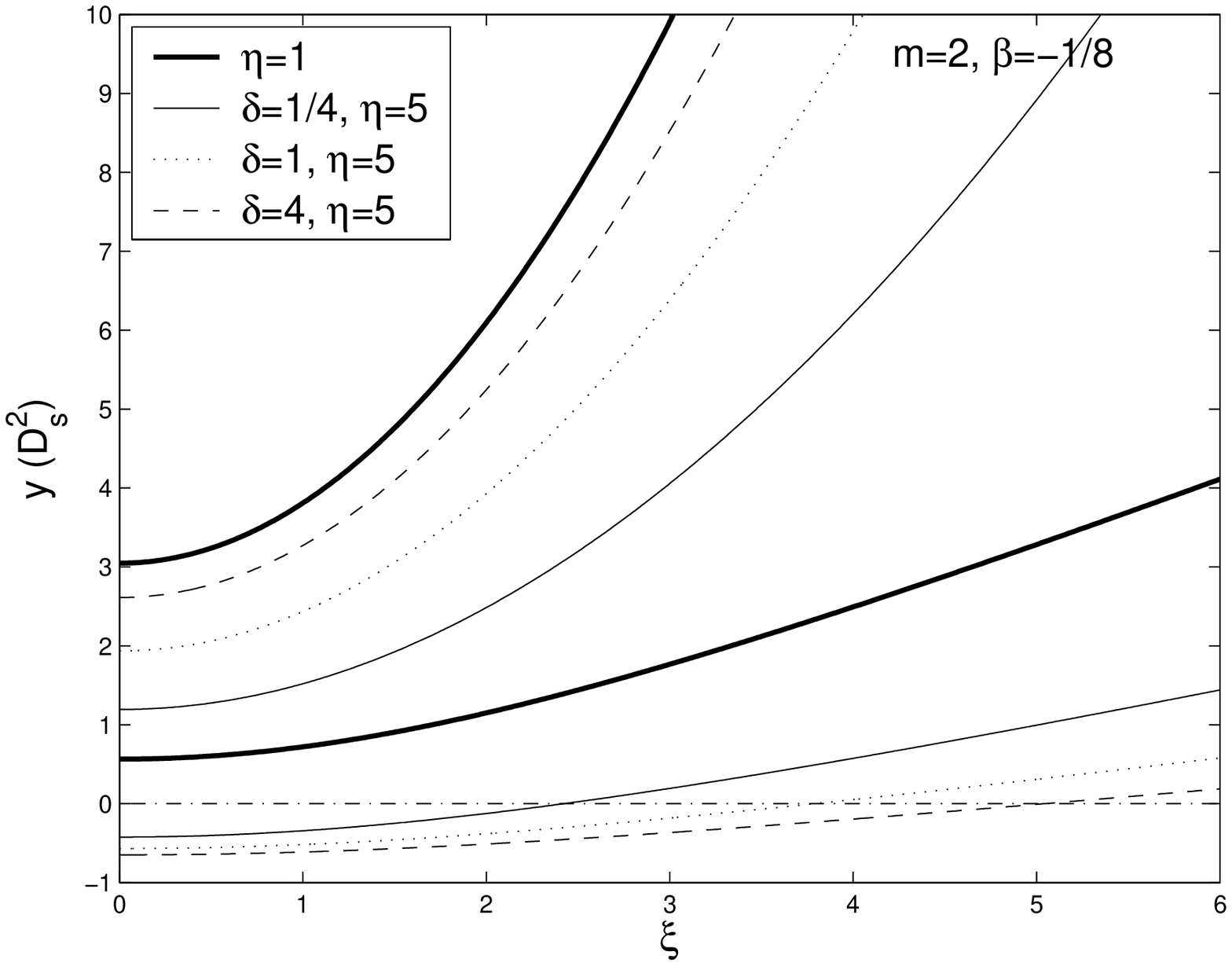}}
\subfigure[$m=2$, $\beta=1/4$]{
\includegraphics[scale=0.42]{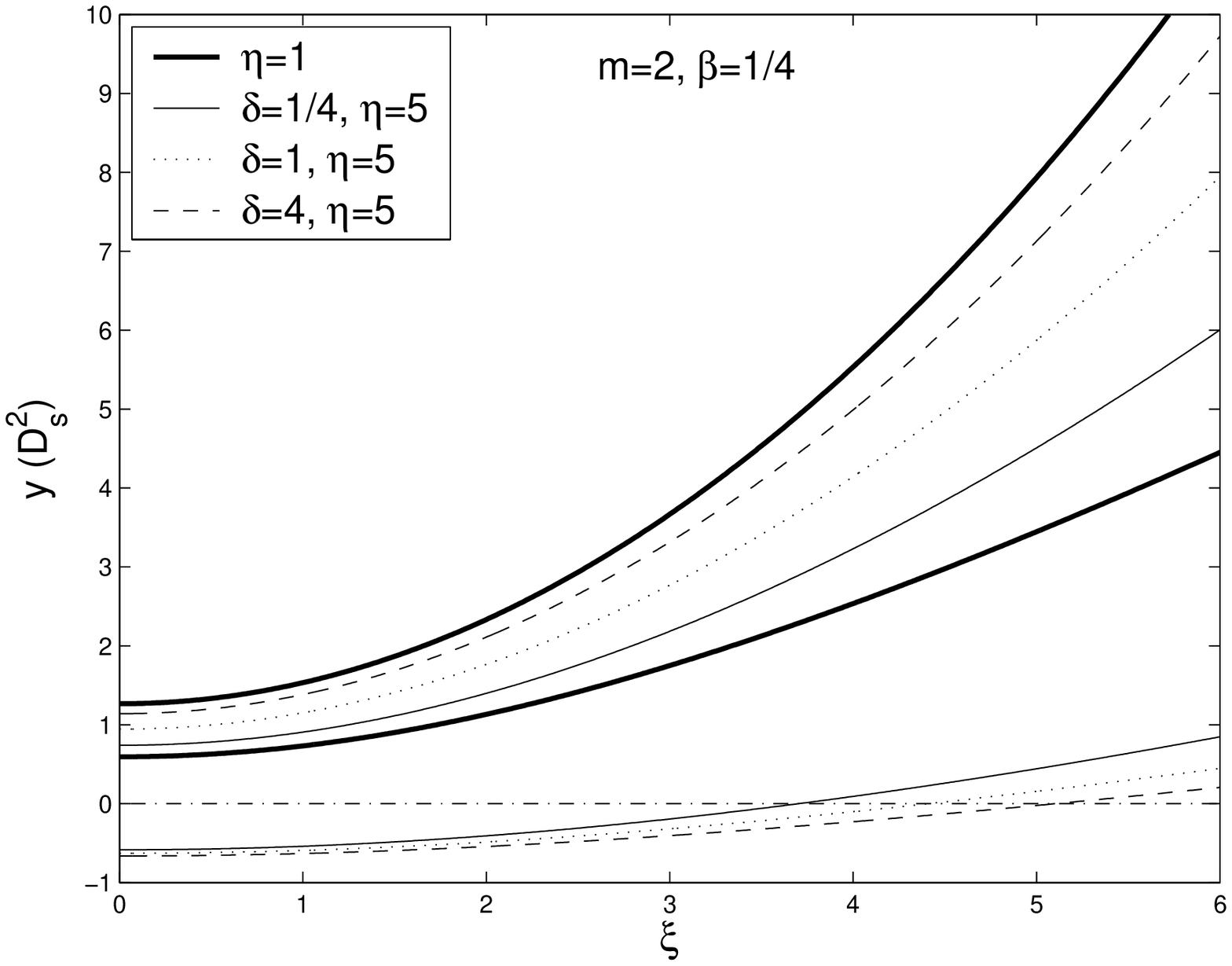}}
\caption{Two $D_s^2$ solution branches (given by the same
linetype) of stationary dispersion relation (\ref{spiral}) for
logarithmic spirals with $m=2$, $\beta=-1/8,\ 1/4$, $\delta=1/4,\
1,\ 4$ and $\eta=1,\ 5$. For $\eta=1$, these solutions are
independent of $\delta$. The lower branches are the counterparts
of the single-disc case. Note the variation in ordering of two
solution branches as $\eta$ changes.}
\end{figure*}

\subsubsection{Logarithmic Spiral Configurations with $m\ge 2$}

It turns out to be much simpler for $m\ge 2$ cases because when
$m\ge 2$, we always have inequalities ${\cal A}_m>0$, ${\cal
B}_m>0$, $0<{\cal C}{\cal N}_m<1$, ${\cal H}_m>0$ and hence
$C_2>0$ valid for all ranges of parameters under consideration.
This greatly simplifies the analysis, as $y_1$ and $y_2$ remain to
be upper and lower branches, respectively. Meanwhile, we have
$Y_1^A>Y_1^B>0$ and $Y_{\infty}^A>0>Y_{\infty}^B=-1$.

For $\eta=1$, we determine
\begin{equation}
\begin{split}
&y_1=Y_1^A=\frac{{\cal A}_m}{{\cal B}_m}>0\ ,\\
&y_2=Y_1^B=\frac{(1-{\cal C}{\cal N}_m){\cal A}_m}
{{\cal H}_m}>0\ ;\\
\end{split}
\end{equation}
while for $\eta\rightarrow\infty$, we have
\begin{equation}
\begin{split}
&y_1=Y_{\infty}^A=\frac{{\cal A}_m[{\cal H}_m\delta+(1-{\cal C}
{\cal N}_m){\cal B}_m]}{{\cal B}_m{\cal H}_m(1+\delta)}>0\ ,\\
&y_2=Y_{\infty}^B=-1\ .
\end{split}
\end{equation}
Therefore, the upper $y_1$ branch remains always positive,
while the lower $y_2$ branch first remains positive for
small $\eta$ and then becomes negative for $\eta$ greater
than a critical $\eta_c$ given explicitly by
\begin{equation}\label{spiralmge2etac}
\eta_c\equiv 1+\frac{(1-{\cal C}{\cal N}_m){\cal A}_m(1+\delta)}
{[{\cal H}_m\delta+(1-{\cal C}{\cal N}_m){\cal B}_m] }\ .
\end{equation}
This critical $\eta_c$ always exists for any given $\delta$
because $\eta_c$ remains always greater than 1 as dictated
by equation (\ref{spiralmge2etac}).

Entirely similar to the aligned $m\ge 2$ cases, the phase
relationship between surface mass densities $\mu^g$ and
$\mu^s$ for the upper $y_1$ branch of $D_s^2$ solution is
\begin{equation}
-1<\frac{\mu^g}{\mu^s}<-\frac{{\cal C}{\cal N}_m
{\cal A}_m\delta}{({\cal H}_m\delta+{\cal B}_m)}\ ,
\end{equation}
where the left-hand side corresponds to $\eta=1$ and the
right-hand side corresponds to $\eta\rightarrow\infty$.
This branch always has out-of-phase relationship between
surface mass densities $\mu^g$ and $\mu^s$ in the range
of $\beta\in(-1/4,1/2)$.

Meanwhile for the lower $y_2$ branch, the phase relationship
between surface mass densities $\mu^g$ and $\mu^s$ is determined
by
\begin{equation}
\delta<\frac{\mu^g}{\mu^s}
<\frac{(1+\delta )}{{\cal C}{\cal N}_m}-1\ ,
\end{equation}
where the left-hand side corresponds to $\eta=1$ and
the right-hand side corresponds to $\eta=\eta_c$ where
$D_s^2=y_2=0$. This branch always has in-phase
relationship between surface mass densities $\mu^g$
and $\mu^s$ in the prescribed $\beta$ and $\eta$ ranges.

For purpose of illustration, we present in Fig. 11 a
few solution examples in terms of $y\equiv D_s^2)$ as
functions of `radial wavenumber' $\xi$ for specific
parameters $m=2$, $\beta=-1/8$ and $1/4$,
$\delta=1/4,\ 1$ and $4$ and $\eta=1$ and $5$.

\section{Summary and Discussions}

The analysis presented in this paper is a generalization and
extension of the previous work by Syer \& Tremaine (1996), Shu
et al. (2000), Lou \& Shen (2003) and Shen \& Lou (2003). We
have constructed both aligned and logarithmic spiral, scale-free,
coplanar stationary perturbation configurations in a composite
system of two gravitationally coupled discs. While highly
idealized, we have in mind, at least conceptually, is a system
of spiral galaxy consisting of one stellar disc and one gaseous
disc, with a barotropic equation of state. This problem may then
have relevance to distributions of stellar mass and gas materials
in a disc galaxy.
Qualitatively, the two branches of solutions derived in this paper
suggest two possible coupled perturbation modes (not necessarily
stationary; see Lou \& Fan 1998b) where surface mass density
perturbations in the stellar disc and in the gaseous disc exhibit
either in-phase or out-of-phase correlations. These two distinctly
different classes of perturbation modes are mathematically
allowable, although there might be some kind of prevalence related
to initial conditions or other uncertainties. For observational
diagnostics of disc galaxies, one can obtain non-axisymmetric
stellar structures in the optical band (e.g. Rix \& Zaritsky 1995)
and derive H{\eightrm I} maps for the gaseous disc component (e.g.
Richter \& Sancisi 1994). These two maps in different wave bands
may be compared to see whether the stellar and H{\eightrm I}
gaseous arms are roughly coincident or apparently interlaced. The
real situation may be even more complicated than this. Active star
formation processes in the optical arms will further consume
H{\eightrm I} gas and the places where H{\eightrm I} gas clumps
will trigger more star formation activities. Depending on the
level of these interrelated processes, a phase shift between
optical arms and H{\eightrm I} arms may not be easily interpreted
in terms of the out-of-phase perturbation modes. Perhaps, the most
cogent evidence for out-of-phase density perturbations would be a
lopsided disc galaxy where the gaseous and stellar disc components
are comparable and the lopsidednesses for the two components are
opposite. In a broader perspective, the ideal two-fluid approach
adopted here may be applicable to other two-component disc system
where the two components can be treated as ideal fluids with
different temperatures, e.g., a composite disc system of stars and
dusts or a composite disc system composed of young massive stars
and relatively old stars, or even to composite disc systems with
more components. These different `hot' and `cold' fluid disc
components are coupled in the overall disc dynamics and contribute
to various structures in multi-band observations.

In order to satisfy the scale-free conditions in our model, both the
stellar and gaseous discs in an axisymmetric equilibrium state have
rotation curves $v\propto r^{-\beta}$ and surface mass densities
$\Sigma_0\propto r^{-1-2\beta}$ with a barotropic index
$n=(1+4\beta)/(1+2\beta)$ for the parameter regime of
$\beta\in(-1/4,1/2)$. For both cases of aligned and logarithmic spiral
perturbations, we derive sensible values of $D_s^2$ to support such
neutral or stationary density wave modes in an inertial frame of
reference. There are two classes of stationary density wave modes in a
composite system of two coupled discs in general; this is in contrast
to one class of stationary density wave modes in a single disc system.
We now summarize our main results below.

\begin{enumerate}
\item[(i)] {\it Aligned Stationary Configurations }

For aligned configurations, we focus on the coplanar
disturbances whose surface mass densities carry the
same cylindrical radial variations of the background
equilibrium state.

The aligned axisymmetric $m=0$ case represents merely
a rescaling from one axisymmetric equilibrium to a
neighbouring one (Shu et al. 2000; Lou 2002; Lou \&
Fan 2002; Lou \& Shen 2003; Lou \& Zou 2004), except
that the rescaling here happens in both discs
simultaneously.

In contrast to the eccentric $m=1$ case in a composite system
of two gravitationally coupled SIDs with $\beta=0$ (Lou \& Shen
2003), the aligned $m=1$ configurations in two coupled scale-free
discs with $\beta\neq 0$ are not trivial. Only one branch of
$D_s^2$ solution is physically sensible. For $\beta>-(2^{1/2}-1)/2$,
this branch of $D_s^2$ solution stands as the counterpart of the
single-disc case and the surface mass density perturbations in the
two discs are in-phase. For $\beta<-(2^{1/2}-1)/2$, this branch of
$D_s^2$ solution has no counterpart of the single-disc case
and surface mass density perturbations in the two discs are
out-of-phase. There may or may not exist a critical $\eta_c$,
determined by expression (\ref{alignetac1}), beyond which the $D_s^2$
solution becomes negative and thus unphysical, depending on the
value of $\delta$. Specific classifications and analyses have
been presented in subsection 3.1.1.

For $m\ge 2$ cases, we have derived two branches of solution
for possible values of the rotation speed parameter $D_s^2$
such that aligned stationary perturbation configurations are
sustained in a composite disc system. Of the two $D_s^2$
branches, one is always the upper branch and thus physical
for being positive and the coplanar surface mass density
perturbations in two discs are out-of-phase (see Lou \& Fan
1998). This branch of $D_s^2$ solution has no counterpart in
the case of a single disc. Meanwhile, the other branch of $D_s^2$
solution stands as the counterpart of the single-disc case with
coplanar surface mass density perturbations in the two discs
being in-phase. Furthermore, this second branch of $D_s^2$ solution
decreases with increasing $\eta$ and may become negative and thus
unphysical when $\eta$ exceeds a critical value $\eta_c$ that
varies with $\beta$, $m$ and $\delta$ as seen from definition
(\ref{alignetacm2}). In contrast to the aligned $m=1$ case, this
critical $\eta_c$ always exists for any values of $\delta$.
Details and specific examples can be found in subsection 3.1.2.

\item[(ii)] {\it Stationary Configurations of Logarithmic Spirals }

For unaligned or spiral coplanar perturbations, we consider
global logarithmic spiral configurations (Kalnajs 1971;
Syer \& Tremaine 1996; Shu et al. 2000; Lou 2002; Lou \& Fan
2002; Lou \& Shen 2003; Lou \& Zou 2004).

For the axisymmetric $m=0$ case with radial oscillations, we have
determined the marginal stability curves of $D_s^2$ versus the
`radial wavenumber' $\xi$ for various values of $\beta$. The
limiting case of $\eta=1$ reduces to the case of single disc case
as if the secondary mode due to the gravitationally coupling
between the two discs were absent. The axisymmetric stability
criterion is expressed in terms of the stellar rotation parameter
$D_s^2$. Those systems that rotate too slowly will succumb to
large-scale instabilities in the collapse regime, while those
systems that rotate too fast will fall into the ring-fragmentation
regime for short-wavelength instabilities (Safronov 1960; Toomre
1964; Shu et al. 2000; Lou 2002; Lou \& Fan 1998, 2002; Lou \&
Shen 2003; Shen \& Lou 2003; Lou \& Zou 2004). The stable range of
$D_s^2$ against overall axisymmetric instabilities is expanded for
larger $\beta$ values. When $\beta>\beta_c\sim 0.436$, the
large-scale collapse regime will disappear completely. On the
other hand when $\beta<-0.130$, the system cannot be stable at
all. The $D_s^2$ criterion for axisymmetric instabilities with
radial oscillations presented here is entirely equivalent to the
$w$ parameter used by Syer \& Tremaine (1996) for the the case of
a full single disc. The composite disc system becomes less stable
than the single-disc system. The overall stable $D_s^2$ range will
diminish for larger $\delta$ or $\eta$, while the large-scale
collapse instability by itself tends to be suppressed. Specific
examples and some analysis techniques are presented in subsection
3.2.1 and Appendixes C and D.

For the logarithmic spiral case of $m=1$ , the lower branch of
$D_s^2$ solution is always unphysical for being negative and the
limiting case of $\eta=1$ corresponds to just the single-disc
case. For a given `radial wavenumber' $\xi$, there may or may
not exist a critical $\eta_c$ beyond which the $D_s^2$ solution
becomes negative, depending on the value of $\delta$. This case
is almost identical with the aligned case (III) described in
subsection 3.1.1, except for a substitution of ${\cal P}_1$ with
${\cal N}_1$ and redefinitions for ${\cal A}_1$ and ${\cal H}_1$.
Details and specific examples can be found in subsection 3.2.2.

For logarithmic spiral cases with $m\ge 2$ , we have obtained
analytical results almost in the same forms of the aligned cases.
There are two possible values of rotation parameter $D_s^2$ that
can sustain stationary logarithmic spiral configurations in a
composite system. Of these two perturbation modes, one has no
counterpart in the single-disc case (Lou \& Fan 1998) with the
surface mass density perturbations in the two discs being
out-of-phase, while the other is the counterpart of the
single-disc case and has in-phase surface mass density
perturbations in the two discs. The out-of-phase mode always
exists while the in-phase mode disappears when $\eta$ exceeds
a certain critical value $\eta_c$ that depends on $\beta$, $m$,
$\delta$ and $\xi$ by equation (\ref{spiralmge2etac}). Specific
examples can be found in subsection 3.2.3.

\item[(iii)] {\it Non-Axisymmetric Instabilities }

For axisymmetric instabilities it is well established that neutral
modes mark the onset of instabilities. We have analytically
constructed non-axisymmetric neutral (i.e. stationary) modes for
both aligned and logarithmic spiral cases. Do these neutral modes
or stationary configurations mark the non-axisymmetric spiral
instabilities? Based on transmissions and over-reflections of
leading and/or trailing spiral density waves across corotation
in a time-dependent analysis, Shu et al. (2000) speculated that
the condition for stationary logarithmic spiral configurations
in a SID would determine whether spiral density waves could be
swing-amplified (Goldreich \& Lynden-Bell 1965; Fan \& Lou 1997).
The criterion of Shu et al. (2000) for swing amplification is
consistent with that of Goodman \& Evans (1999) in their
normal-mode analysis. It is appears worthwhile to pursue the
criterion for the onset of non-axisymmetric instabilities in
terms of a normal-mode analysis for a composite system that
will be performed in a separate paper.

\item[(iv)]  {\it Partial Discs }

All computations and discussions above deal with full discs. One
can also impose an axisymmetric gravitational potential associated
with a background dark matter halo of axisymmetry and ignore
disturbances in the dark matter halo caused by coplanar surface
mass density perturbations in the composite disc
system,\footnote{This simplifying approximation is crudely
justifiable for high velocity dispersions in a dark matter halo.
By numerical simulations, velocity dispersions of dark matter halo
are presumably of the order of a few hundred kilometers per
second. } that is, the composite system is composed of two partial
discs (Syer \& Tremaine 1996; Shu et al. 2000; Lou 2002; Lou \&
Fan 2002; Lou \& Shen 2003; Shen \& Lou 2003; Lou \& Zou 2004). By
defining a dimensionless factor $0<{\cal F}<1$ for the ratio of
the gravitational potential arising from the two discs together to
that of the whole system including the dark matter halo, one may
follow the same procedure for full discs to analyze the problem of
a composite system of two coupled partial discs. Practically, what
one needs to do is to replace all ${\cal P}_m$ and ${\cal N}_m$
with ${\cal F}{\cal P}_m$ and ${\cal F}{\cal N}_m$, respectively.
The dynamical effect of this background dark matter halo tends to
suppress axisymmetric instabilities (Syer \& Tremaine 1996; Shu et
al. 2000; Lou 2002; Lou \& Fan 2002; Lou \& Shen 2003; Shen \& Lou
2003; Lou \& Zou 2004).

\item[(v)]  {\it Magnetized Discs }

By synchrotron radio observations, one can infer spiral
magnetic field structures in nearby spiral galaxies
(e.g., Lou \& Fan 2003 and references therein). It is
believed that this is generically true for distant spiral
galaxies as well. The presence of magnetic fields in
spiral galaxies should affect global star formation
rates and thus influence the evolution of disc galaxies.
Technically, the interesting problem of including magnetic
fields can become quite involved. More specifically, a
stellar disc is gravitationally coupled with a magnetized
gaseous disc. There are two relatively simple geometries
to model configurations of magnetic fields. The first one
is the so-called isopedically magnetized configurations
(Shu \& Li 1997; Shu et al. 2000; Lou \& Wu 2004). The
second one is to consider coplanar azimuthal magnetic
fields with their strengths scaled as powers of
cylindrical radius $r$ (Lou 2002; Lou \& Fan 2002; Lou
\& Zou 2004). Now with a more general scale-free disc system
investigated in this paper, it would be very interesting to
model magnetic fields for both isopedic and azimuthal
configurations in a more general composite disc system,
a problem also to be presented in a separate paper.

\end{enumerate}

\section*{Acknowledgments}
This research has been supported in part by the ASCI Center for
Astrophysical Thermonuclear Flashes at the University of Chicago
under Department of Energy contract B341495, by the Special Funds
for Major State Basic Science Research Projects of China, by the
Tsinghua Center for Astrophysics, by the Collaborative Research
Fund from the NSF of China (NSFC) for Young Outstanding Overseas
Chinese Scholars (NSFC 10028306) at the National Astronomical
Observatory, Chinese Academy of Sciences, by NSFC grant 10373009
at the Tsinghua University, and by the Yangtze Endowment from the
Ministry of Education through the Tsinghua University. Affiliated
institutions of Y.Q.L. share this contribution.

\appendix
\section{Two Real $D_s^2$ Solutions}

We here show that equation (\ref{aligned}) always
has two different real solutions. The determinant
of equation (\ref{aligned}) reads
\begin{equation}
\begin{split}
\Delta\equiv C_1^2-4C_2C_0=\bigg(\frac{{\cal A}_m
+{\cal B}_m}{1+\delta}\bigg)^2\wp\
\end{split}
\end{equation}
where
\begin{equation}
\wp\equiv c_2\eta^2+c_1\eta+c_0
=c_2\bigg(\eta+\frac{c_1}{2c_2}\bigg)^2
-\frac{c_1^2-4c_0c_2}{4c_2}\ ,
\end{equation}
and
\begin{equation}
\begin{split}
&c_2\equiv ({\cal B}_m+{\cal H}_m\delta)^2\ ,\\
&c_1\equiv 2\delta({\cal H}_m-{\cal B}_m)^2
-2{\cal B}_m{\cal H}_m(1+\delta)^2\ ,\\
&c_0\equiv ({\cal H}_m+{\cal B}_m\delta)^2\ .
\end{split}
\end{equation}
It follows further that
\begin{equation}
\Delta_1\equiv c_1^2-4c_0c_2=-16{\cal B}_m{\cal H}_m
({\cal H}_m-{\cal B}_m)^2\delta(1+\delta)^2\ .
\end{equation}
Now in the open interval of $\beta\in(-1/4,1/2)$ and for $m\ge 2$,
we have $c_2>0$, $\Delta_1<0$ and thus $\wp>0$ for all $\eta\ge 1$.
We have thus proven that $\Delta>0$. While the proof procedure
is slightly different for the $m=1$ case, we also can show that
$\Delta>0$. It follows that equation (\ref{aligned}) always has
two different real solutions for $D_s^2$ corresponding to the
upper and lower branches, respectively.

The same proof procedure can be repeated for the logarithmic
spiral cases corresponding to stationary dispersion relation
(\ref{spiral}),
which always has $\Delta>0$ for $m\ge 1$. Moreover in
the same spirit, we can show $\Delta\ge 0$ for the
axisymmetric $m=0$ case. The equal sign corresponds
to $({\cal A}_0^{\prime}+{\cal B}_0)=0$.

\section{Monotonic Functions}

We here provide proofs that the two $D_s^2$ solutions of
dispersion relation (\ref{aligned}) for the aligned case
and of dispersion relation (\ref{spiral}) for logarithmic
spiral cases are monotonic functions of $\eta$. Therefore,
one can make use of the explicit solutions at $\eta=1$ and
in the limit of $\eta\rightarrow\infty$ to well bracket
the $D_s^2$ solution ranges.

First, we rewrite the coefficients $C_2$, $C_1$ and $C_0$
as defined by either expressions (\ref{alignedcoeff}) for
aligned perturbations or expressions (\ref{spiralcoef})
for logarithmic spiral perturbations, namely
\begin{equation}
\begin{split}
&C_2=a_2\eta\ ,\\
&C_1=a_1\eta+b_1\ ,\\
&C_0=a_0\eta+b_0\ ,
\end{split}
\end{equation}
where coefficients $a_2$, $a_1$, $b_1$, $a_0$ and $b_0$ are
determined by directly comparing expressions (\ref{alignedcoeff})
or (\ref{spiralcoef}) of the actual coefficients $C_2$, $C_1$
and $C_0$ that appear in the main text.

With the two $D_s^2$ solutions explicitly given
by $y_{1,2}=(-C_1\pm\Delta^{1/2})/2C_2$, we have
\begin{equation}
\frac{dy_{1,2}}{d\eta}=\frac{\Delta^{1/2}
(C_1C_2^{\prime}-C_2C_1^{\prime})
+(C_2\Delta^{\prime}-\Delta C_2^{\prime})}
{4C_2^2\Delta^{1/2}}\ ,
\end{equation}
where $\prime$ denotes the derivative with respect to
$\eta$ and $\Delta\equiv C_1^2-4C_2C_0$ is the determinant
that has been proven non-negative in Appendix A.

We next consider to solve equation
$y'_{1,2}\equiv dy_{1,2}/d\eta=0$ that is
equivalent to the following condition
\begin{equation}
4\eta^2a_2(b_0^2a_2-a_1b_1b_0+a_0b_1^2)=0\ .
\end{equation}
By straightforward substitutions of the expressions
of $a_2$, $a_1$, $b_1$, $a_0$, $b_0$, this condition
turns out to be
\begin{equation}
4\eta^2{\cal B}_m{\cal H}_m\frac{({\cal A}_m+{\cal B}_m)^4
({\cal H}_m-{\cal B}_m)^2\delta}{(1+\delta)^2}=0\ ,
\end{equation}
which gives only one solution $\eta=0$ for any given sets
of $\{m,\ \beta,\ \delta\}$ for aligned perturbations or
$\{m,\ \beta,\ \delta,\ \xi\}$ for logarithmic spiral
perturbations. Since $dy_{1,2}/d\eta$ are continuous
functions of $\eta$, it then follows that for $\eta>1$,
$dy_{1,2}/d\eta$ remain either always positive or always
negative for any specified sets of
$\{m,\ \beta,\ \delta\}$ for aligned perturbations or
$\{m,\ \beta,\ \delta,\ \xi\}$ for logarithmic spiral
perturbations. In summary, the two $D_s^2$ solutions
$y_{1,2}$ must be monotonic functions of $\eta$ once
$\{m,\ \beta,\ \delta\}$ for aligned perturbations or
$\{m,\ \beta,\ \delta,\ \xi\}$ for logarithmic spiral
perturbations are specified.

\section{Properties of ${\cal H}_{\it m}$ }

We here study the variation of ${\cal H}_m$ with respect to
$\xi$ in the logarithmic spiral case. For $m>0$, we consider
\begin{equation}
\begin{split}
&{\cal A}_m=m^2+\xi^2+1/4+2\beta\ ,\\
&{\cal B}_m=(1+2\beta)(m^2-2+2\beta)\ ,\\
&{\cal C}=(1+2\beta)/(2\beta{\cal P}_0)\ ,\\
&{\cal H}_m={\cal C}{\cal N}_m{\cal A}_m+{\cal B}_m\ .
\end{split}
\end{equation}
It is then straightforward to show
\begin{equation}
\frac{d{\cal H}_m}{d\xi}={\cal C}
{\cal N}_m(\Psi_m{\cal A}_m+2\xi)\ ,
\end{equation}
where
\begin{equation}
\begin{split}
\Psi_m(\xi)&=\frac{i}{2}[\psi(m/2+i\xi/2+1/4)
+\psi(m/2-i\xi/2+3/4)\\
&-\psi(m/2-i\xi/2+1/4) -\psi(m/2+i\xi/2+3/4)]\ ,
\end{split}
\end{equation}
where $\psi$ is the digamma function (or $\psi-$function)
defined by $\psi(x)\equiv d[\ln \Gamma(x)]/dx$ with
$\Gamma(x)$ being the $\Gamma-$function.
In the above derivation, we
have used the following relation
\begin{equation}
\frac{d{\cal N}_m}{d\xi}={\cal N}_m\Psi_m\ .
\end{equation}

It is also straightforward to show that $\Psi_m(\xi>0)$
is negative, $\Psi_m(0)=0$ and therefore
\begin{equation}
\frac{d{\cal H}_m}{d\xi}\bigg|_{\xi=0}=0\ .
\end{equation}
Furthermore, we have
\begin{equation}
\begin{split}
&\frac{d^2{\cal H}_m}{d\xi^2}\bigg|_{\xi=0}>0\
\hbox{for $\beta\in(-1/4,1/2)$}\ ,\\
&\frac{d{\cal H}_m}{d\xi}\bigg|_{\xi>0}>0\
\hbox{for $\beta\in(-1/4,1/2)$}\ .
\end{split}
\end{equation}
Therefore, ${\cal H}_m$ increases monotonically with
increasing $\xi$ and attains its minimum value at
$\xi=0$, that is,
\begin{equation}
{\cal H}_{m\_min}={\cal H}_m|_{\xi=0}
={\cal C}{\cal N}_m(0){\cal A}_m(0)+{\cal B}_m\ .
\end{equation}

For the $m=0$ case with radial oscillations,
we should make the following two replacements
\begin{equation}
\begin{split}
&{\cal A}_0\rightarrow{\cal A^{\prime}}_0=\xi^2+1/4\ ,\\
&{\cal H}_0\rightarrow{\cal C}
{\cal N}_0{\cal A^{\prime}}_0+{\cal B}_0\ ,
\end{split}
\end{equation}
and repeat the same procedure for $m>0$ cases. The preceding
results remain valid. In summary, for $m\ge 0$, ${\cal H}_m$
increases monotonically with increasing $\xi$ and attains
its minimum value at $\xi=0$. More specifically, we have
\begin{equation}
\begin{split}
&{\cal H}_{0\_min}={\cal H}_0|_{\xi=0}={\cal C}{\cal
N}_0(0)(1/4)+(1+2\beta)(-2+2\beta)<0\ ,\\
&{\cal H}_{1\_min}={\cal H}_1|_{\xi=0}={\cal C}{\cal
N}_1(0)(5/4+2\beta)+4\beta^2-1>0\ ,\\
&{\cal H}_{m\_min}={\cal H}_m|_{\xi=0}={\cal C}{\cal N}_m(0)
{\cal A}_m(0)+{\cal B}_m>0\ \hbox{for $m\ge 2$\ .}
\end{split}
\end{equation}

\section\\
\begin{figure}
\centering
\includegraphics[scale=0.42]{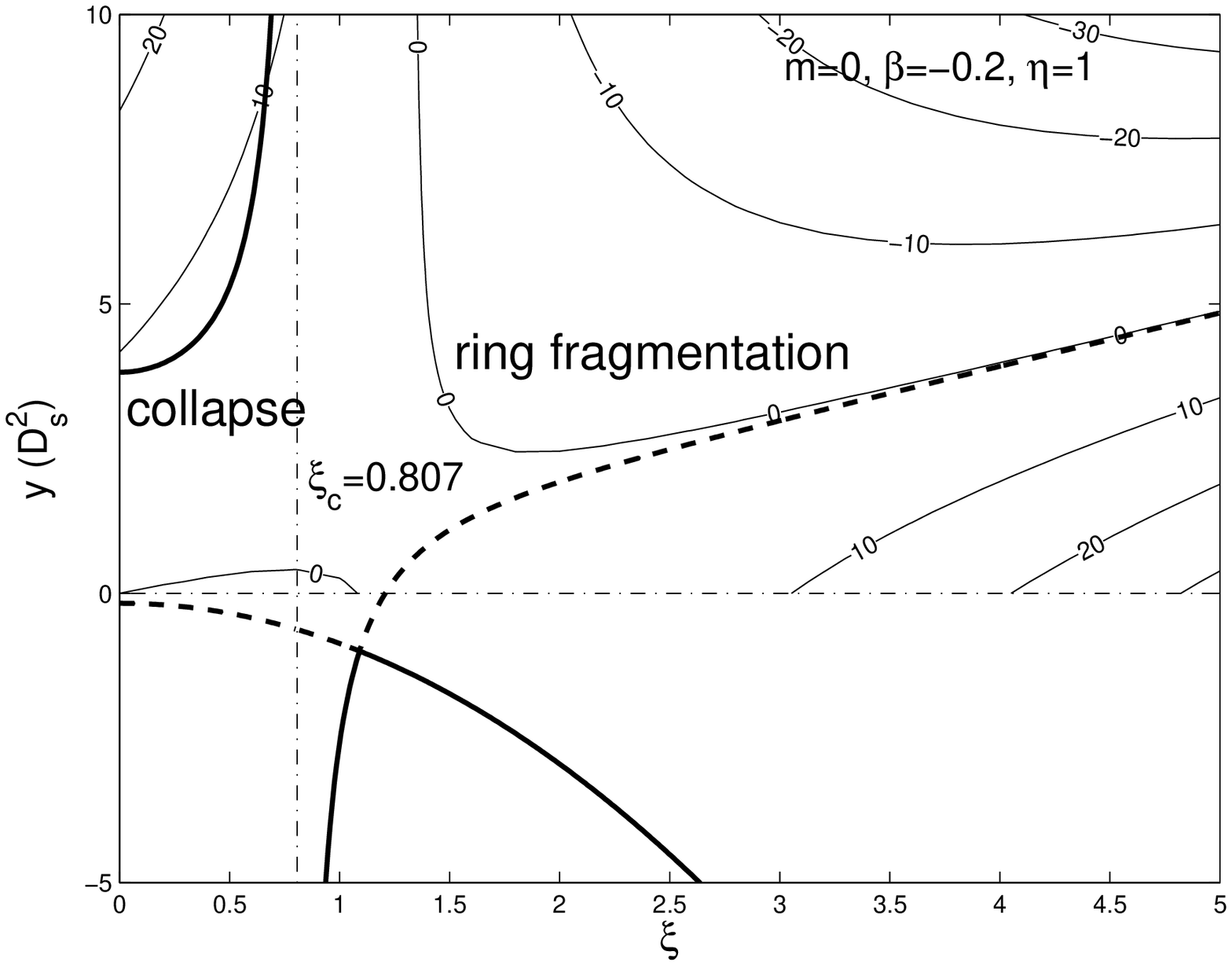}
\caption{Two branches of $D_s^2$ solutions to stationary
dispersion relation (\ref{spiral}) for $m=0$, $\beta=-0.2$ and
$\eta=1$. The divergent point is $\xi_{c}=0.807$. In this case of
$\eta=1$, the value of $\delta$ can be arbitrary. The system is
inevitable to bear ring-fragmentation instabilities and therefore
there is no stable range of $D_s^2$. We also show the deviation
from contours of local WKBJ analysis, which are consistent with
the global analysis at large `radial wavenumber' $\xi$.}
\end{figure}

We here discuss more specifically the $m=0$ case
with radial oscillations, for which we have
\begin{equation}
\begin{split}
&{\cal A^{\prime}}_0=\xi^2+1/4>0\ ,\\
&{\cal B}_0=(1+2\beta)(-2+2\beta)<0\ ,\\
&{\cal C}=(1+2\beta)/(2\beta{\cal P}_0)\ ,\\
&{\cal H}_0={\cal C}{\cal N}_0{\cal A^{\prime}}_0+{\cal B}_0\ .
\end{split}
\end{equation}
For $\eta=1$ (equivalent to the case of a single
disc), the physical $D_s^2$ solution is
\begin{equation}
Y_1^B=\frac{(1-{\cal C}{\cal N}_0)
{\cal A^{\prime}}_0}{{\cal H}_0}>0 \ ;
\end{equation}
the other $D_s^2$ solution is negative and thus unphysical. It
is noted that for $m\ge 1$, we have $0<{\cal C}{\cal N}_m<1$
for all $\xi>0$ and $\beta\in(-1/4,1/2)$. In contrast, the
situation of $m=0$ is different, that is, ${\cal C}{\cal N}_0$
can be either greater or smaller than 1 and this complicates
the analysis.

According to Appendix C, we infer
\begin{equation}
\frac{d({\cal C}{\cal N}_0)}{d\xi}
={\cal C}{\cal N}_0\Psi_0<0\
\qquad\hbox{for }\qquad \xi>0\ ,
\end{equation}
and is equal to zero at $\xi=0$. This implies that
${\cal C}{\cal N}_0$ decreases monotonically with increasing
$\xi$ and reaches the maximum value at $\xi=0$, namely
\begin{equation}
({\cal C}{\cal N}_0)_{max}={\cal C}{\cal N}_0(0)\bigg\{
\begin{array}{l l}
>1\ \qquad\hbox{if $\beta<\beta_c\sim 0.436$}\ , \\
<1\ \qquad\hbox{if $\beta>\beta_c\sim 0.436$}\ .
\end{array}
\end{equation}
Therefore for $\beta<\beta_c\sim 0.436$, there is a critical
$\xi_{c^{\prime}}$ at which $1-{\cal C}{\cal N}_0$ vanishes.
If this $\xi_{c^{\prime}}$ coincides with $\xi_c$ at which
${\cal H}_0$ vanishes, then there is no divergent point for
$Y_1^B$. For this reason, we now check the possibility of
$\xi_{c^{\prime}}=\xi_c$.

If $\xi_{c^{\prime}}=\xi_c$ happens for a specific $\beta_{co}$,
the following two equations must be satisfied simultaneously,
\begin{equation}
\begin{split}
&(1-{\cal C}{\cal N}_0)|_{\xi_{c^{\prime}}=\xi_c}=0\ ,\\
&({\cal C}{\cal N}_0{\cal A^{\prime}}_0
+{\cal B}_0)|_{\xi_{c^{\prime}}=\xi_c}=0\ ,
\end{split}
\end{equation}
which gives the relation of $\xi_c$ in terms of $\beta_{co}$,
\begin{equation}
{\cal A^{\prime}}_0+{\cal B}_0
=\xi_c^2+1/4+(1+2\beta)(-2+2\beta)=0
\end{equation}
and thus
\begin{equation}\label{xic}
\xi_c=(7/4+2\beta-4\beta^2)^{1/2}\ .
\end{equation}
Inserting solution (\ref{xic}) into the first
equation of (D5), we obtain numerically
\begin{equation}
\beta_{co}\sim -0.130,\ \qquad\qquad\xi_{c^{\prime}}=\xi_c\sim 1.193\ .
\end{equation}
From the above analysis, we realize that for
$\beta_{co}<\beta<1/2$, the axisymmetric marginal stability curves
are typical as discussed in subsection 3.2.1. In contrast, for
$-1/4<\beta<\beta_{co}$, the axisymmetric marginal stability curves
are different and there is no stable range of $D_s^2$ against overall
axisymmetric instabilities, a result also implied in Syer \& Tremaine
(1996) (see their fig. 2). To see this more clearly, we consider a
specific case of $\beta=-0.2$. The divergent point of $Y_1^B$ is
located at $\xi_c=0.807$. The stationary configuration for $\eta=1$
(equivalent to the case of a single disc) is shown in Fig. D1.

\end{document}